\documentclass[12pt]{article}
\pdfoutput=1

\usepackage{draft}
\usepackage{hyperref}
\usepackage{graphicx, color, subfig}
\usepackage[nosort]{cite}
\usepackage{mciteplus}
\usepackage{empheq}
\usepackage{anyfontsize}
\usepackage{dsfont}
\usepackage{float}
\usepackage{tabu}
\usepackage{multirow}
\usepackage{hhline}

\def\be#1\ee{\begin{align}#1\end{align}}

\allowdisplaybreaks

\usepackage{silence}
\newsavebox{\ns}
\newsavebox{\dbrane}
\newsavebox{\dbshort}

\def\be{\begin{equation}}
\def\ee{\end{equation}}
\def\bea{\begin{eqnarray}}
\def\eea{\end{eqnarray}}

\newcommand\R{\mathbb{R}}
\newcommand\Z{\mathbb{Z}}

\newlength{\sswidth}

\newcommand\cB{\mathcal{B}}

\newcommand\cV{\mathcal{V}}

\newcommand\cY{\mathcal{Y}}

\usepackage{color}
\usepackage{latexsym}
\usepackage{epsfig, amssymb, euscript, mathrsfs, cite}
\usepackage{arydshln} 
\usepackage{amsmath}
\definecolor{MyDarkBlue}{rgb}{0.15,0.15,0.45}

\newcommand\cX{\mathcal{X}}
\renewcommand\cY{\mathcal{Y}}
\renewcommand\cZ{\mathcal{Z}}

\renewcommand{\tr}{\operatorname{tr}} 
\renewcommand{\Tr}{\operatorname{Tr}} 

\usepackage[all]{xy}
\newcommand{\sh}{\operatorname{sh}}
\newcommand{\ch}{\operatorname{ch}}
\newcommand{\Th}{\operatorname{th}}

\begin{document}

\unitlength = .8mm

\pagenumbering{roman}
\begin{titlepage}

\preprint{PUPT-2605}

\begin{center}

\hfill \\
\vskip 0.5cm

\title{Nonabelian Mirror Symmetry \\ Beyond the Chiral Ring}

\author{Yale Fan$^\ast$ and Yifan Wang$^{\ast, \dag, \ddag}$}

\vspace{-.15in}
\address{\footnotesize $^\ast$Department of Physics, Princeton University, Princeton, NJ 08544, USA}

\vspace{-.15in}
\address{\footnotesize $^\dag$Center of Mathematical Sciences and Applications, Harvard University, Cambridge, MA 02138, USA}

\vspace{-.15in}
\address{\footnotesize $^\ddag$Department of Physics, Harvard University, Cambridge, MA 02138, USA}


\end{center}

\abstract{
Mirror symmetry is a type of infrared duality in 3D quantum field theory that relates the low-energy dynamics of two distinct ultraviolet descriptions. Though first discovered in the supersymmetric context, it has far-reaching implications for understanding nonperturbative physics in general 3D quantum field theories. We study mirror symmetry in 3D $\cN=4$ supersymmetric field theories whose Higgs or Coulomb branches realize $D$- and $E$-type Kleinian singularities in the $ADE$ classification, generalizing previous work on the $A$-type case. Such theories include the $SU(2)$ gauge theory coupled to fundamental matter in the $D$-type case and non-Lagrangian generalizations thereof in the $E$-type case. In these cases, the mirror description is given by a quiver gauge theory of affine $D$- or $E$-type. We investigate the mirror map at the level of the recently identified 1D protected subsector described by topological quantum mechanics, which implements a deformation quantization of the corresponding $ADE$ singularity. We give an explicit dictionary between the monopole operators and their dual mesonic operators in the $D$-type case. Along the way, we extract various operator product expansion (OPE) coefficients for the quantized Higgs and Coulomb branches. We conclude by offering some perspectives on how the topological subsectors of the $E$-type quivers might shed light on their non-Lagrangian duals.
}

\vfill

\end{titlepage}
\pagenumbering{arabic}

\eject

\begingroup
\hypersetup{linkcolor=black}
\tableofcontents
\endgroup

\section{Introduction}

Three-dimensional gauge theories are strongly coupled at low energies due to the positive mass dimension of the Yang-Mills coupling. Consequently, they exhibit a wide range of interesting nonperturbative phenomena, including monopole operators and con\-fine\-ment/de\-con\-fine\-ment transitions. A powerful tool for elucidating complicated gauge dynamics is duality, which states that two distinct ultraviolet (UV) field theory descriptions give rise to the same theory in the deep infrared (IR) \cite{Giombi:2011kc, Aharony:2012nh, Aharony:2012ns, Aharony:2015mjs, Karch:2016sxi, Murugan:2016zal, Seiberg:2016gmd}. A key merit of duality is that there often exists a manifestly weakly coupled dual description. Thus duality provides an efficient language for tackling the problem of strong coupling.

While duality, by definition, requires a map between all observables of the dual quantum field theories, most known dualities are motivated by matching quantities that are insensitive to dynamical details of the theories, such as 't Hooft anomalies.\footnote{For a class of large-$N$ Chern-Simons-matter theories, the relevant dualities have been checked at the level of local correlation functions \cite{Giombi:2011kc, Aharony:2012nh, Aharony:2012ns, Aharony:2015mjs}.} On one hand, this is precisely what makes duality an efficient and elegant way to extract physical information.  On the other hand, this procedure can be misleading in cases where it fails to pinpoint the fate of the renormalization group (RG) flow (see \cite{Intriligator:1994rx, Brodie:1998vv, Intriligator:2005if} for examples in 4D). Fortunately, for a subclass of dualities known as mirror symmetry of supersymmetric gauge theories \cite{Intriligator:1996ex, deBoer:1996mp, deBoer:1996ck, Hanany:1996ie}, we have increased analytic control over the dynamics thanks to supersymmetry even while many features of generic three-dimensional gauge dynamics remain.

Supersymmetry, especially the localization method, allows us to extract nontrivial dynamical data from quantum field theories, such as their protected operator spectrum, low-energy effective action, and supersymmetric partition functions, which all play important roles in testing and refining the duality maps. Once a supersymmetric dual pair passes such tests, one can consider supersymmetry-breaking deformations to generate a larger class of dualities. Indeed, many recently formulated 3D dualities are motivated by mirror symmetry and supported by SUSY-breaking deformations thereof, including the important bosonization dualities of \cite{Kachru:2016rui, Kachru:2016aon}.

Much recent progress has been made in better understanding 3D gauge theories with $\cN=4$ supersymmetry, which is the original context in which mirror symmetry was discovered \cite{Intriligator:1996ex}.  These theories are characterized by an $SU(2)_H\times SU(2)_C$ R-symmetry, and they have a vacuum moduli space consisting of (singular) hyperk\"ahler manifolds that can be labeled as Coulomb branch $\cM_C$, Higgs branch $\cM_H$, or mixed branches, depending on which combination of R-symmetries is broken. The Coulomb and Higgs branches appear very different at first sight: the Coulomb and mixed branches embody the complicated dynamics of 3D gauge theories, whereas the Higgs branch is protected by supersymmetric non-renormalization theorems and thus has a rigid structure \cite{Intriligator:1996ex}. The nontriviality of mirror symmetry amounts to the statement that there exist mirror-dual pairs of 3D $\cN=4$ theories where the roles of the Coulomb and Higgs branches, as well as classical and quantum effects, are interchanged. A particularly simple example is that of 3D $\cN=4$ $U(1)$ super-QED with one charged hypermultiplet, which is dual to a free hypermultiplet. For suitable matter content, a gauge theory can flow to a superconformal fixed point in the IR whose operator spectrum naturally has a description in terms of the elementary degrees of freedom in the UV gauge theory. In this case, mirror symmetry amounts to interchanging the descriptions of the CFT operators. Namely, under mirror symmetry, an order-type \textit{mesonic} operator written in terms of the fundamental fields in one UV description is mapped to a disorder-type operator such as a \textit{monopole} in the dual description. A particular class of operators in the SCFT is that of chiral ring operators, which are half-BPS and whose vacuum expectation values give rise to the Coulomb branch, Higgs branch, and mixed branches. The matching of the chiral rings (equivalently, the moduli spaces)  \cite{Cremonesi:2013lqa,Cremonesi:2014kwa,Hanany:2016pfm,Hanany:2016gbz,Hanany:2017ooe,Hanany:2018xth} provides a first check for mirror duality proposals beyond anomaly matching.

A more refined protected subsector in 3D $\cN=4$ SCFTs was discovered in \cite{Chester:2014mea, Beem:2016cbd}.  It is described by a one-dimensional topological quantum mechanics (TQM) associated to either the Higgs or Coulomb branch.\footnote{A variation of the $\Omega$-background \cite{Nekrasov:2002qd,Nekrasov:2003rj,Nekrasov:2010ka} leads to related deformation quantizations of the Higgs and Coulomb branches \cite{Yagi:2014toa,Bullimore:2015lsa,Bullimore:2016hdc}. It would be interesting to spell out the explicit relation to the TQM sector.} The relevant operators are twisted translations of the Higgs (resp.\ Coulomb) branch chiral primaries by $SU(2)_H$ (resp.\ $SU(2)_C$) R-symmetry rotations along a line in $\mR^3$.  Their correlation functions depend only on the ordering of the insertions. The TQM contains nontrivial information about the operator product expansion (OPE) data of the full SCFT, which can be computed systematically from supersymmetric localization after mapping the TQM to a great $S^1$ on $S^3$ \cite{Dedushenko:2016jxl, Dedushenko:2017avn, Dedushenko:2018icp}, and plays an important role in determining the full OPE data of the SCFT using the conformal bootstrap technique  \cite{Chester:2014mea, Chester:2014fya, Agmon:2017xes, Agmon:2019imm, Chang:2019dzt}. In recent work \cite{Etingof:2019guc}, these TQMs are formalized as noncommutative associative algebras equipped with an \textit{even} and \textit{positive} short star product --- equivalently, a (twisted) trace or bilinear form. The latter is essential for mapping the TQM data to CFT correlators. The action of mirror symmetry in the TQM sectors has been studied to a limited extent in \cite{Dedushenko:2016jxl, Dedushenko:2017avn, Dedushenko:2018icp}, focusing mainly on the case where the corresponding SCFTs arise from abelian gauge theories such as SQED and abelian quivers.\footnote{Specifically, the abelian $A$-type mirror symmetries were analyzed in \cite{Dedushenko:2017avn}, and a simple $\mathcal{N} = 8$ nonabelian $A$-type mirror symmetry was analyzed in \cite{Dedushenko:2018icp}.  The $D_3$ case was also discussed in Appendix F.2 of \cite{Dedushenko:2018icp}, but this belongs to the $A$-series ($D_3\cong A_3$).} The bootstrap analysis for the particular case of SQED$_2$, or equivalently the $T[SU(2)]$ theory, was carried out in \cite{Chang:2019dzt}, where nontrivial evidence for the (self-)mirror symmetry beyond the TQM sector was found.

In this paper, we initiate the systematic study of nonabelian mirror symmetry in the TQM sectors of 3D $\cN=4$ SCFTs.  Beautifully, the most well-studied examples of mirror symmetry fall into an $ADE$ classification \cite{Intriligator:1996ex}, with those of $A$- and $D$-type admitting higher-rank generalizations \cite{deBoer:1996mp, deBoer:1996ck}.  We focus on the simple class of theories $\cT_\mf{g}$ that have rank-one Coulomb branches given by the $ADE$ singularities
\ie
\cM_{\rm C}(\cT_\mf{g})={\mC^2/\Gamma_{\mf g}}
\label{ADECB}
\fe
where $\mf g$ labels an $ADE$ Lie algebra and $\Gamma_{\mf g}$ is the corresponding discrete subgroup of $SU(2)$ under the McKay correspondence, which can equivalently be represented as a hypersurface singularity in $\mC^3$ (see Table~\ref{table:ade}). The latter description makes explicit the Coulomb branch chiral ring of the theory $\cT_\mf{g}$, which is nothing but the coordinate ring of the hypersurface singularity.

\begin{table}[htb]
	\centering
	\begin{tabular}{|c|c|c|}
		\hline
		$\mf g$ & $\Gamma_{\mf g}$ & $f_{\mf g}(X,Y,Z)$
		\\\hline
		$A_{n-1}$ & $\mZ_{n}$ & $X^2+Y^2+Z^n$
		\\\hline
		$D_{n+1}$ & $Q_{4(n+1)}$ & $X^2+ZY^2+Z^n$
		\\\hline
		$E_6$ & $2T$ &$X^2+Y^3+Z^4$
		\\\hline
		$E_7$ & $2O$ & $X^2+Y^3+YZ^3$
		\\\hline
		$E_8$ & $2I$ &$X^2+Y^3+Z^5$
		\\\hline
	\end{tabular}
	\caption{$ADE$ Lie algebras along with their corresponding $SU(2)$ discrete subgroups and hypersurface singularities. Here, $Q_{4 n}$ and $2T,2O,2I$ are $SU(2)$ lifts of the familiar dihedral ($D_{2n}$), tetrahedral ($A_4$), octahedral ($S_4$), and icosahedral ($A_5$) subgroups of $SO(3)$.}
	\label{table:ade}
\end{table}

From the quotient structure of the Coulomb branch, it is obvious that the free 3D $\cN=4$ theory with a single twisted hyper in which the discrete symmetry $\Gamma_{\mf g}\subset SU(2)_F$ is gauged realizes this Coulomb branch as its vacuum moduli space (similarly for its mirror in terms of a free hypermultiplet). A more interesting theory with Coulomb branch \eqref{ADECB} in the $A_{n-1}$ case is super-QED with $n$ hypermultiplets of unit charge, which we denote by SQED$_n$. Similarly, for the $D_n$ case, an interacting theory with Coulomb branch \eqref{ADECB} is $SU(2)$ SQCD with $n$ fundamental hypermultiplets. The exceptional cases of \eqref{ADECB} do not appear to have gauge theory realizations: however, there are mirror dual descriptions which instead realize \eqref{ADECB} as their Higgs branch
\ie
\cM_{\rm H}(\cT_\mf{g}^{\rm mirror})={\mC^2/\Gamma_{\mf g}}.
\label{ADECBm}
\fe
In general, they are given by 3D $\cN=4$ quiver gauge theories of affine $ADE$ type. The associative algebras associated to the $ADE$ singularities are in general given by the spherical symplectic reflection algebras of complex dimension two \cite{Etingof:2019guc}. This $ADE$ series of theories $\cT_\mf{g}$ (resp.\ $\cT_\mf{g}^{\rm mirror}$) has, in addition, a Higgs branch (resp.\ Coulomb branch) given by 
\ie
\cM_{\rm H}(\cT_\mf{g})=\cM_{\rm C}(\cT_\mf{g}^{\rm mirror})={\overline\cO_{\rm min}(\mf g)}
\label{ADEHB}
\fe
where ${\cO_{\rm min}(\mf g)}$ denotes the  minimal nilpotent orbit of $\mf{g}$. This Higgs branch has a rigid structure in the TQM sector thanks to the $\mf{g}$ flavor symmetry. The $\mf{g}$-equivariant deformation quantization of these  hyperk\"ahler cones was solved in \cite{Chang:2019dzt}, where unique star products were obtained except in the case of $A_1 = \mathfrak{sl}_2$ \cite{Beem:2016cbd}.  Returning to the Coulomb branches in \eqref{ADECB}, the deformation quantization of the $A_{n-1}$ case was studied in \cite{Beem:2016cbd}, where the extra $U(1)$ flavor symmetry played an important role in simplifying the analysis. The mirror symmetry between SQED and the cyclic abelian quiver was then spelled out in \cite{Dedushenko:2017avn}.

The primary goal of this paper is to carry out the analysis of mirror symmetry at the level of the TQM for the $D$- and $E$-type cases, where the relevant (mirror) gauge theories are nonabelian. In these cases, there are no continuous global symmetries at our disposal, although there do exist discrete $\mZ_2$ symmetries for $D_n$ (enhanced to $S_3$ for $D_4$) and $E_6$, which still place some constraints on the TQM.  
We start by solving the algebraic problem of deformation quantization  for $D_n$ singularities. We then compute the correlators in the TQM from both the SQCD description and the affine $D$-type quiver description. By studying the explicit form of the matrix models and insertions that are obtained from supersymmetric localization, we establish the precise mirror map for the operators in the TQM that preserves the short star product. For the $E$-type cases, we solve the deformation quantization of the $E_{6,7,8}$ singularities and present some preliminary observations from the affine quiver side, leaving a complete analysis to future work. We also include results for free theories that realize \eqref{ADEHB} via discrete gauging.

Here is an outline of the rest of the paper. We start by providing some relevant background on TQM sectors in 3D $\cN=4$ theories in Section \ref{tqmreview}. We then give a brief review of mirror symmetry for the abelian $A$-type theories in Section \ref{areview}. We move on to deformation quantizations of $D$-type singularities in Section \ref{defdn} and explain how they are realized in 3D $\cN=4$ gauge theories. In Section \ref{dmirror}, by explicitly computing TQM correlators, we infer the mirror map for TQM operators that quantize the $D_n$ singularity: our results are summarized in \eqref{mirrormapXandYandZ} for $n > 4$ and \eqref{mirrormap} for $n = 4$. We carry out a similar deformation quantization of $E$-type singularities in Section \ref{defen} and, in Section \ref{emirror}, present some motivating remarks toward understanding the non-Lagrangian theories whose Coulomb branches realize \eqref{ADECB} with $\mathfrak{g} = E_n$ through the lens of the TQM. Some details of our Higgs branch TQM computations, which tend to be more convoluted than their Coulomb branch counterparts, are gathered in Appendix \ref{detailstqm}.  In Appendix \ref{ADcoulomb}, we consider additional quantizations of the $A$- and $D$-type singularities via field theory (outside the context of mirror symmetry), generalizing some examples from \cite{Dedushenko:2017avn, Dedushenko:2018icp}. In Appendix \ref{DEhiggs}, we give a self-contained exposition of the Higgs branch chiral rings of the affine $D$- and $E$-type quivers, filling some gaps in the literature.

\paragraph{Notation.}

Throughout this paper, we adhere to the following notational conventions:
\begin{itemize}
\item Straight $O$ denotes an abstract chiral ring generator.
\item Hatted $\hat{O}$ denotes an abstract quantum algebra generator.
\item Curly $\mathcal{O}$ denotes the realization of $\hat{O}$ as an SCFT operator (with suitable mixing).
\end{itemize}
We also introduce the shorthand
\begin{equation}
\sh(x)\equiv 2\sinh(\pi x), \quad \ch(x)\equiv 2\cosh(\pi x), \quad \Th(x)\equiv \frac{\sh(x)}{\ch(x)}.
\end{equation}

\section{Topological Quantum Mechanics} \label{tqmreview}

In this section, we briefly review the prescriptions of \cite{Dedushenko:2016jxl, Dedushenko:2017avn, Dedushenko:2018icp} for computing observables within certain protected operator algebras of 3D $\mathcal{N} = 4$ theories on the sphere.\footnote{See \cite{Gaiotto:2019mmf} for a complementary perspective on these protected correlation functions.}  Combining these formalisms gives a way to derive precise maps between half-BPS operators across nonabelian 3D mirror symmetry, and to compute previously unknown quantizations of Higgs and Coulomb branch chiral rings.

We consider 3D $\cN=4$ gauge theories of cotangent type, namely with gauge group $G$ and matter representation $R\oplus \overline{R}$. We denote by $\mathfrak{g}$ the Lie algebra of $G$, $\mathfrak{t}$ a fixed Cartan subalgebra of $\mathfrak{g}$, $\cW$ the Weyl group, $\Delta$ the set of roots, $\Lambda_W$ the weight lattice, and $\Lambda^\vee_W$ the coweight lattice.

3D $\mathcal{N} = 4$ SCFTs have two one-dimensional protected subsectors that each take the form of a topological quantum mechanics (TQM) \cite{Chester:2014mea, Beem:2016cbd}.  The associative operator algebra of the TQM is a deformation quantization of either the Higgs or Coulomb branch chiral ring, and as such, it encodes detailed information about the geometry of the vacuum manifold. When the SCFT arises from an RG flow with a Lagrangian description in the UV, the Higgs branch sector is directly accessible by supersymmetric localization \cite{Dedushenko:2016jxl}, but the Coulomb branch sector includes monopole operators, which are disorder operators that cannot be represented in terms of the Lagrangian fields \cite{Dedushenko:2017avn, Dedushenko:2018icp}.  For this reason, the known methods for computing OPE data within these two sectors look qualitatively different.

Each 1D sector can be described as the equivariant cohomology of an appropriate supercharge.  The corresponding cohomology classes are called \emph{twisted} Higgs or Coulomb branch operators (HBOs or CBOs).\footnote{Mixed-branch operators are not in the cohomology of either supercharge.}  They are realized as Higgs or Coulomb branch chiral ring operators which, when translated along a chosen line in $\mathbb{R}^3$ or a chosen great circle $S_\varphi^1$ on $S^3$, are simultaneously twisted by $SU(2)_H$ or $SU(2)_C$ rotations.  The OPE within each sector takes the form of a noncommutative star product
\begin{equation}
\mathcal{O}_i\star \mathcal{O}_j = \sum_k \zeta^{\Delta_i+\Delta_j-\Delta_k}c_{ij}{}^k\mathcal{O}_k
\label{starproduct}
\end{equation}
where, for theories placed on $S^3$, the quantization parameter $\zeta$ is the inverse radius of the sphere: $\zeta = 1/r$.  In addition to associativity, the star product inherits several conditions from the physical SCFT, namely \cite{Beem:2016cbd}: truncation or \textit{shortness} (the sum in \eqref{starproduct} terminates after the term of order $\zeta^{2\min(\Delta_i, \Delta_j)}$) due to the $SU(2)_H$ or $SU(2)_C$ selection rule, \textit{evenness} (swapping $\mathcal{O}_i$ and $\mathcal{O}_j$ in \eqref{starproduct} takes $\zeta\to -\zeta$) inherited from the symmetry properties of the 3D OPE, and \textit{positivity} from unitarity (reflection positivity) of the 3D SCFT.

\subsection{Higgs Branch Formalism}

Assuming a UV Lagrangian, the operators that comprise the Higgs branch topological sector are gauge-invariant polynomials in antiperiodic scalars $Q(\varphi), \tilde{Q}(\varphi)$ on $S^1_\varphi$, which are twisted versions of the hypermultiplet scalars $q_a, \tilde{q}_a$ transforming in the fundamental of $\mathfrak{su}(2)_H$ and in $R, \overline{R}$ of $G$.  The correlation functions of these twisted HBOs $\mathcal{O}_i(\varphi)$ can be computed within a 1D Gaussian theory \cite{Dedushenko:2016jxl} with path integral
\begin{equation}
Z_\sigma \equiv \int DQ\, D\tilde{Q}\, \exp \left[ 4 \pi r \int d\varphi\, \tilde{Q} (\partial_\varphi + \sigma) Q \right],
\label{ZsigmaDef}
\end{equation}
in terms of which the $S^3$ partition function is
\begin{equation}
Z_{S^3} = \frac{1}{|\cW|}\int_{\mathfrak{t}}\, d\mu(\sigma), \quad d\mu(\sigma) \equiv d\sigma \, \operatorname{det}'_\text{adj} (\sh(\sigma))\, Z_\sigma = d\sigma\, \frac{\operatorname{det}'_\text{adj} (\sh(\sigma))}{\operatorname{det}_R (\ch(\sigma))}.
\end{equation}
Namely, an $n$-point correlation function $\langle {\cal O}_1(\varphi_1) \cdots {\cal O}_n (\varphi_n) \rangle$ on $S^3$ can be written as
\begin{equation}
\langle {\cal O}_1(\varphi_1) \cdots {\cal O}_n (\varphi_n) \rangle = \frac{1}{|\cW|Z_{S^3}} \int_{\mathfrak{t}}\, d\mu(\sigma) \, \langle {\cal O}_1(\varphi_1) \cdots {\cal O}_n (\varphi_n) \rangle_\sigma
\end{equation}
in terms of an auxiliary correlator $\langle {\cal O}_1(\varphi_1) \cdots {\cal O}_n (\varphi_n) \rangle_\sigma$ at fixed $\sigma$.  The latter is computed via Wick contractions with the 1D propagator
\begin{equation}
\langle Q(\varphi_1) \tilde{Q}(\varphi_2) \rangle_\sigma \equiv G_\sigma(\varphi_{12}) \equiv -\frac{\operatorname{sgn}\varphi_{12} + \Th(\sigma)}{8 \pi r}e^{-\sigma\varphi_{12}}, \quad \varphi_{12} \equiv \varphi_1 - \varphi_2,
\end{equation}
derived from \eqref{ZsigmaDef}.\footnote{Wick contractions between elementary operators at coincident points are performed using
\begin{equation}
\langle Q(\varphi) \tilde{Q}(\varphi) \rangle_\sigma \equiv G_\sigma(0) = -\frac{\Th(\sigma)}{8 \pi r}
\end{equation}
to resolve normal-ordering ambiguities.}

\subsection{Coulomb Branch Formalism} \label{CBformalism}

The operators in the Coulomb branch topological sector, in terms of a UV gauge theory Lagrangian, consist of a scalar $\Phi(\varphi)$ (a twisted combination of the vector multiplet scalars $\Phi_{\dot{a}\dot{b}}$ transforming in the adjoint of $\mathfrak{su}(2)_C$ and of $G$), bare monopoles $\mathcal{M}^b(\varphi)$, and dressed monopoles $P(\Phi)\mathcal{M}^b(\varphi)$.  The coweight $b$ breaks the gauge group at the insertion point to $G_b$, the centralizer of $b$, and the corresponding monopole may be dressed by a $G_b$-invariant polynomial $P(\Phi)$ in $\Phi(\varphi)$ \cite{Bullimore:2015lsa}.

In \cite{Dedushenko:2017avn, Dedushenko:2018icp}, a method for computing all observables within the Coulomb branch TQM was obtained for 3D $\mathcal{N} = 4$ gauge theories of cotangent type by constructing a set of ``shift operators,'' acting on functions of $\sigma\in \mathfrak{t}$ and $B\in\Lambda_W^\vee$, whose algebra is a representation of the 1D OPE.\footnote{All expressions are given in the ``North'' picture.  See \cite{Dedushenko:2017avn, Dedushenko:2018icp} for details.}  We find that $\Phi(\varphi)$ is represented by a simple multiplication operator
\begin{equation}
\Phi=\frac{1}{r}\left(\sigma + \frac{i}{2}B \right)\in\mathfrak{t}_{\mathbb{C}}=\mathfrak{t}\otimes\mathbb{C}.
\end{equation}
The shift operator describing a dressed monopole has a more intricate definition: it is constructed as
\begin{equation}
P(\Phi)\cM^b = \frac{1}{|\cW_b|}\sum_{w\in\cW} P(w^{-1}\cdot\Phi)\widetilde{M}^{w\cdot b}
\end{equation}
where $\cW_b$ is the stabilizer of $b$ in $\cW$, with the Weyl sum reflecting the fact that a physical magnetic charge is labeled by the Weyl orbit of a coweight $b$.  For a given coweight $b$, we define the abelianized (non-Weyl-averaged) monopole shift operator
\begin{equation}
\widetilde{M}^b = M^b + \sum_{|v|< |b|}Z^\text{ab}_{b\to v}(\Phi)M^v,
\end{equation}
where the sum is taken over all coweights shorter than $b$ and the rational functions $Z^\text{ab}_{b\to v}(\Phi)$, dubbed \emph{abelianized bubbling coefficients} in \cite{Dedushenko:2018icp}, account for nonperturbative effects in nonabelian gauge theories in which the GNO charge of a singular monopole is screened away from the insertion point by smooth monopoles of vanishing size \cite{Kapustin:2006pk}.\footnote{It was proposed in \cite{Dedushenko:2018icp} that the abelianized bubbling coefficients are fixed by algebraic consistency of the OPE within the Coulomb branch topological sector.}  Finally, $M^b$ is an abelianized monopole shift operator that represents a bare monopole singularity in the absence of monopole bubbling:
\begin{equation}
M^b = \frac{\prod_{\rho\in R}\left[\frac{(-1)^{(\rho\cdot b)_+}}{r^{|\rho\cdot b|/2}} \left(\frac{1}{2} + ir\rho\cdot \Phi\right)_{(\rho\cdot b)_+} \right]}{\prod_{\alpha\in\Delta}\left[\frac{(-1)^{(\alpha\cdot b)_+}}{r^{|\alpha\cdot b|/2}} \left(ir\alpha\cdot \Phi\right)_{(\alpha\cdot b)_+} \right]} e^{-b\cdot(\frac{i}2 \partial_\sigma +\partial_B)},
\end{equation}
where $(x)_+\equiv \max(x, 0)$, $(x)_n\equiv \Gamma(x+n)/\Gamma(x)$, and powers of $r$ encode scaling dimensions.

With the above shift operators in hand, the $S^3$ correlator of twisted CBOs $\cO_i(\varphi_i)$, inserted at points $\varphi_i$ along $S^1_\varphi$ with $0 < \varphi_1 < \cdots < \varphi_n < \pi$, can be computed as
\begin{equation}
\langle\mathcal{O}_1(\varphi_1)\cdots \mathcal{O}_n(\varphi_n)\rangle_{S^3} = \frac{1}{|\mathcal{W}|Z_{S^3}}\sum_B\int d\sigma\, \mu(\sigma, B)\Psi_0(\sigma, B)\mathcal{O}_1\cdots \mathcal{O}_n\Psi_0(\sigma, B)
\label{CBmatrixmodel}
\end{equation}
where the operators on the right are understood to be the shift operators corresponding to $\cO_i$ and $\langle 1\rangle_{S^3} = 1$.  Above, we have introduced the empty hemisphere wavefunction
\begin{equation}
\Psi_0(\sigma, B)\equiv \delta_{B, 0}\frac{\prod_{\rho\in R} \frac{1}{\sqrt{2\pi}}\Gamma(\frac{1}{2} - i\rho\cdot \sigma)}{\prod_{\alpha\in \Delta} \frac{1}{\sqrt{2\pi}}\Gamma(1 - i\alpha\cdot \sigma)}
\end{equation}
as well as the gluing measure
\begin{equation}
\mu(\sigma, B) = \prod_{\alpha\in\Delta^+}(-1)^{\alpha\cdot B}\left[\left( \frac{\alpha\cdot \sigma}{r} \right)^2 + \left( \frac{\alpha\cdot B}{2r} \right)^2\right]
 \prod_{\rho\in R} (-1)^{\frac{|\rho\cdot B|-\rho\cdot B}{2}} \frac{\Gamma\left(\frac12 + i\rho\cdot\sigma +\frac{|\rho\cdot B|}{2} \right)}{\Gamma\left(\frac12 - i\rho\cdot\sigma +\frac{|\rho\cdot B|}{2} \right)}.
\label{gluingmeasure}
\end{equation}
While the matrix model \eqref{CBmatrixmodel} converges only for theories with a sufficiently large matter representation (i.e., ``good'' and ``ugly'' theories \cite{Gaiotto:2008ak}), the shift operators can always be used to compute star products in the Coulomb branch TQM.

Finally, in the commutative limit $r\to\infty$, the algebra of shift operators reduces to the Coulomb branch chiral ring and we recover the abelianization description of the Coulomb branch proposed in \cite{Bullimore:2015lsa}.  In this limit, the operators $e^{-b\cdot (\frac{i}{2}\partial_\sigma + \partial_B)}$ turn into generators $e[b]$ of the group ring $\mathbb{C}[\Lambda_W^\vee]$, which act trivially on functions of $\Phi$ but satisfy the relations
\begin{equation}
e[b_1]e[b_2]=e[b_1+b_2].
\end{equation}
We find that $M^b$ itself has a well-defined $r\to\infty$ limit,\footnote{An important caveat is that the expression \eqref{commlimit} holds for semisimple $G$.  Otherwise, it would have some residual $r$-dependence. \label{caveat}}
\begin{equation}
\lim_{r\to\infty} M^b\equiv M^b_\infty=\frac{\prod_{\rho\in R} \left(-i\rho\cdot \Phi \right)^{(\rho\cdot b)_+}}{\prod_{\alpha\in\Delta} \left(-i \alpha\cdot \Phi \right)^{(\alpha\cdot b)_+}} e[b],
\label{commlimit}
\end{equation}
as do the abelianized bubbling coefficients $Z_{b\to v}^{\rm ab}(\Phi)$.

\section{Review: \texorpdfstring{$A$}{A}-Series} \label{areview}

\subsection{Deformation Quantization of \texorpdfstring{$\mC^2/\Gamma_{A_n}$}{An}}

We begin by reviewing the deformation quantization of $A_{n-1}$ singularities, independently of quantum field theory realizations. See \cite{Beem:2016cbd} for discussions of $A_{1, 2, 3}$ and \cite{Dedushenko:2017avn} for discussions of $A_n$.

For general $A_{n-1}$ singularities defined as
\ie
\cM_{A_{n-1}}:~f(X,Y,Z)=X^2+ Y^2+Z^{n}=0,
\fe
the coordinate ring together with the holomorphic symplectic two-form
\ie
\omega={dX\wedge dY \wedge dZ\over df}
\label{hs2f}
\fe
gives rise to a $\mZ_{\geq 0}$-graded Poisson algebra where the generators have degrees
\ie
{\rm deg}(X,Y,Z)=(n,n,2)
\fe
and the Poisson bracket (equivalently, $\omega$) has degree
\ie
{\rm deg}(\omega)=2.
\fe
The (filtered) deformation quantization of this graded Poisson algebra  is easy to work out (see \cite{Smith80aclass}).
  The quantum algebra is given by the central quotient 
\ie
\cA_{A_{n-1}}={\mC[\hat X, \hat Y,\hat Z]\over \la \Omega_{A_{n-1}} \ra }
\fe
where the noncommutative algebra $\mC[\hat X, \hat Y,\hat Z]$ is defined by the commutators
\ie
&[\hat X,\hat Y]=i\zeta P(\hat Z),
\\
&[\hat X,\hat Z]=2i\zeta \hat Y,
\\
&[\hat Y,\hat Z]=-2i\zeta  \hat X,
\fe
the deformation parameter $\zeta$ has degree
\ie
{\rm deg}(\zeta)=2,
\fe
and the center is generated by
\ie
\Omega_{A_{n-1}}=
Q(\hat Z + 2\zeta) + Q(\hat Z) - 2(\hat X^2 + \hat Y^2).
\fe
Physically, $\Omega_{A_{n-1}}(\hat X, \hat Y, \hat Z)=0$ is the quantum chiral ring relation in the TQM.

Here, $P(t)$ and $Q(t)$ are polynomials of degree $n-1$ and $n$ with leading terms $n t^{n-1}$ and $t^{n}$, respectively.  They satisfy
\ie
Q(\hat Z + 2\zeta) - Q(\hat Z) = 2\zeta P(\hat Z).
\fe
Thus $Q(t)$ is fixed by $P(t)$ except for the constant term. Expanding $P(t)$ as
\ie
P(t)=n t^{n-1}+\sum_{i=1}^{n-1} \A_i \zeta^{n-i} t^{i-1}
\fe
and denoting the constant term of $Q(t)$ by $\A_0$, we see that the space of quantizations of the $A_{n-1}$ singularity is $n$-dimensional and parametrized by $\{\A_0, \A_1,\dots, \A_{n-1}\}$.

Imposing the evenness condition amounts to picking out terms in $P(t)$ that have even degree in $\zeta$.  Thus we end up with a space of even quantizations $\cA_{A_{n-1}}$ of dimension $\lfloor {n\over 2} \rfloor$.

\subsection{\texorpdfstring{$A$}{A}-Type Mirror Symmetry}

A detailed TQM analysis of the abelian mirror duality between the affine $A_{N - 1}$ quiver gauge theory and SQED$_N$ was given in \cite{Dedushenko:2017avn}.  Here, we summarize the results for the ``rank-one'' side of this duality, namely that between the Higgs branch of the former theory and the Coulomb branch of the latter.

We denote by $SU(2)_R$ the relevant $SU(2)$ R-symmetry (either for the Higgs or Coulomb branch) of the TQM sector. The degree of an element $\cO$ in the quantum algebra is related to the R-symmetry spin (taking values in half-integers) by
\ie
R(\cO)={1\over 2}{\rm deg}(\cO),
\fe
since the holomorphic symplectic form $\omega$ must transform as an $SU(2)_R$ triplet (corresponding to the three independent complex structures of the hyperk\"ahler cone). 
Moreover, superconformal representation theory requires that the scaling dimensions of the corresponding operators satisfy
\ie
\Delta(\cO)=R(\cO).
\fe 
From this, we conclude that $\hat X,\hat Y,\hat Z$ must be associated to chiral ring operators of dimension ${n\over 2},{n\over 2},1$, respectively, in the physical theory. Below, we give their explicit realizations in terms of mesonic and monopole operators in 3D $\cN=4$ theories.

The necklace quiver gauge theory has gauge group $U(1)^N/U(1)$, bifundamental hypermultiplets $(Q_I, \tilde{Q}_I)$ for $I = 1, \ldots, N$, and Higgs branch $\mathbb{C}^2 / \Z_N$.  The Higgs branch chiral ring generators are
\begin{equation}
{\cal X} = Q_1 Q_2 \cdots Q_N, \quad {\cal Y} = \tilde{Q}_1 \tilde{Q}_2 \cdots \tilde{Q}_N, \quad {\cal Z} = \tilde{Q}_1 Q_1 = \cdots = \tilde{Q}_N Q_N.
\end{equation}
On the other hand, the Coulomb branch TQM operators of SQED$_N$ are products of the twisted vector multiplet scalar $\Phi$ and monopole operators of charge $b \in \Z$.  The corresponding shift operators act on functions of $\sigma \in \R$ and $B \in \Z$.\footnote{See Appendix \ref{U1coulomb} for details and generalizations.}  The Coulomb branch of this theory is also isomorphic to $\mathbb{C}^2 / \Z_N$, and its chiral ring is generated by
\begin{equation}
{\cal X} = \frac{1}{(4 \pi)^{N/2} } {\cal M}^{-1}, \quad {\cal Y} = \frac{1}{(4 \pi)^{N/2} }  {\cal M}^{1}, \quad {\cal Z} = -\frac{i}{4 \pi} \Phi.
\end{equation}
On either side of the duality, the above operators obey ${\cal X} \star {\cal Y} = {\cal Z}^N + O(1/r)$, have identical correlation functions, and generate all other gauge-invariant operators in the corresponding TQM.  Correlation functions of composite operators can also be matched using the OPE.

In this example, the $U(1)_\text{top}$ symmetry prohibits operator mixing and thus we have unambiguous identifications
\ie
\cX=\hat X, \quad \cY=\hat Y, \quad \cZ=\hat Z
\fe
in both mirror-dual descriptions. This simplifying feature will no longer be present in the $D$ case.

\section{Deformation Quantization of \texorpdfstring{$\mC^2/\Gamma_{D_n}$}{Dn}} \label{defdn}

For general $D_{n+1}$ singularities defined as
\ie
\cM_{D_{n+1}}:~f(X,Y,Z)=X^2+ ZY^2+Z^n=0,
\fe
the degrees of the generators are given by
\ie
{\rm deg}(X,Y,Z) =(2n, 2n-2,4).
\fe
The deformation quantization is again easy to work out (see \cite{Levy}).  The quantum algebra is given by the central quotient 
\ie
\cA_{D_{n+1}}={\mC[\hat X, \hat Y,\hat Z]\over \la \Omega_{D_{n+1}} \ra }
\fe
where the algebra $\mC[\hat X, \hat Y,\hat Z]$ is defined by the commutators
\ie
&[\hat X,\hat Y]=\zeta \hat Y^2+\zeta P(\hat Z),
\\
&[\hat X,\hat Z]=-2\zeta \hat Z \hat Y -2\zeta^2 \hat X+(-1)^{n+1}\C \zeta^{n+2},
\\
&[\hat Y,\hat Z]=2\zeta \hat X,
\label{DtypeDQ}
\fe
and the center is generated by
\ie
\Omega_{D_{n+1}}=
Q(\hat Z)+\hat X^2+\hat Z \hat Y^2+2\zeta \hat X \hat Y+(-1)^n\C \zeta^{n+1} \hat Y.
\fe
Here, $P(t)$ and $Q(t)$ are polynomials of degree $n-1$ and $n$ with leading terms $nt^{n-1}$ and $t^n$, respectively.  They satisfy
\ie
Q(-t(t/\zeta^2-1))-Q(-t(t/\zeta^2+1))=(t-\zeta^2)P(-t(t/\zeta^2-1))+(t+\zeta^2)P(-t(t/\zeta^2+1)).
\fe
Thus $Q(t)$ is fixed by $P(t)$ except for the constant term. Expanding $P(t)$ as
\ie
P(t)=n t^{n-1}+\sum_{i=1}^{n-1} \A_i \zeta^{2(n-i)} t^{i-1}
\fe
and denoting the constant term of $Q(t)$ by $\A_0$, we see that the space of quantizations of the $D_{n+1}$ singularity is ($n+1$)-dimensional and parametrized by $\{\A_0, \A_1,\dots, \A_{n-1},\C \}$. Imposing the evenness condition, we see that $\C=0$ for $n$ even and is unconstrained for $n$ odd. Thus we conclude that the space of even quantizations for $D_{n+1}$ singularities is $n$-dimensional for $n$ even and $(n+1)$-dimensional for $n$ odd.

To pin down the TQM, one needs to further specify the short product structure (which is equivalent to specifying a trace) of the associative algebra $\cA_{D_{n+1}}$. We will analyze how combining discrete symmetry and physical input from 3D $\cN=4$ SCFTs allows us to determine the short product and to provide the deformed mirror map for dual observables.

\subsection{\texorpdfstring{$n = 4$}{n = 4}}

\subsubsection{Periods and Associativity}

We would like to quantize the $D_4$ singularity 
\ie
\cM_{D_4}:~f(X,Y,Z)=X^2+ ZY^2+Z^3=0,
\fe
which merits special attention due to its extra symmetry.  We start by writing down the most general deformed commutators compatible with the Jacobi identity \eqref{DtypeDQ}:
\ie
{}[\hat X,\hat Y]&=\zeta (\hat Y^2 +3\hat Z^2+\zeta^2(2A+8)\hat Z+\zeta^4(2A+B+8)), \\
[\hat X,\hat Z]&=-2 \zeta \hat Z\hat Y-2 \zeta^2 \hat X+\C \zeta^{5}, \\
[\hat Y,\hat Z]&=2\zeta \hat X,
\label{com4}
\fe
as well as the central element
\ie
\Omega_{D_4}=
\hat Z^3+A  \zeta^2\hat Z^2 + B\zeta^4\hat Z + C \zeta^6+\hat X^2+\hat Z \hat Y^2+2\zeta \hat X \hat Y-\C \zeta^4\hat Y.
\fe
Notice that in \eqref{com4}, the leading-order terms in $\zeta$ are simply the Poisson bracket associated with the singularity, coming from the symplectic two-form $\omega$.  The deformation quantization of the $D_4$ singularity falls into isomorphism classes \cite{namikawa2010} (see also \cite{Etingof:2019guc}) that are parametrized by so-called ``periods'' taking values in
\ie
{H^2(\cM^{\rm reg}_{D_4},\mC)\over W(D_4)},
\fe
which is simply the root lattice of $D_4$ modulo the Weyl group. Here, the periods that label the quantizations are $\{A,B,C,\C\}$.
 
The $D_4$ singularity has an $S_3$ symmetry (preserving the holomorphic symplectic two-form $\omega$) that becomes manifest in the coordinates
\ie
 U=\frac{1}{2}\left(Z + \frac{Y}{\sqrt{3}}\right),\quad
  V=\frac{1}{2}\left(Z - \frac{Y}{\sqrt{3}}\right),\quad
 W=\frac{1}{2}X,
\fe
in terms of which the singularity becomes
\ie
\cM_{D_4}:~f(U,V,W)=U^3+V^3+W^2=0.
\fe
This is invariant under
\ie
\mZ_2: (U,V,W)\mapsto (V,U,-W),\quad \mZ_3: (U,V,W)\mapsto (U e^{4\pi i \over 3},V e^{2\pi i \over 3}, W),
\label{symgen}
\fe
generating an $S_3$ symmetry that preserves the hyperk\"ahler structure.

If we insist on having an $S_3$ symmetry upon deformation quantization,\footnote{From the perspective of 3D $\cN=4$ SCFT, this is equivalent to insisting that $S_3$ be a global symmetry of the Higgs or Coulomb branch operator algebra.} then the periods are constrained. Up to a redefinition, the most general $S_3$-preserving (deformed) commutators are
\ie
&[\hat U,\hat V]={2\over \sqrt{3}}\zeta \hat W,\quad
[\hat U,\hat W]=-{\sqrt{3}} \zeta \hat V^2-{4+A\over 2\sqrt{3}}\zeta^3 \hat U ,\quad
[\hat V,\hat W]={\sqrt{3}} \zeta \hat U^2+{4+A\over 2\sqrt{3}}\zeta^3 \hat V.
\fe
From here, we can work out the most general even short star product structures
\begin{align}
\hat U\star\hat U&=\textstyle\vphantom{\frac{\zeta}{\zeta}} \widehat{U^2}+ \A_1   \zeta^2 \hat V, \nonumber
\\
\hat V\star\hat V&=\textstyle\vphantom{\frac{\zeta}{\zeta}} \widehat{V^2}+ \A_1   \zeta^2\hat  U, \nonumber
\\
\hat U\star\hat V&=\textstyle\widehat{UV}+{\zeta\over \sqrt{3}} \hat W  +\A_2 \zeta^4, \nonumber
\\
\hat U \star \hat W&=\textstyle\widehat{UW}- {\sqrt{3}\zeta\over 2}  \widehat{V^2} -{(6\A_1+A+4) \over  4\sqrt{3}} \hat U\zeta^3, \nonumber
\\
\hat V\star \hat W&=\textstyle\widehat{VW}+ {\sqrt{3}\zeta\over 2}   \widehat{U^2}+{(6\A_1+A+4) \over 4\sqrt{3}}  \hat V\zeta^3, \nonumber
\\
\hat W \star \hat W&= \textstyle -\widehat{U^3}-\widehat{V^3}-{A+4(\A_1+\A_4)\over 2}\zeta^2 \widehat{UV}+{(6\A_1+A+4)\A_2\over 4}\zeta^6 \vphantom{\frac{\zeta}{\zeta}}, \nonumber
\\
\hat U \star \widehat{U^2}&=\textstyle\vphantom{\frac{\zeta}{\zeta}} \widehat{U^3}+\A_4 \zeta^2 \widehat{UV} -{\A_1 \over \sqrt{3}} \zeta ^3 \hat W, \label{D4algebraS3}
\\
\hat V \star \widehat{V^2}&=\textstyle\vphantom{\frac{\zeta}{\zeta}} \widehat{V^3}+\A_4 \zeta^2 \widehat{UV}+  {\A_1 \over \sqrt{3}} \zeta ^3 \hat W, \nonumber
\\
\hat U \star \widehat{V^2}&=\textstyle\vphantom{\frac{\zeta}{\zeta}} \widehat{UV^2} -B_1 \widehat{VW} \zeta+\zeta^2 B_2 \widehat{U^2}+B_3 \zeta^4 \hat V, \nonumber
\\
\hat V \star \widehat{U^2}&=\textstyle\vphantom{\frac{\zeta}{\zeta}} \widehat{VU^2} +B_1 \widehat{UW} \zeta+\zeta^2 B_2 \widehat{V^2}+B_3 \zeta^4 \hat U, \nonumber
\\
\hat U \star \widehat{UV}&=\textstyle\vphantom{\frac{\zeta}{\zeta}} \widehat{VU^2} -A_1 \widehat{UW} \zeta+\zeta^2 A_2 \widehat{V^2}+A_3 \zeta^4 \hat U, \nonumber
 \\
\hat V \star \widehat{UV}&=\textstyle\vphantom{\frac{\zeta}{\zeta}} \widehat{UV^2} +A_1 \widehat{VW} \zeta+\zeta^2 A_2 \widehat{U^2}+A_3 \zeta^4 \hat V, \nonumber
\end{align}
where all of the operators $\widehat{U^\A V^\B W^\D}$ are normal-ordered products that are assumed (with suitable shifts) to have vanishing one-point functions, and nonvanishing two-point functions only with their conjugates (conjugation being defined by the $\mZ_2$ generator in \eqref{symgen}).

The parameters that appear above in the star product are further constrained by associativity as follows:
\begin{gather}
A_1=-{1\over \sqrt{3}},\quad A_2={A(2\A_1-1)+2(\A_1+6\A_2-2)\A_4\over 2\A_1 (4+A+6\A_1)}, \nonumber \\[5 pt]
A_3={ A(1-2 \A_1)-2  (\A_1+6 \A_2-2) \over 12}, \\[5 pt]
B_1=-{2\over \sqrt{3}},\quad
B_2=-{1\over 2}-A_1+A_2,\quad
B_3=-{1\over 6}\A_1(4+A+6\A_1). \nonumber
\end{gather}
Consequently, the only free parameters are $\A_1$, $\A_2$, $\A_4$, and $A$.

We can determine these parameters for specific deformation quantizations. One nontrivial example comes from the Coulomb branch of $\cN=4$ $SU(2)$ SQCD with four fundamental hypermultiplets, or by mirror symmetry, the Higgs branch of the affine $D_4$ quiver theory. In either case, by explicit computation, we find
\begin{equation}
\A_1=\frac{32 \pi ^4-2835}{42 \pi ^4-2835}, \quad
\A_2=\frac{2 \pi ^4-135}{180 \pi ^4}, \quad
A=-6, \quad
\ell=3^{1\over 4}\zeta^{-1}
\label{specificsu2}
\end{equation}
as well as
\begin{equation}
\alpha_4 = \frac{20\pi^4(1376\pi^8 - 178185\pi^4 + 4365900)}{231(2\pi^4 - 135)(128\pi^8 - 12180\pi^4 - 14175)},
\end{equation}
where 
\ie
\ell=-4\pi r
\fe
is the natural deformation parameter that arises in the derivation of the 1D TQM from the 3D $\cN=4$ SCFT on $S^3$ ($r$ is the sphere radius).\footnote{In deriving these results, we use the results of Appendix \ref{D4details}, particularly \eqref{similarresults}.} Combined with \eqref{D4algebraS3}, these data determine a large class of correlators in the 1D TQM for either the SQCD or the affine $D_4$ quiver theory.

\subsubsection{Realizations in Lagrangian 3D SCFTs}

Let us define $\cR$ to be the ring of holomorphic functions on the $D_4$ singularity,
\ie
\cM_{D_4}:~X^2+ ZY^2+Z^3=0 \, \mbox{ or } \, U^3+V^3+W^2=0.
\fe
We now discuss 3D $\cN=4$ SCFTs that realize $\cR$ as their Higgs or Coulomb branch chiral ring. In the next section, we will extract the 1D TQM that gives the deformation quantization of $\cR$ by the corresponding SCFT.

We denote by $SU(2)_R$ the relevant $SU(2)$ R-symmetry under which $\cR$ is charged. We can fix the $SU(2)_R$ representations of $\cR$ using the fact that the holomorphic symplectic form $\omega$ must transform as an $SU(2)_R$ triplet:
\ie
R[\omega]=1.
\fe  
Consequently,
\ie
R[U]=R[V]=R[X]=R[Y]=2,\quad R[W]=R[Z]=3.
\fe
Moreover, superconformal representation theory fixes the scaling dimensions to be $\Delta=R$.

\paragraph{\texorpdfstring{$\Gamma_{D_4}$}{D4} gauged free hyper.}

In this case, the generators $\cX,\cY,\cZ$ can be written in terms of the complex scalars $Q,\tilde Q$ in a single hypermultiplet\footnote{Alternatively, we denote the scalars in a hyper by $Q_{aA}$ where $a$ is the $SU(2)_H$ index and $i$ is the $SU(2)$ flavor index. They obey the reality condition
\begin{equation}
(q^{a }_i)^*=q_{a}^i=\epsilon_{ab}\epsilon^{ij} q^{b}_j, \quad \epsilon^{12}=\epsilon_{12}=1,
\end{equation}
where $Q={q_2^1}$ and $\tilde Q={q_1^1}$.} 
which is acted upon by the gauge symmetry $Q_8=\Gamma_{D_4} \subset SU(2)$ as
\ie
r_{\mZ_4}=\begin{pmatrix} i && 0 \\ 0 && -i  \end{pmatrix},\quad
s_{\mZ_2}=\begin{pmatrix} 0 && 1 \\ -1 && 0  \end{pmatrix}.
\fe
Hence a basis of gauge-invariant operators is given by
\ie
\cZ=-2Q^2\tilde Q^2,\quad
\cY=i(Q^4+\tilde Q^4),\quad
\cX=\sqrt{2} i Q\tilde Q (Q^4-\tilde Q^4),
\fe
which satisfy the constraint
\ie
\cX^2+\cZ\cY^2+\cZ^3=0.
\fe
Now the normalizer of $Q_8$ in $SU(2)$ is $\cO_{48}$, the binary octahedral group of order 48. Thus the global symmetry group of the discretely gauged hypermultiplet is\footnote{In general, the global symmetry of $G$ gauged hypers is given by the quotient group ${N_{USp(2n_H)}(G)/G}$.}
\ie
 S_3={\cO_{48}/ \Gamma_{D_4}},
\fe
the permutation group of order 6.

More explicitly, $\cO_{48}$ is defined by generators and relations
\ie
\cO_{48}=\Gamma_{E_7}=\la A,B,C \, | \, A^2=B^3=C^4=(C^2A)^2=-I_2,~(C^2B)^3=(AB)^6=I_2 \ra
\fe
where
\ie
A=\begin{pmatrix} 0 && 1 \\ -1 && 0  \end{pmatrix},\quad
B={1\over 2}\begin{pmatrix} 1+i && -1+i \\ 1+i && 1-i  \end{pmatrix},\quad
C={1\over \sqrt{2}}\begin{pmatrix} 1+i && 0 \\ 0 && 1-i  \end{pmatrix}.
\fe
The $\mZ_2$ and $\mZ_3$ generators of $ S_3$ are identified with $C$ and $B$ in the quotient, respectively, which act as
\begin{equation}
C: (\cX,\cY,\cZ)\mapsto(-\cX,-\cY,\cZ),\quad
B: (\cX,\cY,\cZ) \mapsto \left(\cX,-{\cY+3 i \cZ\over 2},-{\cZ+i \cY\over 2}\right).
\label{s3action}
\end{equation}

\paragraph{Affine \texorpdfstring{$D_4$}{D4} quiver.}

In this case, the gauge theory is described as a 3D $\cN=4$ quiver with $SU(2)\times U(1)^4$ gauge group and four bifundamental hypermultiplets $(Q_A,\tilde Q_A)$ where $A=1,2,3,4$.  The quiver Lagrangian has an obvious $S_4$ global symmetry that acts naturally on the $A$ index, but its action is not faithful on the Higgs branch chiral ring after we take into account the D-term relations. In fact, the Higgs branch chiral ring $\cR$ is organized into faithful representations of $S_3$.  Without loss of generality, we choose an $S_3$ subgroup of $S_4$ to be the one permuting $A=1,2,3$.  Then by identifying the generators
\begin{equation}
s_{\mZ_3}=(123),\quad r_{\mZ_2}=(12),
\label{s3generators}
\end{equation}
we find that
\begin{align}
\cZ &= \sqrt{3}(\tilde{Q}_1 Q_3\tilde{Q}_3 Q_1 + \tilde{Q}_2 Q_3\tilde{Q}_3 Q_2), \nonumber \\
\cY &= \sqrt{3}i(\tilde{Q}_1 Q_3\tilde{Q}_3 Q_1 - \tilde{Q}_2 Q_3\tilde{Q}_3 Q_2), \label{ourbasis} \\
\cX &= 2\cdot 3^{3/4}i\tilde{Q}_1 Q_2\tilde{Q}_2 Q_3\tilde{Q}_3 Q_1 \nonumber
\end{align}
(where the contraction of $SU(2)$ gauge indices is pairwise from the left) transform under $S_3$ in the expected manner and satisfy
\begin{equation}
\cX^2 + \cZ\cY^2 + \cZ^3 = 0.
\end{equation}
Alternatively, we have
\begin{align}
\cU &= e^{i\pi/6}\tilde{Q}_1 Q_3\tilde{Q}_3 Q_1 + e^{-i\pi/6}\tilde{Q}_2 Q_3\tilde{Q}_3 Q_2, \nonumber \\
\cV &= e^{-i\pi/6}\tilde{Q}_1 Q_3\tilde{Q}_3 Q_1 + e^{i\pi/6}\tilde{Q}_2 Q_3\tilde{Q}_3 Q_2, \\
\cW &= 3^{3/4}i\tilde{Q}_1 Q_2\tilde{Q}_2 Q_3\tilde{Q}_3 Q_1. \nonumber
\end{align}
See Appendix \ref{D4chiralring} for a derivation.

\paragraph{\texorpdfstring{$SU(2)$}{SU(2)} SQCD with \texorpdfstring{$N_f = 4$}{Nf = 4}.}

The mirror dual of the affine $D_4$ quiver theory is known to be $SU(2)$ SQCD with four fundamental hypermultiplets. The Coulomb branch chiral ring of the latter theory now realizes $\cR$.  More explicitly, the relevant Coulomb branch operators consist of monopole operators $M^{\pm 2}$ and the Cartan scalar $\Phi$, which are subject to the quantum ring relation (one-loop effect)
\ie
4M^2 M^{-2}=\Phi^4
\fe
and transform under the $\mZ_2$ Weyl group as
\ie
\Phi \to -\Phi,\quad M^{\pm 2}\to M^{\mp 2}.
\fe
$\cR$ is generated by the gauge-invariant combinations
\ie
\cZ=  \Phi^2,\quad 
\cY=i (M^{2}+M^{-2}),\quad 
\cX=\Phi(M^{2}-M^{-2}).
\fe
(We discuss the algebra of these operators in more detail in the next section, using a slightly different normalization.)

\subsection{\texorpdfstring{$n > 4$}{n > 4}}

For $n > 4$, we focus on two specific Lagrangian realizations of the quantized $D_n$ singularity, namely those that participate in 3D mirror symmetry.

\subsubsection{Higgs Branch of Affine \texorpdfstring{$D_n$}{Dn} Quiver}

This quiver takes the shape of an affine $D_n$ Dynkin diagram. We label the four $U(1)\times U(2)$ bifundamental hypermultiplets by $Q_{A},\tilde Q_A$, as before. But now we have, in addition, $n-4$ $U(2)\times U(2)$ bifundamental hypers $K_I,\tilde K_I$ connecting the $n-3$ $U(2)$ gauge nodes.  Nontrivial gauge-invariant elements of the Higgs branch chiral ring correspond to closed paths ending on one of the univalent $U(1)$ nodes.

For this theory, we use the conventions
\begin{equation}
\begin{aligned}
\xymatrix{
& U(1)_1 \ar@{-}[d] & & & & U(1)_2 \ar@{-}[d] \\
U(1)_3 \ar@{-}[r] & U(2) \ar@{-}[r]^1 & U(2) \ar@{-}[r]^2 & \cdots \ar@{-}[r]^{n - 5} & U(2) \ar@{-}[r]^{n - 4} & U(2) \ar@{-}[r] & U(1)_4
}
\end{aligned}
\end{equation}
where the overall quotient by $U(1)$ is implemented in the matrix model by making the first $U(2)$ node $SU(2)$.  We label the hypermultiplets by
\begin{align*}
&(Q_A)_i, (\tilde{Q}_A)^i \text{ for } A = 1, \ldots, 4, \\
&(K_I)_i{}^j, (\tilde{K}_I)_i{}^j \text{ for } I = 1, \ldots, n - 4
\end{align*}
(so that the first index of $K_I$ and the last index of $\tilde{K}_I$ are associated with node $I$), where $i, j = 1, 2$.  With contractions implicit, the Higgs branch chiral ring generators are
\begin{align}
\cZ &= -\tilde{Q}_1 Q_3\tilde{Q}_3 Q_1, \label{hbdnfirst} \\
\cY &= 2\tilde{Q}_3 K_1\cdots K_{n-4}Q_2\tilde{Q}_2\tilde{K}_{n-4}\cdots \tilde{K}_1 Q_3 + (-\cZ)^{n/2 - 1}, \\
\cX &= 2\tilde{Q}_1 K_1\cdots K_{n-4}Q_2\tilde{Q}_2\tilde{K}_{n-4}\cdots \tilde{K}_1 Q_3\tilde{Q}_3 Q_1
\end{align}
for $n\in 2\mathbb{Z}$ (as in \cite{Collinucci:2016hpz, Collinucci:2017bwv}) and
\begin{align}
\cZ &= -\tilde{Q}_1 Q_3\tilde{Q}_3 Q_1, \\
\cY &= 2\tilde{Q}_3 K_1\cdots K_{n-4}Q_2\tilde{Q}_2\tilde{K}_{n-4}\cdots \tilde{K}_1 Q_3, \\
\cX &= 2\tilde{Q}_1 K_1\cdots K_{n-4}Q_2\tilde{Q}_2\tilde{K}_{n-4}\cdots \tilde{K}_1 Q_3\tilde{Q}_3 Q_1 - (-\cZ)^{(n - 1)/2} \label{hbdnlast}
\end{align}
for $n\in 2\mathbb{Z} + 1$, which satisfy the ring relation
\begin{equation}
\cX^2 + \cZ\cY^2- \cZ^{n-1} = 0.
\label{HBchiralringrelation}
\end{equation}
The $\mZ_2$ symmetry
\begin{equation}
\mathbb{Z}_2 : (\cX, \cY, \cZ)\mapsto (-\cX, -\cY, \cZ)
\label{DnZ2}
\end{equation}
is induced by the $1\leftrightarrow 3$ flip of the quiver, or $(Q_1,\tilde Q_1) \leftrightarrow(Q_3,\tilde Q_3)$ at the Lagrangian level.  See Appendix \ref{Dnchiralring} for derivations.

So far, our discussion has been at the level of the (``classical'') chiral ring.  In the next section, we will see through TQM computations how these operators become ``quantized.''

\subsubsection{Coulomb Branch of \texorpdfstring{$SU(2)$}{SU(2)} SQCD with \texorpdfstring{$N_f = n$}{Nf = n}}

Consider $SU(2)$ SQCD with $N_f\geq 0$ fundamental flavors.  Using the slightly more compact notation from Section \ref{CBformalism}, the Coulomb branch chiral ring is generated by the Weyl-invariant operators $\mathcal{M}^2$, $\Phi\mathcal{M}^2$, $\Phi^2$ with
\begin{equation}
\Delta(\mathcal{M}^2) = N_f - 2, \quad \Delta(\Phi\mathcal{M}^2) = N_f - 1, \quad \Delta(\Phi^2) = 2,
\label{su2dimensions}
\end{equation}
where $\mathcal{M}^2$ is the monopole of minimal charge and $\Phi\mathcal{M}^2$ is a dressed monopole. (Again, we discuss the algebra of these operators in more detail in the next section.)

\section{\texorpdfstring{$D$}{D}-Type Mirror Symmetry} \label{dmirror}

Let us now see how the kinematical considerations of the previous section translate into dynamical information about quantum field theories.  Specifically, by computing TQM correlators in the theories whose Higgs and Coulomb branches are exchanged by $D$-type mirror symmetry, we derive the mirror map at the level of \emph{quantized} Higgs and Coulomb branch chiral rings.  As before, we examine the cases $n = 4$ and $n > 4$ separately due to the extra symmetry of the former case.

\subsection{Higgs Branch of Affine \texorpdfstring{$D_n$}{Dn} Quiver}

\subsubsection{\texorpdfstring{$n = 4$}{n = 4}} \label{D4higgsbranch}

We summarize the relevant Higgs branch TQM OPE data for the gauge-invariant operators $\tilde{Q}_1 Q_3\tilde{Q}_3 Q_1$ and $\tilde{Q}_2 Q_3\tilde{Q}_3 Q_2$, extracted from localization computations (see Appendix \ref{D4details}).  The one-point functions are
\begin{equation}
\langle\tilde{Q}_1 Q_3\tilde{Q}_3 Q_1\rangle = \langle\tilde{Q}_2 Q_3\tilde{Q}_3 Q_2\rangle = \frac{1}{96\pi^2 r^2}.
\end{equation}
The diagonal two-point functions are
\begin{align}
\langle\tilde{Q}_1 Q_3\tilde{Q}_3 Q_1\star \tilde{Q}_1 Q_3\tilde{Q}_3 Q_1\rangle = \langle\tilde{Q}_2 Q_3\tilde{Q}_3 Q_2\star \tilde{Q}_2 Q_3\tilde{Q}_3 Q_2\rangle &= \frac{\pi^4 - 30}{5120\pi^8 r^4} \\
\implies \langle\tilde{Q}_1 Q_3\tilde{Q}_3 Q_1\star \tilde{Q}_1 Q_3\tilde{Q}_3 Q_1\rangle_c = \langle\tilde{Q}_2 Q_3\tilde{Q}_3 Q_2\star \tilde{Q}_2 Q_3\tilde{Q}_3 Q_2\rangle_c &= \frac{2\pi^4 - 135}{23040\pi^8 r^4},
\end{align}
where the $c$ subscript denotes a connected correlator.  The mixed two-point function is
\begin{align}
\langle\tilde{Q}_1 Q_3\tilde{Q}_3 Q_1\star \tilde{Q}_2 Q_3\tilde{Q}_3 Q_2\rangle &= \frac{\pi^4 + 45}{15360\pi^8 r^4} \\
\implies \langle\tilde{Q}_1 Q_3\tilde{Q}_3 Q_1\star \tilde{Q}_2 Q_3\tilde{Q}_3 Q_2\rangle_c &= -\frac{2\pi^4 - 135}{46080\pi^8 r^4}.
\end{align}
We then set
\begin{align}
\cU_0 &= e^{-i\pi/6}\tilde{Q}_1 Q_3\tilde{Q}_3 Q_1 + e^{i\pi/6}\tilde{Q}_2 Q_3\tilde{Q}_3 Q_2, \label{U0} \\
\cV_0 &= e^{i\pi/6}\tilde{Q}_1 Q_3\tilde{Q}_3 Q_1 + e^{-i\pi/6}\tilde{Q}_2 Q_3\tilde{Q}_3 Q_2, \\
\cW_0 &= 3^{3/4}i\tilde{Q}_1 Q_2\tilde{Q}_2 Q_3\tilde{Q}_3 Q_1, \label{W0}
\end{align}
where $\langle \cU_0\rangle = \langle \cV_0\rangle = \frac{\sqrt{3}}{96\pi^2 r^2}$. (The connected $n$-point function of operators $\mathcal{O}_0$ is the $n$-point function of the normalized operators $\mathcal{O}\equiv \mathcal{O}_0 - \langle\mathcal{O}_0\rangle$ with $\langle\mathcal{O}\rangle = 0$.) We find that
\begin{gather}
\la \cU \ra =\la \cV\ra=\la \cW\ra =\la \cU\star \cU \ra =\la \cV\star \cV\ra=\la \cU\star \cW\ra=\la \cV\star \cW\ra =0, \nonumber \\
\la \cU\star \cV \ra=\frac{2 \pi ^4-135}{60 \pi ^4 \ell ^4}, \quad \la \cW\star \cW\ra =\frac{3 \sqrt{3} \left(\pi ^4-105\right)}{140 \pi ^4 \ell ^6},
\label{2pfHB}
\end{gather}
as required by the $S_3$ global symmetry of the theory, all of which can be checked explicitly using the Higgs branch TQM path integral \eqref{aD4TQM}.

\subsubsection{\texorpdfstring{$n > 4$}{n > 4}}

To illustrate how Higgs branch computations work for arbitrary $n$, we start with the $S^3$ partition function of the affine $D_n$ quiver theory:
\begin{align}
Z_{D_n} &= \frac{1}{2^{n-4}}\int \prod_{A=1}^4 d\sigma_A\prod_{I=1}^{n-3}\prod_{i=1}^2 du_I^i\, \delta(u_1^1 + u_1^2)\left[\prod_{I=1}^{n-3} \sh(u_I^1 - u_I^2)^2\right]Z_{\sigma, u} \label{zdn}, \\
Z_{\sigma, u} &\equiv \frac{1}{\prod_{i=1}^2 [\prod_{A = 1, 3}\ch(\sigma_A - u_1^i)\prod_{A = 2, 4}\ch(\sigma_A - u_{n-3}^i)]\prod_{I=1}^{n-4}\prod_{i, j = 1}^2 \ch(u_I^i - u_{I+1}^j)}
\end{align}
(note the $1/2^{n-4}$ prefactor rather than $1/2^{n-3}$),\footnote{As explained in Appendix \ref{Dndetails}, accounting for this factor of two is crucial for mirror symmetry to work at the level of $S^3$ partition functions, namely for matching the partition function to that of $SU(2)$ SQCD.  This corrects a number of errors in \cite{Benvenuti:2011ga}.

Roughly speaking, the reason is that the gauge group of the quiver theory is $[U(2)^{n-3}\times U(1)^4]/U(1)$, and one of the $SU(2)$ gauge nodes is really $SO(3)$ due to the identification.  Since the volume of $SO(3)$ is halved relative to that of $SU(2)$, we have a Weyl factor of $1/2$ only for $n-4$ of the $n-3$ nonabelian gauge nodes. \label{factorfootnote}} which reduces to the expected
\begin{equation}
Z_{D_4} = \int du\prod_{A=1}^4 d\sigma_A\, \frac{\sh(2u)^2}{\prod_{A=1}^4 \ch(\sigma_A\pm u)}
\label{zd4}
\end{equation}
for $n = 4$.

Consider insertions of the operator $\cZ = -\tilde{Q}_1 Q_3\tilde{Q}_3 Q_1$.  In Appendix \ref{Dndetails}, we prove that these can be rewritten as insertions of a function of the Coulomb branch scalar VEV $s$ into the rank-one $SU(2)$ SQCD Coulomb branch matrix model.  The $SU(2)$ matrix model takes the form
\begin{equation}
Z_{SU(2) + n}[f(s)] = \frac{1}{4r^2}\int ds\, \frac{\sh(s)^2}{\ch(s/2)^{2n}}f(s),
\end{equation}
where ``$SU(2) + n$'' is shorthand for ``$SU(2)$ with $n$ flavors'' (see \eqref{su2insertion}).  By manipulations on the Higgs branch side of the $D_n$ theory, we find that
\begin{equation}
\langle \cZ\rangle = \frac{Z_{SU(2) + n}[(s^2 - 1)/(8\pi r)^2]}{Z_{SU(2) + n}}, \quad \langle \cZ\star \cZ\rangle = \frac{Z_{SU(2) + n}[(s^2 - 1)^2/(8\pi r)^4]}{Z_{SU(2) + n}},
\label{resultsZandZZ}
\end{equation}
which provides strong evidence for the mirror map
\begin{equation}
\cZ\leftrightarrow \left(\frac{1}{8\pi}\right)^2\left(\Phi^2 - \frac{1}{r^2}\right)
\label{mirrormapZ}
\end{equation}
for all $n > 4$.

To go further, we now set up the computation of more general Higgs branch correlation functions for $D_n$.  We can write
\begin{equation}
Z_{\sigma, u} = \int \left(\prod_{A=1}^4 DQ_A D\tilde{Q}_A\right)\left(\prod_{I=1}^{n-4} DK_I D\tilde{K}_I\right)e^{4\pi r\int d\varphi\, L}
\end{equation}
where
\begin{align}
L &= \sum_{A = 1, 3} (\tilde{Q}_A)^i(\delta_i^j(\partial_\varphi + \sigma_A) - (u_1)_i{}^j)(Q_A)_j + \sum_{A = 2, 4} (\tilde{Q}_A)^i(\delta_i^j(\partial_\varphi + \sigma_A) - (u_{n-3})_i{}^j)(Q_A)_j \nonumber \\
&\phantom{==} + \sum_{I=1}^{n-4} (\tilde{K}_I)_i{}^j(\delta_j^{i'}\delta_{j'}^i\partial_\varphi + (u_I)_j{}^{i'}\delta_{j'}^i - \delta_j^{i'}(u_{I+1})_{j'}{}^i)(K_I)_{i'}{}^{j'}
\end{align}
and $u_I = \operatorname{diag}(u_I^1, u_I^2)$.  Hence we have
\begin{align}
\langle(Q_A)_i(\varphi_1)(\tilde{Q}_B)^j(\varphi_2)\rangle_{\sigma, u} &= -\delta_{AB}\delta_i^j\frac{\operatorname{sgn}(\varphi_{12}) + \Th(\sigma_A - u_1^i)}{8\pi r}e^{-(\sigma_A - u_1^i)\varphi_{12}} \mbox{ } (A, B = 1, 3), \nonumber \\
\langle(Q_A)_i(\varphi_1)(\tilde{Q}_A)^j(\varphi_2)\rangle_{\sigma, u} &= -\delta_{AB}\delta_i^j\frac{\operatorname{sgn}(\varphi_{12}) + \Th(\sigma_A - u_{n-3}^i)}{8\pi r}e^{-(\sigma_A - u_{n-3}^i)\varphi_{12}} \mbox{ } (A, B = 2, 4), \nonumber \\
\langle(K_I)_i{}^j(\varphi_1)(\tilde{K}_I)_{i'}{}^{j'}(\varphi_2)\rangle_{\sigma, u} &= -\delta_i^{j'}\delta_{i'}^j\frac{\operatorname{sgn}(\varphi_{12}) + \Th(u_I^i - u_{I+1}^{i'})}{8\pi r}e^{-(u_I^i - u_{I+1}^{i'})\varphi_{12}}.
\end{align}
As an example, consider
\begin{equation}
\langle[(\tilde{Q}_1)^{i_1}(Q_3)_{i_1}(\tilde{Q}_3)^{j_1}(Q_1)_{j_1}](\varphi_1)\cdots [(\tilde{Q}_1)^{i_p}(Q_3)_{i_p}(\tilde{Q}_3)^{j_p}(Q_1)_{j_p}](\varphi_p)\rangle_{\sigma, u}
\end{equation}
(by convention, we take $\varphi_1 < \cdots < \varphi_p$).  The exponential factors cancel in all full contractions.  There are clearly $(p!)^2$ different contractions ($p!$ for each of $A = 1$ and $A = 3$):
\begin{align}
&\langle(\tilde{Q}_1 Q_3\tilde{Q}_3 Q_1)^p\rangle_{\sigma, u} = \left(\frac{1}{8\pi r}\right)^{2p}\sum_{i_1 = 1}^2\cdots \sum_{i_p = 1}^2\sum_{j_1 = 1}^2\cdots \sum_{j_p = 1}^2\sum_{\rho\in S_p}\sum_{\rho'\in S_p} \nonumber \\
&\phantom{==} \prod_{k=1}^p \delta_{j_{\rho(k)}}^{i_k}(-\operatorname{sgn}(\varphi_{k\rho(k)}) + \Th(\sigma_1 - u_1^{i_k}))\prod_{\ell=1}^p \delta_{i_\ell}^{j_{\rho(\ell)}}(\operatorname{sgn}(\varphi_{\ell\rho(\ell)}) + \Th(\sigma_3 - u_1^{i_\ell})). \label{power1331}
\end{align}
This reduces to previous results in Appendix \ref{Dndetails} for $p = 1, 2$.  On the other hand, the following involves only one contraction:
\begin{align}
&\langle\tilde{Q}_3 K_1\cdots K_{n-4}Q_2\tilde{Q}_2\tilde{K}_{n-4}\cdots \tilde{K}_1 Q_3\rangle_{\sigma, u} \label{nosimpforY} \\
&\phantom{==} = \left(-\frac{1}{8\pi r}\right)^{n-2}\sum_{i_1, \ldots, i_{n-3}} \Th(\sigma_3 - u_1^{i_1})\Th(\sigma_2 - u_{n-3}^{i_{n-3}})\prod_{I=1}^{n-4} \Th(u_I^{i_I} - u_{I+1}^{i_{I+1}}). \nonumber
\end{align}
Similarly,
\begin{align}
&\langle\tilde{Q}_1 K_1\cdots K_{n-4}Q_2\tilde{Q}_2\tilde{K}_{n-4}\cdots \tilde{K}_1 Q_3\tilde{Q}_3 Q_1\rangle_{\sigma, u} \label{nosimpforX} \\
&\phantom{==} = \left(-\frac{1}{8\pi r}\right)^{n-1}\sum_{i_1, \ldots, i_{n-3}} \Th(\sigma_1 - u_1^{i_1})\Th(\sigma_3 - u_1^{i_1})\Th(\sigma_2 - u_{n-3}^{i_{n-3}})\prod_{I=1}^{n-4} \Th(u_I^{i_I} - u_{I+1}^{i_{I+1}}). \nonumber
\end{align}
To simplify these insertions in the matrix model \eqref{zdn}, let us relabel $\sigma_{1, 3} = u_0^{1, 2}$ and $\sigma_{2, 4} = u_{n-2}^{1, 2}$.  As explained in Appendix \ref{Dndetails}, using the Cauchy determinant formula and swapping $u_I^1$ and $u_I^2$ for $I = 1, \ldots, n - 2$ results in a simplified matrix model \eqref{afterswapping} with no hyperbolic functions in the numerator.  Note that \eqref{nosimpforY} and \eqref{nosimpforX} are symmetric under swapping $u_I^1$ and $u_I^2$ for $I = 1, \ldots, n - 3$ and $I = 0, \ldots, n - 3$, respectively.  So to evaluate these insertions in the original matrix model \eqref{zdn}, it suffices to insert them into \eqref{afterswapping} after symmetrizing over $u_{n-2}^1$ and $u_{n-2}^2$:
\begin{align}
&\langle\tilde{Q}_3 K_1\cdots K_{n-4}Q_2\tilde{Q}_2\tilde{K}_{n-4}\cdots \tilde{K}_1 Q_3\rangle_u \\
&\phantom{==} \sim \frac{1}{2}\left(-\frac{1}{8\pi r}\right)^{n-2}\sum_{i_1, \ldots, i_{n-2}} \Th(u_0^2 - u_1^{i_1})\Th(u_{n-2}^{i_{n-2}} - u_{n-3}^{i_{n-3}})\prod_{I=1}^{n-4} \Th(u_I^{i_I} - u_{I+1}^{i_{I+1}}), \nonumber \\
&\langle\tilde{Q}_1 K_1\cdots K_{n-4}Q_2\tilde{Q}_2\tilde{K}_{n-4}\cdots \tilde{K}_1 Q_3\tilde{Q}_3 Q_1\rangle_u \\
&\phantom{==} \sim \frac{1}{2}\left(-\frac{1}{8\pi r}\right)^{n-1}\sum_{i_1, \ldots, i_{n-2}} \Th(u_0^1 - u_1^{i_1})\Th(u_0^2 - u_1^{i_1})\Th(u_{n-2}^{i_{n-2}} - u_{n-3}^{i_{n-3}})\prod_{I=1}^{n-4} \Th(u_I^{i_I} - u_{I+1}^{i_{I+1}}). \nonumber
\end{align}
From the symmetries of the resulting integral, one can see many of the expected equivalences between chiral ring representatives at the level of $\langle\rangle_u$.

For illustration, consider $n = 4$.  By the $\mathbb{Z}_2$ symmetry, we expect
\begin{equation}
\langle\tilde{Q}_3 Q_2\tilde{Q}_2 Q_3\rangle\neq 0, \quad \langle\tilde{Q}_1 Q_2\tilde{Q}_2 Q_3\tilde{Q}_3 Q_1\rangle = 0.
\end{equation}
To demonstrate the latter (which is not obvious in the matrix model), we use that
\begin{equation}
\langle\tilde{Q}_1 Q_2\tilde{Q}_2 Q_3\tilde{Q}_3 Q_1\rangle_u\sim \frac{1}{2}\left(-\frac{1}{8\pi r}\right)^3\sum_{i, j} \Th(u_0^1 - u_1^i)\Th(u_0^2 - u_1^i)\Th(u_2^j - u_1^i)
\end{equation}
when inserted into
\begin{equation}
Z_{D_4} = 4\int \left(\prod_{I=0}^2\prod_{i=1}^2 du_I^i\right)\frac{\delta(u_1^1 + u_1^2)}{\sh(u_0^1 - u_0^2)\sh(u_2^1 - u_2^2)\prod_{i=1}^2 [\ch(u_0^i - u_1^i)\ch(u_1^i - u_2^i)]}.
\label{afterswappingd4}
\end{equation}
By simultaneously swapping $u_0^1\leftrightarrow u_0^2$, $u_1^1\leftrightarrow u_1^2$, $u_2^1\leftrightarrow u_2^2$, we see that insertions of
\[
\Th(u_0^1 - u_1^1)\Th(u_0^2 - u_1^1)\Th(u_2^1 - u_1^1), \quad \Th(u_0^1 - u_1^2)\Th(u_0^2 - u_1^2)\Th(u_2^2 - u_1^2)
\]
into \eqref{afterswappingd4} are the same, and insertions of
\[
\Th(u_0^1 - u_1^2)\Th(u_0^2 - u_1^2)\Th(u_2^1 - u_1^2), \quad \Th(u_0^1 - u_1^1)\Th(u_0^2 - u_1^1)\Th(u_2^2 - u_1^1)
\]
into \eqref{afterswappingd4} are the same.  So we have the \emph{unnormalized} correlator
\begin{align}
Z_{D_4}&[\langle\tilde{Q}_1 Q_2\tilde{Q}_2 Q_3\tilde{Q}_3 Q_1\rangle_u] \nonumber \\
&= 4\left(-\frac{1}{8\pi r}\right)^3\int \left(\prod_{I=0}^2\prod_{i=1}^2 du_I^i\right)\frac{\delta(u_1^1 + u_1^2)}{\sh(u_0^1 - u_0^2)\sh(u_2^1 - u_2^2)\prod_{i=1}^2 [\ch(u_0^i - u_1^i)\ch(u_1^i - u_2^i)]} \nonumber \\
&\phantom{==} \times \sum_i \Th(u_0^1 - u_1^1)\Th(u_0^2 - u_1^1)\Th(u_2^i - u_1^1),
\end{align}
where we have written the integrand in such a way that the insertion contains only $-u_1^1$ in the arguments.  This is useful because one can then write
\begin{align}
Z_{D_4}&[\langle\tilde{Q}_1 Q_2\tilde{Q}_2 Q_3\tilde{Q}_3 Q_1\rangle_u] \nonumber \\
&= 4\left(-\frac{1}{8\pi r}\right)^3\int \frac{du\, du_0^1\, du_0^2\, du_2^1\, du_2^2\Th(u_0^1)\Th(u_0^2)(\Th(u_2^1) + \Th(u_2^2))}{\sh(u_0^1 - u_0^2)\sh(u_2^1 - u_2^2)\ch(u_0^1)\ch(u_0^2 + 2u)\ch(u_2^1)\ch(u_2^2 + 2u)}
\end{align}
and use standard Fourier transform identities, including \eqref{convenient}, to reduce this expression to a single-variable integral (along the lines of Appendix \ref{Dndetails}) that vanishes because the integrand is odd.

The lesson that we draw from the above discussion is that Higgs branch computations for general $n$ are hard.  For example, while the vanishing of $\langle \cX\rangle$ and $\langle \cY\rangle$ follows simply from the $\mathbb{Z}_2$ symmetry for all $D_n$, this fact seems to be highly non-obvious in the Higgs branch matrix model: according to \eqref{hbdnfirst}--\eqref{hbdnlast}, $\langle \cX\rangle$ and $\langle \cY\rangle$ are given by inserting some linear combination of \eqref{nosimpforX} and \eqref{power1331} or \eqref{nosimpforY} and \eqref{power1331} into \eqref{zdn}, respectively, depending on whether $n$ is even or odd.  Thus the $\mathbb{Z}_2$ symmetry of the affine quiver Lagrangian is no longer manifest when we insert $\cX$ and $\cY$ into the matrix model.  Nonetheless, the non-symmetric part of the integrand of the matrix model with $\cX$ or $\cY$ insertions should be a total derivative.  In other words, the F-term relations that imply that $\cX$ and $\cY$ are $\mathbb{Z}_2$-odd (see Appendix \ref{Dnchiralring}) become integration-by-parts identities in the matrix model.

In the end, we would like to derive the mirror map for all of the chiral ring generators, including quantum corrections (which we have, so far, for the generator of smallest dimension in \eqref{mirrormapZ}).  While such a task seems daunting from the point of view of Higgs branch correlators, we can evade most difficulties by passing to the Coulomb branch of the mirror theory, which we do next.  The Coulomb branch analysis is significantly simpler because the gluing formula \eqref{CBmatrixmodel}--\eqref{gluingmeasure} contains a delta function that forces vanishing of magnetic flux, from which one sees that correlators of an odd number of monopoles vanish without even doing any integration.

In fact, symmetries already take us a long way toward rounding out the mirror map.  Using our knowledge of how $\cZ$ maps to the Coulomb branch of the mirror dual, including normalization and quantum corrections, we can deduce the normalization of the mirror map for $\cX$ and $\cY$ by demanding that the chiral ring relation be satisfied (this obviates the need to compute, e.g., $\langle \cX\star \cX\rangle$ on the Higgs branch side).  Furthermore, we know that the quantum correction to $\cY$ vanishes by the $\mathbb{Z}_2$ symmetry, while $\cX$ can only mix with $\cY$.  This completely fixes the quantum mirror map for $\cY$.  Finally, we use the Coulomb branch results of the next subsection regarding the orthogonality of bare and dressed monopoles (particularly \eqref{orthogonality}) to write down the remaining entries in the quantized $D$-type mirror symmetry dictionary.  Combined with \eqref{mirrormapZ}, we arrive at the complete map
\begin{equation}
\boxed{\widehat{\cX}\leftrightarrow \frac{i}{(4\pi)^{n - 1}}\left(\Phi\mathcal{M}^2 - \frac{i}{r}\mathcal{M}^2\right), \quad \cY\leftrightarrow \frac{2\mathcal{M}^2}{(4\pi)^{n - 2}}, \quad \cZ\leftrightarrow \left(\frac{1}{8\pi}\right)^2\left(\Phi^2 - \frac{1}{r^2}\right),}
\label{mirrormapXandYandZ}
\end{equation}
where the Higgs branch operators $\cX, \cY, \cZ$ are given in \eqref{hbdnfirst}--\eqref{hbdnlast} and we have accounted for operator mixing by setting
\begin{equation}
\widehat{\cX}\equiv \cX - \frac{\langle \cX\star \cY\rangle}{\langle \cY\star \cY\rangle}\cY,
\label{unmixing}
\end{equation}
which satisfies $\langle\widehat{\cX}\star \cY\rangle = 0$.\footnote{Strictly speaking, there is a sign ambiguity in (one of) the first two entries of \eqref{mirrormapXandYandZ}.  This is because knowing that $\mathcal{Z}\leftrightarrow \mathscr{C}\mathcal{Z}_C$ in the chiral ring (i.e., ignoring $1/r$ corrections), where $\mathscr{C} = 1/(8\pi)^2\in \mathbb{R}_{>0}$ and $C$ subscripts denote Coulomb branch operators, implies only that
\begin{equation}
\mathcal{X}\leftrightarrow \pm i\mathscr{C}^{(n - 1)/2}\mathcal{X}_C, \quad \mathcal{Y}\leftrightarrow \pm' i\mathscr{C}^{n/2 - 1}\mathcal{Y}_C
\end{equation}
at the level of the chiral ring, where $\pm$ and $\pm'$ are distinct signs.  This follows from the Higgs and Coulomb branch chiral ring relations \eqref{HBchiralringrelation} and \eqref{CBchiralringrelation}.  While the overall sign is inherently ambiguous due to the $\mathbb{Z}_2$ global symmetry $(\mathcal{X}, \mathcal{Y})\to (-\mathcal{X}, -\mathcal{Y})$, the relative sign can be fixed by computing suitable nonvanishing mixed correlators involving $\mathcal{X}$ and $\mathcal{Y}$ (e.g., numerically).}  The two-point functions in \eqref{unmixing} can in principle be computed explicitly from the matrix model for the affine quiver.\footnote{In fact, $\langle\mathcal{Y}\star \mathcal{Y}\rangle$ can also be read off from the SQCD Coulomb branch matrix model via \eqref{M2M2}.}

\subsection{Coulomb Branch of \texorpdfstring{$SU(2)$}{SU(2)} SQCD with \texorpdfstring{$N_f$}{Nf} Flavors}

On the $SU(2)$ SQCD side, the basic non-Weyl-invariant shift operators are
\begin{equation}
M^{\pm 2} = \left(-\frac{1}{2}\right)^{N_f}\frac{1}{r^{N_f - 2}}\frac{(1\pm ir\Phi)^{N_f - 1}}{(\pm ir\Phi)}e^{\mp 2(\frac{i}{2}\partial_\sigma + \partial_B)}, \quad \Phi = \frac{1}{r}\left(\sigma + \frac{i}{2}B\right),
\label{su2shift}
\end{equation}
where $M^{\pm 2}$ are related by the $\mathbb{Z}_2$ Weyl group.  These act on functions $f(\sigma, B)$ where $\sigma\in \mathbb{R}$ and $B\in 2\mathbb{Z}$.  For $N_f\geq 1$, the monopole bubbling terms are necessarily polynomials and can therefore be removed by operator mixing.  In other words, there exists an operator basis in which the bubbling terms for $\mathcal{M}^2$ and $\Phi\mathcal{M}^2$ are zero; all other bases are related to this one by operator mixing.  This means that we can write, without loss of generality,
\begin{equation}
\mathcal{M}^2 = M^2 + M^{-2}, \quad \Phi\mathcal{M}^2 = \Phi(M^2 - M^{-2}).
\end{equation}
These shift operators already allow us to compute the star product in the Coulomb branch TQM.  For $N_f\geq 3$, we can further compute correlators of twisted CBOs as follows.  Define the vacuum wavefunction
\begin{equation}
\Psi_0(\sigma, B) = \delta_{B, 0}\frac{[\frac{1}{2\pi}\Gamma(\frac{1 - i\sigma}{2})\Gamma(\frac{1 + i\sigma}{2})]^{N_f}}{\frac{1}{2\pi}\Gamma(1 - i\sigma)\Gamma(1 + i\sigma)} = \delta_{B, 0}\frac{\sh(\sigma)}{\sigma\ch(\sigma/2)^{N_f}}
\end{equation}
and the gluing measure
\begin{equation}
\mu(\sigma, B) = \frac{1}{r^2}(-1)^{N_f B/2}\left(\sigma^2 + \frac{B^2}{4}\right).
\end{equation}
Since $|\mathcal{W}| = 2$, the partition function is
\begin{equation}
Z = \frac{1}{2}\int \left(\frac{d\sigma}{2}\right) \mu(\sigma, 0)\Psi_0(\sigma, 0)^2 = \frac{1}{4r^2}\int d\sigma\, \frac{\sh(\sigma)^2}{\ch(\sigma/2)^{2N_f}}
\label{su2partition}
\end{equation}
(the $1/2$ in the measure $d\sigma$ accounts for the half-integer normalization of the weights).  Then
\begin{equation}
\langle\mathcal{O}_1\star \cdots\star \mathcal{O}_n\rangle = \frac{1}{2Z}\int \left(\frac{d\sigma}{2}\right) \mu(\sigma, 0)\Psi_0(\sigma, 0)[(\mathcal{O}_1\cdots \mathcal{O}_n\Psi_0)(\sigma, 0)],
\label{su2gluing}
\end{equation}
where the LHS of \eqref{su2gluing} means $\langle\mathcal{O}_1(\varphi_1)\cdots \mathcal{O}_n(\varphi_n)\rangle$ with the $\varphi_i$ in ascending order.  It is also convenient to define
\begin{equation}
Z[f(\sigma)]\equiv \frac{1}{4r^2}\int d\sigma\, f(\sigma)\frac{\sh(\sigma)^2}{\ch(\sigma/2)^{2N_f}}, \quad Z = Z[1]
\label{su2insertion}
\end{equation}
so that, for example, $\langle(\Phi^2)^n\rangle = Z^{-1}Z[(\sigma^2/r^2)^n]$ where $(\Phi^2)^n = \Phi^2\star \cdots \star \Phi^2$ is understood.

\subsubsection{\texorpdfstring{$N_f = 4$}{Nf = 4}} \label{Nf4}

Computing correlators of Coulomb branch chiral primary operators (CPOs) from monopole shift operators is particularly straightforward when $N_f = 4$, as we now show.  This allows for a precise match to the results of Section \ref{D4higgsbranch}.

$SU(2)$ SQCD with $N_f = 4$ is mirror to the affine $D_4$ quiver theory; the Higgs branch of the latter has a global $S_3$ symmetry, which we reproduce.  The mirror map is as follows:
\begin{equation}
\boxed{\cU, \cV\leftrightarrow \frac{\sqrt{3}}{128\pi^2}\left(\Phi^2 - \frac{1}{3r^2}\right)\mp \frac{i}{16\pi^2}\mathcal{M}^2, \quad \cW\leftrightarrow \frac{3^{3/4}}{128\pi^3}\left(\Phi\mathcal{M}^2 - \frac{i}{r}\mathcal{M}^2\right),}
\label{mirrormap}
\end{equation}
where the Higgs branch operators $\cU, \cV, \cW$ are given by \eqref{U0}--\eqref{W0} with one-point functions subtracted.  We present the derivation below, using $C$ subscripts to distinguish Coulomb branch operators from Higgs branch operators.

The Coulomb branch chiral ring generators (in our normalization, following from \eqref{su2shift}) are
\begin{equation}
\mathcal{X}_C = 8\Phi\mathcal{M}^2, \quad \mathcal{Y}_C = -8i\mathcal{M}^2, \quad \mathcal{Z}_C = \Phi^2.
\label{nf4chiralring}
\end{equation}
These operators have dimensions $\Delta(\mathcal{X}_C) = 3$ and $\Delta(\mathcal{Y}_C) = \Delta(\mathcal{Z}_C) = 2$.  At the level of the chiral ring, we have $\mathcal{X}_C^2 + \mathcal{Z}_C\mathcal{Y}_C^2 + \mathcal{Z}_C^3 = 0 \Longleftrightarrow \mathcal{U}_C^3 + \mathcal{V}_C^3 + \mathcal{W}_C^2 = 0$ where
\begin{equation}
\mathcal{U}_C = \frac{1}{2}\left(\mathcal{Z}_C + \frac{\mathcal{Y}_C}{\sqrt{3}}\right), \quad \mathcal{V}_C = \frac{1}{2}\left(\mathcal{Z}_C - \frac{\mathcal{Y}_C}{\sqrt{3}}\right), \quad \mathcal{W}_C = \frac{1}{2}\mathcal{X}_C.
\end{equation}
Using the Coulomb branch formalism,\footnote{Specializing to $N_f = 4$, we have
\begin{equation}
Z[f(\sigma)] = \frac{1}{64r^2}\int d\sigma\, f(\sigma)\frac{\tanh^2(\pi\sigma/2)}{\cosh^4(\pi\sigma/2)}, \quad Z = \frac{1}{120\pi r^2}.
\end{equation}
A useful formula is
\begin{equation}
Z[\sigma^n] = \frac{1}{120(\pi i)^n r^2}\lim_{\tau\to 0}\frac{d^n}{d\tau^n}\left[\frac{\tau(1 - \tau^4)}{\sinh(\pi\tau)}\right] \implies Z[\sigma^2] = \frac{1}{360\pi r^2}, \mbox{ } Z[\sigma^4] = \frac{7\pi^4 - 360}{1800\pi^5 r^2}, \mbox{ } \ldots \mbox{ }.
\end{equation}
This is a special case of \eqref{Zsigmapower} (note that the RHS is real because it vanishes unless $n$ is even).  Some other useful integrals are
\begin{equation}
Z\left[\frac{1}{4 + \sigma^2}\right] = \frac{120 - \pi^4}{120\pi^5 r^2}, \quad Z\left[\frac{1}{16 + \sigma^2}\right] = \frac{2920 - 27\pi^4}{5832\pi^5 r^2}.
\end{equation}
These formulas can be used, for example, to give alternative derivations of \eqref{similarresults}.} we compute the one-point functions
\begin{equation}
\langle\mathcal{M}^2\rangle = \langle\Phi\mathcal{M}^2\rangle = 0, \quad \langle\Phi^2\rangle = \frac{1}{3r^2}
\label{onepoint}
\end{equation}
(the monopole one-point functions vanish automatically in the absence of bubbling).  We also have the two-point functions
\begin{gather}
\langle\Phi^2\star \mathcal{M}^2\rangle = \langle\mathcal{M}^2\star \Phi^2\rangle = \langle\Phi^2\star \Phi\mathcal{M}^2\rangle = \langle\Phi\mathcal{M}^2\star \Phi^2\rangle = 0, \\
\langle\Phi^2\star \Phi^2\rangle = \frac{7\pi^4 - 360}{15\pi^4 r^4}, \quad \langle\mathcal{M}^2\star \mathcal{M}^2\rangle = \frac{2\pi^4 - 135}{120\pi^4 r^4}.
\end{gather}
To define the operators $\mathcal{U}_C$, $\mathcal{V}_C$, $\mathcal{W}_C$ including $1/r$ corrections, note that by dimensional analysis, $\mathcal{U}_C$ and $\mathcal{V}_C$ can only mix with the identity, while $\mathcal{W}_C$ can mix with $\mathcal{U}_C$, $\mathcal{V}_C$, and the identity.  In addition, we would like all correlation functions of $\mathcal{U}_C$, $\mathcal{V}_C$, $\mathcal{W}_C$, and composites thereof to respect the $S_3$ symmetry.  In particular, we must have
\begin{equation}
\langle\mathcal{U}_C\rangle = \langle\mathcal{V}_C\rangle = \langle\mathcal{W}_C\rangle = 0.
\end{equation}
Using \eqref{onepoint}, the requirement that $\langle\mathcal{U}_C\rangle = \langle\mathcal{V}_C\rangle = 0$ fixes
\begin{equation}
\mathcal{U}_C = \frac{1}{2}\left(\Phi^2 - \frac{1}{3r^2}\right) - \frac{4i}{\sqrt{3}}\mathcal{M}^2, \quad \mathcal{V}_C = \frac{1}{2}\left(\Phi^2 - \frac{1}{3r^2}\right) + \frac{4i}{\sqrt{3}}\mathcal{M}^2
\label{curlyUV}
\end{equation}
(note that conjugation flips the sign of the monopole).  Next, requiring that $\langle\mathcal{W}_C\rangle = 0$ shows that $\mathcal{W}_C = 4\Phi\mathcal{M}^2 + O(1/r)$ cannot mix with the identity, so
\begin{equation}
\mathcal{W}_C = 4\Phi\mathcal{M}^2 + \frac{u\mathcal{U}_C + v\mathcal{V}_C}{r}
\end{equation}
for some dimensionless constants $u, v$.  To respect the $S_3$ symmetry, we must also impose that $\langle\mathcal{U}_C\star \mathcal{W}_C\rangle = \langle\mathcal{V}_C\star \mathcal{W}_C\rangle = 0$.  More simply, we have the following ansatz and requirements for $\mathcal{W}_C$:
\begin{equation}
\mathcal{W}_C = 4\Phi\mathcal{M}^2 + \frac{a}{r}\mathcal{M}^2 + \frac{b}{r}\left(\Phi^2 - \frac{1}{3r^2}\right), \quad \left\langle\mathcal{M}^2\star \mathcal{W}_C\right\rangle = \left\langle\left(\Phi^2 - \frac{1}{3r^2}\right)\star \mathcal{W}_C\right\rangle = 0.
\end{equation}
Using\footnote{This equation can be derived as follows.  Let $I_{M^2 M^{-2}}$ and $I_{M^{-2} M^2}$ denote the insertions in the Coulomb branch matrix model corresponding to $M^2 M^{-2}$ and $M^{-2} M^2$.  Then the correlators $\langle\mathcal{M}^2\star \mathcal{M}^2\rangle$, $\langle\Phi\mathcal{M}^2\star \mathcal{M}^2\rangle$, $\langle\mathcal{M}^2\star \Phi\mathcal{M}^2\rangle$ correspond to the insertions
\begin{equation}
I_{M^2 M^{-2}} + I_{M^{-2} M^2}, \quad \frac{\sigma}{r}(I_{M^2 M^{-2}} - I_{M^{-2} M^2}), \quad \frac{\sigma + 2i}{r}I_{M^{-2} M^2} - \frac{\sigma - 2i}{r}I_{M^2 M^{-2}},
\end{equation}
respectively, which implies that
\begin{gather}
\langle\mathcal{M}^2\star \Phi\mathcal{M}^2\rangle + \langle\Phi\mathcal{M}^2\star \mathcal{M}^2\rangle = \frac{2i}{r}\langle\mathcal{M}^2\star \mathcal{M}^2\rangle \\
\Longleftrightarrow \left\langle\mathcal{M}^2\star \left(\Phi\mathcal{M}^2 - \frac{i}{r}\mathcal{M}^2\right)\right\rangle + \left\langle\left(\Phi\mathcal{M}^2 - \frac{i}{r}\mathcal{M}^2\right)\star \mathcal{M}^2\right\rangle = 0.
\end{gather}
The commutativity $\langle\mathcal{M}^2\star \Phi\mathcal{M}^2\rangle = \langle\Phi\mathcal{M}^2\star \mathcal{M}^2\rangle$ is required for consistency of the deformation quantization.}
\begin{equation}
\langle\mathcal{M}^2\star \Phi\mathcal{M}^2\rangle = \langle\Phi\mathcal{M}^2\star \mathcal{M}^2\rangle = \frac{i}{r}\langle\mathcal{M}^2\star \mathcal{M}^2\rangle,
\label{orthogonality}
\end{equation}
which holds for all $N_f$, fixes $a = -4i$ and $b = 0$:
\begin{equation}
\mathcal{W}_C = 4\left(\Phi\mathcal{M}^2 - \frac{i}{r}\mathcal{M}^2\right).
\label{curlyW}
\end{equation}
Having fixed the exact definitions of $\mathcal{U}_C, \mathcal{V}_C, \mathcal{W}_C$ in \eqref{curlyUV} and \eqref{curlyW}, we check that
\begin{equation}
\langle\mathcal{U}_C\star \mathcal{W}_C\rangle = \langle\mathcal{V}_C\star \mathcal{W}_C\rangle = \langle\mathcal{U}_C\star \mathcal{U}_C\rangle = \langle\mathcal{V}_C\star \mathcal{V}_C\rangle = 0, \quad \langle\mathcal{U}_C\star \mathcal{V}_C\rangle\neq 0, \quad \langle\mathcal{W}_C\star \mathcal{W}_C\rangle\neq 0,
\end{equation}
so these correlators respect the full $S_3$ symmetry.\footnote{One can go on to define additional composite operators.  For example, $\widehat{\mathcal{U}_C^2}$ (defined as a shift of $\mathcal{U}_C\star \mathcal{U}_C$) can mix with $\mathcal{V}_C$ and $\widehat{\mathcal{V}_C^2}$ can mix with $\mathcal{U}_C$, which is consistent with the $S_3$ symmetry.}  Specifically, the Higgs branch computation gives (for $\cU = \cU_0 - \langle \cU_0\rangle$, $\cV = \cV_0 - \langle \cV_0\rangle$, etc.)
\begin{equation}
\langle \cU\star \cV\rangle = \alpha_2\zeta^4 = \frac{2\pi^4 - 135}{15360\pi^8 r^4}, \quad \langle \cW\star \cW\rangle = \frac{(6\alpha_1 + A + 4)\alpha_2}{4}\zeta^6 = \frac{3^{3/2}(\pi^4 - 105)}{573440\pi^{10}r^6},
\end{equation}
which we reproduce on the Coulomb branch side by identifying
\begin{equation}
(\cU, \cV, \cW)\leftrightarrow (c\mathcal{U}_C, c\mathcal{V}_C, c^{3/2}\mathcal{W}_C), \quad c\equiv \frac{\sqrt{3}}{64\pi^2},
\end{equation}
thus substantiating the mirror map \eqref{mirrormap}.  Note that such a rescaling by powers of $c\in \mathbb{R}_{> 0}$ preserves the chiral ring relation.

As a further check, \eqref{mirrormap} implies that
\begin{align}
\tilde{Q}_1 Q_3\tilde{Q}_3 Q_1 &\leftrightarrow \frac{1}{128\pi^2}\left(\Phi^2 + \frac{1}{r^2}\right) + \frac{1}{16\pi^2}\mathcal{M}^2, \label{1331} \\
\tilde{Q}_2 Q_3\tilde{Q}_3 Q_2 &\leftrightarrow \frac{1}{128\pi^2}\left(\Phi^2 + \frac{1}{r^2}\right) - \frac{1}{16\pi^2}\mathcal{M}^2. \label{2332}
\end{align}
Using \eqref{1331} and \eqref{2332}, we reproduce all of the Higgs branch correlators of the $\tilde{Q}_i Q_3\tilde{Q}_3 Q_i$ ($i = 1, 2$) using the Coulomb branch formalism.  These identifications make sense because the $\mathbb{Z}_2$ switches $1\leftrightarrow 2$ on the Higgs branch side.

The integral manipulations that led to \eqref{resultsZandZZ} are equally valid when $n = 4$.  So how do we reconcile the conclusion \eqref{mirrormapZ} with the known mirror map \eqref{1331} in this case?  It turns out that there is no contradiction.  By writing
\begin{equation}
\frac{\sh(s)^2}{\ch(s/2)^{2n}} = \frac{1}{\ch(s/2)^{2n - 4}} - \frac{4}{\ch(s/2)^{2n - 2}}
\end{equation}
and using the trick of differentiating $\int ds\, \frac{e^{2\pi ist}}{\ch(s)^\#}$,\footnote{In other words, $Z[\sigma^n]$ can be evaluated analytically by writing the $SU(2)$ SQCD partition function as a sum of SQED partition functions and differentiating with respect to the FI parameter.} we derive below that
\begin{equation}
\langle\Phi^2\rangle = \frac{2}{\pi^2 r^2}\left[\psi^{(1)}(n - 2) + \frac{2}{n - 2}\right]
\end{equation}
in the $SU(2) + n$ theory.  For each $n\geq 3$, there exists a $q_n\in \mathbb{Q}$ such that
\begin{equation}
\psi^{(1)}(n - 2) = q_n + \frac{\pi^2}{6}.
\end{equation}
In fact, we have $q_3 = 0$ and $q_4 = -1$, whereas $q_{n\geq 5}\in \mathbb{Q}\setminus \mathbb{Z}$.  Hence $n = 4$ is special in that $\langle\Phi^2\rangle = 1/3r^2$ is simply a rational number with no factors of $\pi^2$.  In particular, we see that
\begin{equation}
3Z_{SU(2) + 4}[s^2] = Z_{SU(2) + 4}[1],
\label{happyaccident}
\end{equation}
so an insertion of $3s^2$ is equivalent to a trivial insertion!  Hence the results \eqref{resultsZandZZ} cannot be used directly to read off the mirror map when $n = 4$: they are ambiguous.  Specifically, \eqref{happyaccident} implies that the one-point function of the operator \eqref{1331} in $SU(2) + 4$ is equivalent to an insertion of $-(s^2 - 1)/(8\pi r)^2$, despite appearances.  Somewhat miraculously, a similar statement holds for all $p$-point functions despite the mixing with the monopole.  Namely, the insertion corresponding to $p$ copies of \eqref{1331} can always be written as a polynomial of degree $p$ in $s^2$ plus a multiple of $(4 + s^2)^{-1}$, and one can check numerically for any given $p$ that this gives the same result as an insertion of $[-(s^2 - 1)/(8\pi r)^2]^p$.  It would be interesting to construct a proof of this fact.  Finally, for $n\geq 5$, there is no mixing with the monopole and we can read off the mirror map directly from \eqref{resultsZandZZ}.

We finish with some conceptual comments.  At fixed $\varphi$, the twisted CBOs in \eqref{nf4chiralring} represent nontrivial elements of the chiral ring.  However, only after operator mixing is properly accounted for do they correspond to twisted translations of scalar conformal primaries in the CFT (hence CPOs).  Namely, we must choose a basis in which their one-point functions vanish and they are orthogonal to all lower-dimension operators (this is the ``CFT gauge'' of \cite{Beem:2016cbd}).  In this basis, the monopoles correspond to the primary operators constructed in \cite{Borokhov:2002cg}.  Usually, such a basis is obtained by diagonalizing the matrix of two-point functions.  However, to respect the $S_3$ symmetry, that is not what we do here: rather, we impose that composite operators have vanishing one-point functions and nonvanishing two-point functions only with their conjugates, where conjugation is defined by the $\mathbb{Z}_2$ subgroup of $S_3$.

\subsubsection{\texorpdfstring{$N_f > 4$}{Nf > 4}}

For $N_f > 4$, we do not expect any mixing between the scalar and the monopole(s), by dimension-counting.  Correspondingly, we lose the $S_3$ symmetry and are left with only the $\mathbb{Z}_2$ of charge conjugation.

In our normalization, the Coulomb branch chiral ring generators are
\begin{equation}
(\mathcal{X}_C, \mathcal{Y}_C, \mathcal{Z}_C) = (2^{N_f - 1}\Phi\mathcal{M}^2, -i2^{N_f - 1}\mathcal{M}^2, \Phi^2) \implies \mathcal{X}_C^2 + \mathcal{Z}_C\mathcal{Y}_C^2 + \mathcal{Z}_C^{N_f - 1} = 0,
\label{CBchiralringrelation}
\end{equation}
where the above equalities hold at the level of the classical chiral ring.  The dimensions are as in \eqref{su2dimensions}.  The $\mathbb{Z}_2$ symmetry takes $(\mathcal{X}_C, \mathcal{Y}_C, \mathcal{Z}_C)\mapsto (-\mathcal{X}_C, -\mathcal{Y}_C, \mathcal{Z}_C)$.  We wish to determine the ``quantum corrections'' to $\mathcal{X}_C, \mathcal{Y}_C, \mathcal{Z}_C$.  First note that
\begin{equation}
\int d\sigma\, \frac{e^{2\pi i\tau\sigma}}{\ch(\sigma)^N} = \frac{\Gamma(\frac{N}{2} - i\tau)\Gamma(\frac{N}{2} + i\tau)}{2\pi\Gamma(N)} \implies \int \frac{d\sigma}{\ch(\sigma)^N} = \frac{\Gamma(\frac{N}{2})}{2^N\sqrt{\pi}\Gamma(\frac{N+1}{2})}
\label{usefulfourier}
\end{equation}
by the duplication formula $\Gamma(z)\Gamma(z + \frac{1}{2}) = 2^{1-2z}\sqrt{\pi}\Gamma(2z)$, so the partition function \eqref{su2partition} is
\begin{align}
Z &= \frac{1}{2r^2}\int \frac{d\sigma}{\ch(\sigma)^{2(N_f - 2)}} - \frac{2}{r^2}\int \frac{d\sigma}{\ch(\sigma)^{2(N_f - 1)}} \\
&= \frac{1}{r^2}\frac{\Gamma(N_f - 2)}{2^{2(N_f - 1)}\sqrt{\pi}\Gamma(N_f - \frac{1}{2})}.
\end{align}
More generally, it is convenient to write \eqref{su2insertion} as
\begin{equation}
Z[f(\sigma)] = \frac{1}{r^2}\left(\frac{1}{2}z_{2(N_f - 2)}[f(2\sigma)] - 2z_{2(N_f - 1)}[f(2\sigma)]\right), \quad z_N[f(\sigma)]\equiv \int d\sigma\, \frac{f(\sigma)}{\ch(\sigma)^N}.
\end{equation}
By differentiating \eqref{usefulfourier} with respect to $\tau$, we get
\begin{equation}
\int d\sigma\, \frac{(2\pi i\sigma)^p}{\ch(\sigma)^N} = \frac{d^p}{d\tau^p}\left[\frac{e^{\ln\Gamma(\frac{N}{2} - i\tau) + \ln\Gamma(\frac{N}{2} + i\tau)}}{2\pi\Gamma(N)}\right]\Bigg|_{\tau=0},
\end{equation}
which vanishes for odd $p$ and can be written in terms of polygamma functions for even $p$ (it seems challenging to obtain a closed-form expression for this integral, but it can be evaluated for fixed $p$ and arbitrary $N$).\footnote{Likewise, one derives the simple formula
\begin{equation}
Z[\sigma^n] = \frac{1}{(\pi i)^n r^2}\lim_{\tau\to 0}\frac{d^n}{d\tau^n}\left[\frac{(N_f - 2(\tau^2 + 1))\Gamma(N_f - 2 - i\tau)\Gamma(N_f - 2 + i\tau)}{2\pi\Gamma(2(N_f - 1))}\right],
\label{Zsigmapower}
\end{equation}
which can be evaluated on a case-by-case basis.}  In particular, we have
\begin{equation}
\int d\sigma\, \frac{\sigma^2}{\ch(\sigma)^N} = \frac{\Gamma(\frac{N}{2})\psi^{(1)}(\frac{N}{2})}{2^{N+1}\pi^{5/2}\Gamma(\frac{N + 1}{2})}.
\end{equation}
We can then evaluate
\begin{equation}
\langle\Phi^2\rangle = \frac{2}{r^4 Z}(z_{2(N_f - 2)}[\sigma^2] - 4z_{2(N_f - 1)}[\sigma^2]) = \frac{2}{\pi^2 r^2}\left[\psi^{(1)}(N_f - 2) + \frac{2}{N_f - 2}\right],
\end{equation}
where we have used the recurrence relation
\begin{equation}
\psi^{(m)}(z + 1) = \psi^{(m)}(z) + \frac{(-1)^m m!}{z^{m+1}}
\end{equation}
for simplification (the other one-point functions are trivial: $\langle\mathcal{M}^2\rangle = \langle\Phi\mathcal{M}^2\rangle = 0$).  Now note that the $\mathbb{Z}_2$ symmetry requires that $\langle\mathcal{X}_C\rangle = \langle\mathcal{Y}_C\rangle = 0$, but does not restrict $\langle\mathcal{Z}_C\rangle$.  Therefore, including $1/r$ corrections, the most general mixing pattern is
\begin{align}
\mathcal{X}_C &= 2^{N_f - 1}\Phi\mathcal{M}^2 + \frac{x}{r}\mathcal{M}^2 + \frac{x'}{r^{N_f - 3}}(\Phi^2 - \langle\Phi^2\rangle), \\
\mathcal{Y}_C &= -i2^{N_f - 1}\mathcal{M}^2 + \frac{y}{r^{N_f - 4}}(\Phi^2 - \langle\Phi^2\rangle), \\
\mathcal{Z}_C &= \Phi^2 + \frac{z}{r^2}
\end{align}
for $x, x', y, z\in \mathbb{C}$.  To constrain these coefficients, we consider two-point functions.  The $\mathbb{Z}_2$ symmetry requires that
\begin{equation}
\langle\mathcal{X}_C\star \mathcal{Z}_C\rangle = \langle\mathcal{Y}_C\star \mathcal{Z}_C\rangle = 0
\end{equation}
and does not restrict $\langle\mathcal{X}_C\star \mathcal{X}_C\rangle, \langle\mathcal{Y}_C\star \mathcal{Y}_C\rangle, \langle\mathcal{Z}_C\star \mathcal{Z}_C\rangle, \langle\mathcal{X}_C\star \mathcal{Y}_C\rangle$.  We clearly have (by flux conservation) that the two-point functions of $\mathcal{M}^2$ or $\Phi\mathcal{M}^2$ with $\Phi^2$ vanish.  Using
\begin{equation}
\int d\sigma\, \frac{\sigma^4}{\ch(\sigma)^N} = \frac{\Gamma(\frac{N}{2})(6\psi^{(1)}(\frac{N}{2})^2 + \psi^{(3)}(\frac{N}{2}))}{2^{N+3}\pi^{9/2}\Gamma(\frac{N + 1}{2})},
\end{equation}
we compute that
\begin{align}
\langle\Phi^2\star \Phi^2\rangle &= \frac{8}{r^6 Z}\left(z_{2(N_f - 2)}[\sigma^4] - 4z_{2(N_f - 1)}[\sigma^4]\right) \\
&= \frac{2}{\pi^4 r^4}\left[\psi^{(3)}(N_f - 2) + 6\psi^{(1)}(N_f - 2)\left(\psi^{(1)}(N_f - 2) + \frac{4}{N_f - 2}\right)\right].
\end{align}
On general grounds, we have the relation \eqref{orthogonality} where
\begin{equation}
\langle\mathcal{M}^2\star \mathcal{M}^2\rangle = Z^{-1}Z\left[\frac{1}{2^{2N_f}r^{2(N_f - 2)}}\frac{(\sigma + 2i)(\sigma - i)^{2(N_f - 1)} + (\sigma - 2i)(\sigma + i)^{2(N_f - 1)}}{\sigma(\sigma^2 + 4)}\right].
\label{M2M2}
\end{equation}
We also compute that
\begin{equation}
\langle\Phi\mathcal{M}^2\star \Phi\mathcal{M}^2\rangle = Z^{-1}Z\left[-\frac{(\sigma - i)^{2(N_f - 1)} + (\sigma + i)^{2(N_f - 1)}}{2^{2N_f}r^{2(N_f - 1)}}\right].
\end{equation}
The monopole two-point functions are difficult to evaluate analytically, unless one fixes $N_f$.  Requiring $\langle\mathcal{X}_C\star \mathcal{Z}_C\rangle = \langle\mathcal{Y}_C\star \mathcal{Z}_C\rangle = 0$ gives
\begin{equation}
x'\langle\Phi^2\star \Phi^2\rangle_c = y\langle\Phi^2\star \Phi^2\rangle_c = 0 \Longleftrightarrow x' = y = 0
\end{equation}
where $\langle\Phi^2\star \Phi^2\rangle_c = \langle\Phi^2\star \Phi^2\rangle - \langle\Phi^2\rangle^2\neq 0$, so we have determined that
\begin{equation}
\mathcal{X}_C = 2^{N_f - 1}\Phi\mathcal{M}^2 + \frac{x}{r}\mathcal{M}^2, \quad \mathcal{Y}_C = -i2^{N_f - 1}\mathcal{M}^2, \quad \mathcal{Z}_C = \Phi^2 + \frac{z}{r^2}.
\label{xzundetermined}
\end{equation}
But now any correlator containing odd numbers of $\mathcal{X}_C, \mathcal{Y}_C$ obviously vanishes by flux conservation, so higher-point functions automatically respect the $\mathbb{Z}_2$ symmetry and do not fix $x, z$.  That is, the $\mathbb{Z}_2$ symmetry does not completely determine $\mathcal{X}_C, \mathcal{Z}_C$ at the quantum level (we have more freedom than when $N_f = 4$).  However, we can still map $\mathcal{M}^2, \Phi\mathcal{M}^2, \Phi^2$ individually to the Higgs branch side by matching correlators.\footnote{The matching of $\cZ$ across mirror symmetry in \eqref{mirrormapZ} determines $z$ in \eqref{xzundetermined}.  A scheme for fixing $x$ is given in \eqref{mirrormapXandYandZ}.  Namely, imposing that $\mathcal{X}_C$ and $\mathcal{Y}_C$ be orthogonal (a natural choice of basis, in lieu of additional symmetry) gives $x = -i2^{N_f - 1}$, by \eqref{orthogonality}.}

\section{Deformation Quantization of \texorpdfstring{$\mC^2/\Gamma_{E_{6, 7, 8}}$}{E678}} \label{defen}

\subsection{Periods and Associativity}

We now move on to the $E$-type singularities
\ie
&\cM_{E_6}:~f(X,Y,Z)=X^2+Y^3+Z^4=0,
\\
&\cM_{E_7}:~f(X,Y,Z)=X^2+Y^3+Y Z^3=0,
\\
&\cM_{E_8}:~f(X,Y,Z)=X^2+Y^3+Z^5=0,
\label{EtypeCRR}
\fe
which are hyperk\"ahler quotients of the type $\mC^2/\Gamma_{E_{6,7,8}}$.

In 3D $\cN=4$ SCFTs that realize these on the Higgs or Coulomb branch, $X,Y,Z$ are half-BPS chiral primaries of scaling dimension (= $SU(2)_R$ spin)
\ie
&E_6:\Delta =(6,4,3),
\\
&E_7:\Delta =(9,6,4), \label{endimensions}
\\
&E_8:\Delta =(15,10,6).
\fe
The equations in \eqref{EtypeCRR} then correspond to the chiral ring relations.

As usual, the dynamics of the SCFT gives rise to a deformation quantization of the Higgs or Coulomb branch. In particular, the chiral ring relations \eqref{EtypeCRR} are deformed. The truncation property of the Higgs or Coulomb branch algebra (TQM) implies that the deformations are all relevant, the sense of which should be obvious below.

For the $E_6$ singularity, we start by writing down the most general deformed chiral ring relation as $\Omega_{E_6}(\hat X,\hat Y,\hat Z)=0$ with
\ie
\Omega_{E_6}=\hat X^2+\hat Y^3+\hat Z^4+\B_1 \zeta^2\hat Y\hat Z^2 + \B_2\zeta^3 \hat Y\hat Z + \B_3 \zeta^4\hat Z^2 +\B_4\zeta^6 \hat Y +\B_5\zeta^8 \hat Z+ \B_6\zeta^{12},
\fe
where the $\B_i$ are dimensionless parameters.\footnote{Note that we have partially fixed the gauge redundancy in defining the operators to put the deformed chiral ring relation in this form.}  The most general even deformations of the commutators that satisfy the Jacobi identities are given by
\begin{align}
[\hat X,\hat Y]&=
4\zeta \hat Z^3 +\A_1 \zeta^3(\hat Y \hat Z+\hat Z\hat Y)+\A_2 \zeta^{6}\hat Z, \nonumber
\\
[\hat X,\hat Z]
&=-3\zeta \hat Y^2+\A_1 \zeta^{3} \hat Z^2+\A_3  \zeta^{9}, \label{E6ct}
\\
[\hat Y,\hat Z]&=2\zeta  \hat X. \nonumber
\end{align}
Here, we have used the freedom in operator redefinitions to put the last two commutators above in simpler forms.  Furthermore, consistency requires $\Omega$ to be in the center of the algebra $\mC[\hat X,\hat Y,\hat Z]$ with commutators \eqref{E6ct}, so that
\ie
{}[\hat X,\Omega]=[\hat Y,\Omega]=[\hat Z,\Omega]=0.
\label{center}
\fe
This puts constraints on the coefficients $\B_i$. Indeed, all of the $\B_i$ except for one are uniquely determined by $\A_{1,2,3}$ in \eqref{E6ct} as follows:
\begin{align}
\Omega_{E_6} &= \hat X^2+\hat Y^3+\hat Z^4
+(12-\A_1) (\zeta^2 \hat Y \hat Z^2-2\zeta^3 \hat X \hat Z) \nonumber
\\
&\phantom{==} + 4(6-\A_1) \zeta^4 \hat Y^2 +{24\A_1 +\A_2 \over 2 }\zeta^6 \hat Z^2-(\A_2+\A_3) \zeta^8 \hat Y + \C\zeta^{12}. \label{E6c}
\end{align}
The four parameters $\{\A_1,\dots, \A_3, \C\}$ label the even quantizations of the $E_6$ singularity.

The undeformed $E_6$ singularity has a nontrivial $\mZ_2$ symmetry that acts as $X\to -X$, $Z\to -Z$. Note that while the general Coulomb branch algebra presented in \eqref{E6ct} and \eqref{E6c} is conveniently invariant under this $\mZ_2$, the constraints of $\mZ_2$ will show up in specifying the short products on this algebra.

For the $E_7$ case, by solving the Jacobi identities and the center condition \eqref{center}, we find a seven-parameter family of even quantizations labeled by $\{\A_1,\ldots, \A_6,\C \}$, where the commutation relations are
\begin{align}
[\hat X,\hat Y] &= 3\zeta \hat Y\hat Z^2 -6 \zeta^2 \hat X\hat Z +\zeta^3(\A_4 \hat Z^3 -6 \hat Y^2) \nonumber \\
&\phantom{==} -2 \zeta^5 \A_6 \hat Y \hat Z +2 \zeta^6 \A_6 \hat X +\zeta^7 \hat Z^2 \A_3+\zeta^{11} \A_2\hat Z+\zeta^{15} \A_1, \nonumber \\
[\hat X,\hat Z] &= -\zeta (3\hat Y^2+\hat Z^3)+\A_6 \zeta^{5} \hat Z^2+\A_5  \zeta^{13}, \\
[\hat Y,\hat Z] &= 2\zeta \hat X \nonumber
\end{align}
and the center element is
\begin{align}
\Omega_{E_7}&=\hat X^2+\hat Y^3+\hat Y \hat Z^3 \nonumber
\\
&\phantom{==}-3 \zeta \hat X \hat Z^2
+\zeta^2\left(-12 \hat Y^2 \hat Z +{\A_4\over 4} \hat Z^4\right) +18 \zeta^3 \hat X\hat Y 
-(36+3\A_4+\A_6)\zeta^4 (\hat Y \hat Z^2 
-2 \zeta \hat X \hat Z) \nonumber
\\
&\phantom{==}
+{1\over 3} \zeta^6(6(36+3\A_4+2\A_6) \hat Y^2+(\A_3-27\A_4) \hat Z^3)
-2 \zeta^8 (\A_3-12\A_6) (\hat Y \hat Z-\zeta\hat X) \nonumber
\\
&\phantom{==}
+{1\over 2}\zeta^{10}(\A_2 -6(4\A_3+\A_4 \A_6)) \hat Z^2
-\zeta^{12} (\A_2+\A_5) \hat Y 
+\zeta^{14}(\A_1-12 \A_2) \hat Z
+ \C\zeta^{18}.
\end{align}
Similarly, for the $E_8$ singularity, we find an eight-parameter family of even quantizations labeled by $\{\A_1,\A_2,\ldots, \A_7,\C\}$ with commutators
\begin{align}
[\hat X,\hat Y]&=
5\zeta  \hat Z^4
 + 3\A_7\zeta^3(-\hat Y\hat Z^2
+2  \zeta  \hat X\hat Z+2  \zeta^2  \hat Y^2) + 
\zeta^7 \A_4 \hat Z^3 \nonumber
\\
&\phantom{==}
-2 \zeta^9 \A_6 (\hat Y \hat Z -\zeta \hat X)
+\zeta^{13} \A_3  \hat Z^2
+\zeta^{19} \A_2  \hat Z 
+\zeta^{25} \A_1, \nonumber
\\
[\hat X,\hat Z]
&=-3 \zeta  \hat{Y}^2+\zeta ^3\hat Z^3 \alpha _7+\zeta ^9 \hat{Z}^2 \alpha _6+\zeta ^{21} \alpha _5,
\\
[\hat Y,\hat Z]&=2\zeta  \hat X \nonumber
\end{align}
and center element
\begin{align}
\Omega_{E_8}&=
\hat X^2+\hat Y^3+\hat Z^5 -20 \hat Y \hat Z^3 \zeta ^2+60 \hat X \hat Z^2 \zeta ^3+120 \hat Y^2 \hat Z \zeta ^4-120 \hat X \hat Y \zeta ^5
+{960+\A_4\over 4} \zeta^6 \hat Z^4 \nonumber
\\
&\phantom{==}-\zeta^8(3\A_4+\A_6)( \hat Y \hat Z^2 -2 \zeta \hat X\hat Z)
+2 \zeta^{10} (3\A_4+2\A_6)\hat Y^2+\zeta ^{12} \left(\frac{ \alpha _3}{3}+48  \alpha _4-56  \alpha _6\right)\hat Z^3 \nonumber
\\
&\phantom{==}-2\zeta ^{14} (  \alpha _3+60 \alpha _6)(\hat Y \hat Z-\zeta \hat X)
+\zeta ^{18} \left(\frac{ \alpha _2}{2}+48  \alpha _3-3   \alpha _4 \alpha _6\right)\hat Z^2 \nonumber
\\
&\phantom{==}-\zeta ^{20}  t(\alpha _2+\alpha _5)\hat Y +\zeta ^{24}  ( \alpha _1+48 (\alpha _2- \alpha _5))\hat Z+\gamma  \zeta ^{30}.
\end{align}
Note that for $E_{7,8}$, there is no hyperk\"ahler $\mZ_2$ isometry: thus the operators $\hat X, \hat Y, \hat Z$ are all self-conjugate.

\subsection{Realizations in Lagrangian 3D SCFTs}

\subsubsection{Discretely Gauged Free Hyper}

In the following free theories, the Higgs branch chiral ring generators are easily deduced from the known polynomial invariants of the binary tetrahedral, octahedral, and icosahedral groups (see, e.g., \cite{Dumas}).

\paragraph{$\Gamma_{E_6}$ gauged free hyper.}

In this case,
\begin{align}
\cZ &= 3^{3\over 4}\sqrt{2}Q\tilde Q(Q^4-\tilde Q^4), \nonumber
\\
\cY &= -(Q^8+\tilde Q^8+14 Q^4 \tilde Q^4),
\\
\cX &= Q^{12}+\tilde Q^{12}-33  Q^4 \tilde Q^4 (Q^4 +\tilde Q^4), \nonumber
\end{align}
which satisfy $\cX^2+\cY^3+\cZ^4=0$.

\paragraph{$\Gamma_{E_7}$ gauged free hyper.}

In this case,
\begin{align}
\cZ &= - 2^{-{2\over 9} }3^{-{1\over 3}} (Q^8+\tilde Q^8+14 Q^4 \tilde Q^4), \nonumber
\\
\cY &= - 3\times  2^{2\over 3}Q^2  \tilde Q^2 (Q^8+\tilde Q^8-2 Q^4 \tilde Q^4),
\\
\cX &= i(Q \tilde Q(Q^{16}-\tilde Q^{16})-34  Q^5 \tilde Q^5 (Q^8-\tilde Q^8)), \vphantom{2^{\frac{2}{3}}} \nonumber
\end{align}
which satisfy $\cX^2+\cY^3+\cY\cZ^3=0$.

\paragraph{$\Gamma_{E_8}$ gauged free hyper.}

In this case,
\begin{align}
\cZ &= Q\tilde Q(Q^{10}-\tilde Q^{10}+11  Q^5 \tilde Q^5), \vphantom{\frac{}{2}} \nonumber
\\
\cY &= {1\over 12}(Q^{20}-228 Q^{15} \tilde Q^5+494 Q^{10} \tilde Q^{10}+228 Q^5 \tilde Q^{15}+\tilde Q^{20}),
\\
\cX &= {i\over 24\sqrt{3}}(Q^{30}+522 Q^{25} \tilde Q^5-10005 Q^{20} \tilde Q^{10}-10005 Q^{10} \tilde Q^{20}-522 Q^5\tilde Q^{25}+\tilde Q^{30}), \nonumber
\end{align}
which satisfy $\cX^2+\cY^3+\cZ^5=0$.

\subsubsection{Star-Shaped Quivers} \label{starshaped}

The only known realizations of $E_{6,7,8}$ singularities by interacting SCFTs are through the Higgs branches of affine $E_{6,7,8}$ quiver theories.

\paragraph{Affine $E_6$ quiver.}

This theory looks as follows:
\ie
\xymatrix{
	 & & U(1)  \ar@{-}[d]
	\\
	 & & U(2)  \ar@{-}[d]
\\
	U(1)\ar@{-}[r] & U(2) \ar@{-}[r] & SU(3) \ar@{-}[r] & U(2) \ar@{-}[r] & U(1) 
	}
\fe
The quiver has an obvious $S_3$ symmetry acting on the Higgs branch, but at the operator level, only a $\mZ_2$ subgroup acts faithfully.  The latter corresponds to the nontrivial $\mZ_2$ acting as $(\cX,\cY,\cZ)\to (-\cX,\cY,-\cZ)$ on the Higgs branch CPOs.

The affine $E_6$ quiver is obtained by gauging the diagonal $SU(3)$ Higgs branch flavor symmetry of three $T[SU(3)]$ linear quiver theories. Hence the Higgs branch CPOs of the $E_6$ theory are conveniently described as $SU(3)$-invariant combinations of those of $T[SU(3)]$.

Recall the $T[SU(3)]$ quiver
\ie
\xymatrix{
	&	U(1)\ar@{-}[r] & U(2) \ar@{-}[r] &   \boxed{SU(3)}
	}
\fe
and denote the two bifundamental hypers by $(q_i,\tilde q^i)$ and $(Q_i^A,\tilde Q^i_A)$, with $i=1,2$ and $A=1,2,3$ being the fundamental indices for $U(2)$ and $SU(3)$, respectively. The Higgs branch chiral ring is generated by the meson (moment map operator)
\ie
M^A{}_B\equiv Q_i^A\tilde Q^i_B-{1\over 3}Q_i^C\tilde Q^i_C \D^A_B,
\fe 
whose quarks are $U(2)\times SU(3)$ bifundamentals.  It has dimension $\Delta=1$ and transforms in the adjoint representation of $SU(3)$.

Let us denote the generators of the Higgs branch algebra of the three copies of $T[SU(3)]$ by $M_{(a)}{}^A{}_B$. By contracting the $SU(3)$ indices, we can construct the Higgs branch algebra of the $E_6$ theory.  Recall that the dimensions of the CPOs are $\Delta(\cX)=6$, $\Delta(\cY)=4$, and $\Delta(\cZ)=3$.  Thus
\begin{align}
\cX &\propto \tr(M_{(1)}^2 M_{(2)}^2 M_{(3)}^2) + Z^2, \nonumber \\
\cY &\propto \tr(M_{(1)}^2 M_{(2)}^2), \label{E6HBXYZ} \\
\cZ &\propto \tr(M_{(1)}^2 M_{(2)}). \nonumber
\end{align}
The precise expressions are given in Appendix \ref{e6chiralring}.  In \eqref{E6HBXYZ}, we give a particular way to represent the CPOs $\cX,\cY,\cZ$ in terms of the hypermultiplet scalars.  All other representatives differ by terms involving the D-term relations.

\paragraph{Affine $E_7$ quiver.}

Our conventions are
\begin{equation}
\begin{aligned}
\xymatrix{
& & & U(2)_{(3)} \ar@{-}[d] & & \\
U(1)_{(1)} \ar@{-}[r] & U(2) \ar@{-}[r] & U(3) \ar@{-}[r] & SU(4) \ar@{-}[r] & U(3) \ar@{-}[r] & U(2) \ar@{-}[r] & U(1)_{(2)}
}
\end{aligned}
\end{equation}
where subscripts label mesons for each leg.  One can use the same reasoning as for $E_6$ to find the invariants at various $\Delta$; the result, as summarized in \cite{Lindstrom:1999pz} and \cite{Collinucci:2017bwv}, is that the basic invariants are
\begin{align}
\cZ &= \tr(M_{(1)}^3 M_{(3)}), \nonumber \\
\cY &= -\tr(M_{(1)}^3 M_{(2)}^3), \\
\cX &= \tr(M_{(1)}^2 M_{(2)}^3 M_{(1)}^3 M_{(3)}). \nonumber
\end{align}
See Appendix \ref{e7chiralring} for details.

\paragraph{Affine $E_8$ quiver.}

Our conventions are:
\begin{equation}
\begin{aligned}
\xymatrix{
& & & & & U(3)_{(3)} \ar@{-}[d] & \\
U(1)_{(1)} \ar@{-}[r] & U(2) \ar@{-}[r] & U(3) \ar@{-}[r] & U(4) \ar@{-}[r] & U(5) \ar@{-}[r] & SU(6) \ar@{-}[r] & U(4) \ar@{-}[r] & U(2)_{(2)}
}
\end{aligned}
\end{equation}
The basic invariants (again, see \cite{Collinucci:2017bwv}) are
\begin{align}
\cZ &= \tr(M_{(1)}^5 M_{(2)}), \nonumber \\
\cY &= \tr(M_{(1)}^5 M_{(2)}^2 M_{(1)}M_{(2)}^2), \\
\cX &= \tr(M_{(1)}^5 M_{(2)}^2 M_{(1)}M_{(2)}^2 M_{(1)}^3 M_{(2)}^2). \nonumber
\end{align}
See Appendix \ref{e8chiralring} for details.

\section{\texorpdfstring{$E$}{E}-Type Mirror Symmetry} \label{emirror}

The main appeal of the Higgs branch topological sectors in the affine $E$-type quivers is that they might shed light on the non-Lagrangian Coulomb branch algebras (not associated to a nonabelian gauge theory with matter) to which they are mirror dual.  One hope is that applying suitable manipulations and Fourier transform identities to the Higgs branch matrix models for the $E$-series partition functions might give hints about the mirror duals.

Since the $E$-type (and $D$-type) theories can be built from $T[SU(N)]$ theories (which are realized on $S$-duality domain walls of 4D $\mathcal{N} = 4$ $SU(N)$ SYM \cite{Gaiotto:2008ak}) by diagonal gauging, it is natural to use the massive TQM of the constituent $T[SU(N)]$ theories to determine the operator algebras of the full quiver theories.\footnote{See \cite{Benvenuti:2011ga, Gulotta:2011si, Nishioka:2011dq, Chang:2019dzt} for results on the sphere partition functions of the $T[SU(N)]$ (and more generally, the $T_\rho^\sigma[G]$) theories, and in particular Appendix A of \cite{Chang:2019dzt} for comments on the $T_3$ theory \cite{Tachikawa:2015bga} mirror to the affine $E_6$ quiver.  See also \cite{Dey:2014tka} for applications of the technique of gauging linear quivers to the study of mirror symmetry for various balanced quivers.}

\subsection{\texorpdfstring{$E_n$}{En} Matrix Models}

While we leave an in-depth examination of the Higgs branch matrix models of the affine $E_n$ quiver theories to future work, we briefly make some comments on the most tractable case, $E_6$.  The partition function of the affine $E_6$ quiver is given by
\begin{equation}
Z_{E_6} = \frac{1}{2!}\int \prod_{a=1}^3 du_3^a\, \delta(u_3^1 + u_3^2 + u_3^3)\left[\prod_{a < b} \sh(u_3^a - u_3^b)^2\right]Z_{T[SU(3)]}(u_3^a)^3,
\label{ze6}
\end{equation}
where $Z_{T[SU(3)]}(u_3^a)$ with $\sum_{a=1}^3 u_3^a = 0$ is the Higgs branch mass-deformed $T[SU(3)]$ partition function.\footnote{The prefactor of $1/2!$ rather than $1/3!$ is due to our convention of defining the affine $E$-type quivers by making the central node $PSU$ as opposed to $SU$; see Footnote \ref{factorfootnote}.}  The partition function of a single leg can be evaluated explicitly \cite{Benvenuti:2011ga}:
\begin{align}
Z_{T[SU(3)]}(u_3^a) &= \frac{1}{2!}\int du_1\prod_{i=1}^2 du_2^i\, \frac{\sh(u_2^1 - u_2^2)^2}{\prod_{i=1}^2 \ch(u_1 - u_2^i)\prod_{i=1}^2\prod_{a=1}^3 \ch(u_2^i - u_3^a)} \\
&= \frac{1}{2!}\int \prod_{i=1}^2 du_2^i\, \frac{(u_2^1 - u_2^2)\sh(u_2^1 - u_2^2)}{\prod_{i=1}^2\prod_{a=1}^3 \ch(u_2^i - u_3^a)} \label{secondline} \\
&= \frac{1}{2}\prod_{a < b}\frac{u_3^a - u_3^b}{\sh(u_3^a - u_3^b)}. \vphantom{\prod^2}
\end{align}
Thus by integrating over the $T[SU(3)]$ variables $u_{1, 2}$ and then taking Fourier transforms, one can rewrite \eqref{ze6} in a form reminiscent of a rank-one matrix model.  Namely, using
\begin{equation}
\int dy\, e^{2\pi ixy}\frac{y^n}{\sh(y)} = \frac{i}{2}\frac{\partial_x^n\Th(x)}{(2\pi i)^n} \implies \frac{x^3}{\sh(x)} = \int dy\, e^{2\pi ixy}\left[\frac{4 - \ch(2y)}{\ch(y)^4}\right]
\end{equation}
and a cyclic convolution identity from \cite{Dedushenko:2017avn} gives\footnote{Let $\sigma_{j-1,1}\equiv\sigma_{j-1}-\sigma_j$, $\sigma_0\equiv \sigma_N$.  If $F_j(\sigma)$ are functions whose Fourier transforms $\widetilde{F}_j(\tau)$ are defined by 
\begin{equation}
F_j(\sigma) = \int d\tau\, e^{-2 \pi i \sigma \tau} \widetilde{F}_j(\tau), \quad \widetilde{F}_j(\tau) = \int d\sigma\, e^{2 \pi i \sigma \tau} F_j(\sigma),
\end{equation}
then we have
\begin{equation}
\int \left(\prod_{j=1}^N d\sigma_j\right)\delta\left(\frac{1}{N}\sum_{j=1}^N \sigma_j\right)\prod_{j=1}^N F_j (\sigma_{j-1, j}) = \int d\tau\, \prod_{j=1}^N \widetilde{F}_j(\tau).
\label{convolution}
\end{equation}}
\begin{equation}
Z_{E_6} = \frac{1}{16}\int \prod_{a=1}^3 du_3^a\, \delta(u_3^1 + u_3^2 + u_3^3)\prod_{a < b}\frac{(u_3^a - u_3^b)^3}{\sh(u_3^a - u_3^b)} = \frac{1}{48}\int dy\left[\frac{4 - \ch(2y)}{\ch(y)^4}\right]^3.
\end{equation}
To mimic the one-loop determinants in a rank-one Lagrangian theory, one might wish to write the integrand in the form $\frac{\sh\cdots \sh}{\ch\cdots \ch}$, but it remains to be seen whether this rewriting has any physical significance.

Note that the $E_7$ theory contains two copies of $T[SU(4)]$ and one copy of $T_{[2, 2]}[SU(4)]$, whereas the $E_8$ theory contains one copy each of $T[SU(6)]$, $T_{[3, 3]}[SU(6)]$, and $T_{[2, 2, 2]}[SU(6)]$.\footnote{Here, we use the notation of \cite{Nishioka:2011dq} where $T[SU(N)]\equiv T_{[1, \ldots, 1]}^{[1, \ldots, 1]}[SU(N)]$ and $T_\rho[SU(N)]\equiv T_\rho^{[1, \ldots, 1]}[SU(N)]$.}  Hence one can use the same strategy of combining the convolution identity \eqref{convolution} with the results of \cite{Benvenuti:2011ga, Gulotta:2011si} for $T[SU(N)]$ and the results of \cite{Nishioka:2011dq} for the partition functions of the other legs (in the limit of vanishing FI parameters) to rewrite the $E_{7, 8}$ partition functions as one-dimensional integrals.

Returning to $T[SU(3)]$, we have
\begin{equation}
Z_{T[SU(3)]}(u_3^a) = \frac{1}{2!}\int du_1\left(\prod_{i=1}^2 du_2^i\right)\sh(u_2^1 - u_2^2)^2 Z_u(u_3^a)
\end{equation}
where $\sum_{a=1}^3 u_3^a = 0$ and
\begin{equation}
Z_u(u_3^a) = \frac{1}{\prod_{i=1}^2 \ch(u_1 - u_2^i)\prod_{i=1}^2\prod_{a=1}^3 \ch(u_2^i - u_3^a)} = \int Dq\, D\tilde{q}\, DQ\, D\tilde{Q}\, e^{4\pi r\int d\varphi\, L}
\end{equation}
with
\begin{equation}
L = \tilde{q}^i(\delta_i^j(\partial_\varphi + u_1) - (u_2)_i{}^j)q_j + \tilde{Q}_A^i(\delta_i^j\delta_B^A\partial_\varphi + (u_2)_i{}^j\delta_B^A - \delta_i^j(u_3)_B{}^A)Q_j^B
\end{equation}
and $u_2 = \operatorname{diag}(u_2^1, u_2^2)$, $u_3 = \operatorname{diag}(u_3^1, u_3^2, u_3^3)$.  Hence we have
\begin{align}
\langle q_i(\varphi_1)\tilde{q}^j(\varphi_2)\rangle_u &= -\delta_i^j\frac{\operatorname{sgn}(\varphi_{12}) + \Th(u_1 - u_2^i)}{8\pi r}e^{-(u_1 - u_2^i)\varphi_{12}}, \\
\langle Q_i^A(\varphi_1)\tilde{Q}_B^j(\varphi_2)\rangle_u &= -\delta_i^j\delta_B^A\frac{\operatorname{sgn}(\varphi_{12}) + \Th(u_2^i - u_3^A)}{8\pi r}e^{-(u_2^i - u_3^A)\varphi_{12}}.
\end{align}
These two-point functions allow us to compute the OPE within the TQM and hence the quantization of the $E_6$ chiral ring relation, along the lines of Section 6 of \cite{Dedushenko:2017avn}.  Recalling that $(M_{(I)})^A{}_B = (Q_{(I)})_i^A(\tilde{Q}_{(I)})_B^i$, we can also consider insertions of operators built from these mesons into $Z_{T[SU(3)]}(u_3^a)$ written in the simplified form \eqref{secondline}.  We have yet to find a way to write these insertions in an enlightening way.

Finally, the $\mathbb{Z}_2$ symmetry of the $E_6$ theory may help identify which chiral ring generators map to monopoles and which to scalars, assuming that this $\mathbb{Z}_2$ is realized as charge conjugation in the mirror theory.\footnote{The Coulomb branches in the $A$, $D$, and $E_6$ cases all have a $\mathbb{Z}_2$ symmetry ($S_3$ in the case of $D_4$) that commutes with the hyperk\"ahler structure, whereas the $E_7$ and $E_8$ cases do not have any symmetries.  For $A$ and $D$, it is natural to identify the $\mathbb{Z}_2$ with charge conjugation, which acts on monopoles.}  By this logic, $\cX$ and $\cZ$ in the $E_6$ theory (which flip sign under charge conjugation, and whose one-point functions must vanish) should map to monopoles in the non-Lagrangian dual.  This is contrary to the $D$-case, where $\cZ$ maps to a scalar.  In the $E_7$ and $E_8$ cases, we no longer have a $\mathbb{Z}_2$ symmetry, so the circumstantial vanishing of one-point functions can no longer be used as evidence of mapping to monopoles (for instance, one cannot rule out mixing with the Cartan scalar, after subtracting one-point functions).

\subsection{\texorpdfstring{$E_n$}{En} Monopoles}

Putting aside the structure of the (known) matrix models, it is interesting to ask whether the structure of the would-be shift operators themselves reveals any information about the monopole spectrum of the non-Lagrangian duals to the $E_n$ quiver theories.  Some hints that we can use to answer this question are Lagrangian intuition, the commutative limit, and scaling dimensions (for constraining bubbling coefficients).

Let us make a few preliminary comments that can hopefully be clarified in future work.  We make the following assumptions:
\begin{itemize}
\item The fact that the mirror dual theories have rank one means that their monopole charges belong to a one-dimensional vector space.
\item The hypothetical dual gauge group is ``semisimple,'' meaning that the dimensions of (dressed) monopoles are fully accounted for by powers of the vector multiplet scalar in the commutative limit (see Footnote \ref{caveat}).
\item One of the Coulomb branch chiral ring generators is constructed from the vector multiplet scalar and therefore takes the form $\Phi^d$, where $d$ is a positive integer determined by the hypothetical Weyl group.
\end{itemize}
The second assumption is motivated by the fact that the dimensions \eqref{endimensions} of the $E_n$ chiral ring generators are known to be integers, just as the dimensions of monopoles in a Lagrangian theory with semisimple gauge group are integers (otherwise, they could be half-integers, or conceivably even other fractions in a non-Lagrangian theory).  The third assumption is perhaps most plausible in the case of $E_6$, which has a $\mathbb{Z}_2$ symmetry.

We now work out the consequences of these assumptions.  In the commutative limit, a primitive monopole \cite{Dedushenko:2018icp} of dimension $\Delta$ and charge $q$ can only bubble to the identity:
\begin{align}
M_\infty^q = \Phi^\Delta c(q)e[q] &\implies \widetilde{M}_\infty^q = \Phi^\Delta(c(q)e[q] + b(q)) \\
&\implies \Phi^\delta\mathcal{M}_\infty^q = \sum_{w\in \mathcal{W}} w^{-1}(\Phi)^{\Delta + \delta}(c(w(q))e[w(q)] + b(w(q))),
\end{align}
where $b, c$ are complex numbers.  By the rank-one assumption, a given Weyl group element $w$ can only act via multiplication by a constant $c_w$, so
\begin{equation}
\Phi^\delta\mathcal{M}_\infty^q = \sum_{w\in \mathcal{W}} \left(\frac{\Phi}{c_w}\right)^{\Delta + \delta}(c(c_w q)e[c_w q] + b(c_w q)).
\end{equation}
If $\Delta\geq 0$, then the bubbling term is a ``Weyl-invariant'' polynomial (monomial in the commutative limit) and can be removed by a change of basis:
\begin{equation}
\Phi^\delta\mathcal{M}_\infty^q = \sum_{w\in \mathcal{W}} \left(\frac{\Phi}{c_w}\right)^{\Delta + \delta}c(c_w q)e[c_w q].
\end{equation}
Note that $e[c_w q] = e[q]^{c_w}$.  Now recall the relevant singularities (below, we omit the subscript $C$ for ``Coulomb''):
\begin{align}
A_N &: \mathcal{X}^2 + \mathcal{Y}^2 + \mathcal{Z}^{N+1} = 0, \\
D_N &: \mathcal{X}^2 + \mathcal{Z}\mathcal{Y}^2 + \mathcal{Z}^{N-1} = 0, \\
E_6 &: \mathcal{X}^2 + \mathcal{Y}^3 + \mathcal{Z}^4 = 0, \\
E_7 &: \mathcal{X}^2 + \mathcal{Y}^3 + \mathcal{Y}\mathcal{Z}^3 = 0, \\
E_8 &: \mathcal{X}^2 + \mathcal{Y}^3 + \mathcal{Z}^5 = 0.
\end{align}
We wish to determine the Coulomb branch operators in rank-one theories that satisfy these relations.  For $A_N$ and $D_N$, we think of $\mathcal{X}$ and $\mathcal{Y}$ as (dressed) monopoles and $\mathcal{Z}$ as the vector multiplet scalar.  Since overall factors of $\Phi$ must cancel for dimensional reasons, to solve the above relations, we may set $\mathcal{Z} = 1$ and replace $\mathcal{X}$ and $\mathcal{Y}$ by Laurent polynomials $P$ and $Q$ in a single variable $x\sim e[q]$ (by the rank-one assumption):
\begin{equation}
A_N \text{ and } D_N: P(x, x^{-1})^2 + Q(x, x^{-1})^2 + 1 = 0. \label{easy}
\end{equation}
This equation is easily solved by
\begin{equation}
P(x, x^{-1}) = \frac{x - x^{-1}}{2}, \quad Q(x, x^{-1}) = \frac{i(x + x^{-1})}{2}.
\end{equation}
For the $E$-series, there are more possibilities to consider for which operators are scalars and which are monopoles, but let us restrict our attention to the possibility that $\mathcal{Y}$ is the scalar (which is most plausible in the case of $E_6$).  Then we obtain the equation
\begin{equation}
P(x, x^{-1})^m + Q(x, x^{-1})^n + 1 = 0
\label{diophantine}
\end{equation}
where $m, n$ are positive integers and $P, Q$ are single-variable Laurent polynomials with coefficients in $\mathbb{C}$.  The cases of $E_{6, 7, 8}$ correspond to $(m, n) = (2, 4), (2, 3), (2, 5)$, respectively, while the $A$- and $D$-series correspond to $(m, n) = (2, 2)$.  We wish to find nontrivial solutions to the above polynomial Diophantine equations, i.e., solutions with neither $P$ nor $Q$ constant (if no such solutions exist, then the assumptions should be relaxed).\footnote{Recall that the $A$-series has two independent monopoles and trivial Weyl group, while the $D$-series has one independent monopole and nontrivial Weyl group.  For the $E$-series, we assume that two of the generators are monopoles, but assuming only one independent monopole (so that the other is simply a dressed version of it) would imply that $P$ and $Q$ have the same powers of $x$ and differ only in their coefficients; then $P$ and $Q$ (and the corresponding monopole operator) would need infinitely many terms, since the degrees could not match otherwise.  So we are led to postulate two independent monopoles for the $E$-series.}

In general, one can ask for which $m, n$ there exist nontrivial solutions to \eqref{diophantine} (WLOG, we may restrict our attention to $2\leq m\leq n$, where we impose the first inequality because solutions are trivial to obtain if either $m, n$ is 1).  We have not been able to find or rule out nontrivial solutions beyond $(m, n) = (2, 2)$.\footnote{However, the existence of nontrivial solutions for small $(m, n)$ is not immediately ruled out by the $abc$ inequality for polynomials (Mason-Stothers theorem).  We thank J.\ Silverman for this comment.}  One possibility is that we should abandon semisimplicity, so that the monopoles have dimensions not accounted for by $\Phi$.

\section{Summary and Future Directions}

This paper presents the results of precision studies of nonabelian $ADE$ mirror symmetry beyond the chiral ring, using the recently developed TQM techniques in \cite{Dedushenko:2016jxl, Dedushenko:2017avn, Dedushenko:2018icp}.  As a byproduct, we extend the construction of deformation quantizations of \cite{Beem:2016cbd} to the $D$- and $E$-series.  We focus on $D$-type quivers, in particular synthesizing OPE data (structure constants) for the chiral ring generators of the $D$-series, but we also comment on possible implications for the monopole spectrum of the non-Lagrangian theories whose Coulomb branches are $\mathbb{C}^2/\Gamma_{E_{6, 7, 8}}$.  We find the precise map between quantized Higgs branch chiral ring generators in $D$-type quivers and quantized Coulomb branch chiral ring generators in $SU(2)$ SQCD.  Our results provide additional entries in the mirror symmetry dictionary for nonabelian 3D $\mathcal{N} = 4$ gauge theories beyond, e.g., the matching of supersymmetric partition functions\footnote{Matrix models for sphere partition functions of affine $A$-type quiver theories of arbitrary rank have been studied in \cite{Kapustin:2010xq}, leading to a derivation of the mirror map between mass and FI parameters.  The corresponding analysis for $D$-type quivers was performed in \cite{Dey:2011pt, Dey:2013nf}, and in this case, a free-fermion representation for the partition function (with vanishing mass and FI parameters) was derived in \cite{Assel:2015hsa}.} \cite{Kapustin:2010xq, Dey:2011pt, Dey:2013nf, Assel:2015hsa} and chiral rings \cite{Cremonesi:2013lqa}.

It is safe to say that the range of applications of the Higgs and Coulomb branch TQM has yet to be fully explored.  For one thing, it would be interesting to incorporate the additional constraints of $\mathcal{N} = 6$ or $\mathcal{N} = 8$ SUSY \cite{Tachikawa:2019dvq} into the TQM analysis.  For another, the OPE data that we have computed can be fed into the bootstrap machine to study the full CFT spectrum and (self-)mirror symmetry beyond the TQM sector, \`a la \cite{Chang:2019dzt}.  Finally, the connection between these techniques and protected operator algebras in one dimension higher \cite{Beem:2013sza} (several aspects of which have recently been derived from localization \cite{Pan:2019bor, Dedushenko:2019yiw}) via dimensional reduction \cite{Dedushenko:2019mzv, Pan:2019shz, Dedushenko:2019mnd} leads us to wonder whether the TQM contains tractable lessons about line operators in 4D gauge theories.

A technical detail that we have glossed over is the following.  To define the star-shaped quivers of interest, we start with all nodes unitary ($U$) and quotient by the diagonal $U(1)$, as suggested by their brane constructions. (See \cite{Dimofte:2018abu} for Coulomb branch computations in these theories.) As a computational matter, it is convenient to implement the quotient simply by making one of the nodes $SU$.  More precisely, we should make one of the nodes $PSU$.  The distinction between $SU$ and $PSU$ is irrelevant to normalized correlation functions of local operators (in particular, TQM observables).  However, the precise normalization of the partition function depends on which $U$ node we make $PSU$: a $PSU(N)$ node introduces a factor of $N$ in the partition function relative to an $SU(N)$ node because the volumes of these groups differ, and the inverse volume enters into the gauge-fixed path integral.  We found that to match the partition function of the affine $D$-type quiver to that of $SU(2)$ SQCD, it suffices to make one of the $U(2)$ nodes $SO(3)$.  The situation is less clear for the affine $E$-type quivers since their mirrors are non-Lagrangian, but one can in principle match partition functions (including discrete factors) by reducing the 4D index of the $E_n$ Minahan-Nemeschansky theories \cite{Minahan:1996fg, Minahan:1996cj}.  It would be interesting to clarify the general procedure for decoupling the overall $U(1)$ and to understand better the global structure of the gauge group in the affine quiver when comparing to the mirror theory.

\section*{Acknowledgements}

We thank M.\ Dedushenko for useful discussions about short star products and S.\ Pufu for interesting comments that inspired Appendix \ref{ADcoulomb}.  The work of YF was supported in part by the NSF GRFP under Grant No.\ DGE-1656466 and by the Graduate School at Princeton University through a Centennial Fellowship.  The work of YW was supported in part by the US NSF under Grant No.\ PHY-1620059 and by the Simons Foundation Grant No.\ 488653.

\appendix

\section{Details of TQM Computations} \label{detailstqm}

\subsection{Fourier Transform Identities} \label{integralidentities}

The basic Fourier transform identities that we will need are
\begin{equation}
\frac{1}{\ch(x)} = \int dy\, \frac{e^{2\pi ixy}}{\ch(y)}, \quad \frac{1}{\sh(x)} = \frac{i}{2}\int dy\, e^{-2\pi ixy}\Th(y).
\label{basicfourier}
\end{equation}
Other useful identities include
\begin{equation}
\int dy\, \frac{e^{2\pi ixy}}{\ch(y)^2} = \frac{x}{\sh(x)}, \quad \int dy\, \frac{e^{2\pi ixy}}{\ch(y)^3} = \frac{1 + 4x^2}{8\ch(x)}.
\end{equation}
By differentiating \eqref{basicfourier}, we obtain
\begin{equation}
\int dy\, e^{2\pi ixy}\frac{\Th(y)}{\ch(y)} = \frac{2ix}{\ch(x)}.
\label{diffbasicfourier}
\end{equation}
By further differentiating \eqref{diffbasicfourier}, we obtain analogous formulas for $y^n\Th(y)/\ch(y)$, e.g.,
\begin{equation}
\int dy\, e^{2\pi ixy}\frac{y\Th(y)}{\ch(y)} = \frac{1 - \pi x\Th(x)}{\pi\ch(x)}.
\end{equation}
We have in addition that
\begin{equation}
\int dy\, e^{2\pi ixy}\frac{\Th(y)^2}{\ch(y)} = \frac{1 - 4x^2}{2\ch(x)},
\label{otherfourier}
\end{equation}
and by differentiating \eqref{otherfourier}, we obtain analogous formulas for $y^n\Th(y)^2/\ch(y)$, e.g.,
\begin{equation}
\int dy\, e^{2\pi ixy}\frac{y\Th(y)^2}{\ch(y)} = \frac{i(8x + \pi(1 - 4x^2)\Th(x))}{4\pi\ch(x)}.
\end{equation}
One can go on to derive similar identities.  Finally, we note that
\begin{equation}
\int \frac{d\sigma}{\ch(\sigma - u_1)\ch(\sigma - u_2)} = \frac{u_1 - u_2}{\sh(u_1 - u_2)}.
\label{convenient}
\end{equation}

\subsection{\texorpdfstring{$\Gamma_{D_4}$}{D4} Gauged Free Hyper}

Recall that in this theory,
\begin{equation}
\cZ_0=-2Q^2\tilde Q^2,\quad \cY_0=i(Q^4+\tilde Q^4),\quad \cX_0=\sqrt{2} i Q\tilde Q (Q^4-\tilde Q^4).
\end{equation}
Thus
\begin{equation}
\cU_0, \cV_0 = -Q^2\tilde Q^2\pm{i\over 2\sqrt{3}} (Q^4+\tilde Q^4), \quad \cW_0 = {i\over \sqrt{2}} Q\tilde Q (Q^4-\tilde Q^4),
\end{equation}
where we use the 0 subscript to denote a ``bare'' Higgs branch CPO. Canonically normalized CPOs without 0 subscripts have vanishing one-point functions and diagonal two-point functions (in a real basis).
 
We would like to compute correlation functions of the CPOs. To proceed, we need the two-point function of $Q,\tilde Q$, which is
\begin{equation}
\la Q (\varphi_1)\tilde Q (\varphi_2)\ra=-{\operatorname{sgn}(\varphi_{12})\over 8\pi r}={\operatorname{sgn}(\varphi_{12})\over2\ell}
\end{equation}
(recall that $\ell=-4\pi r$ from \cite{Dedushenko:2016jxl}). The correlator at coincident points is 0.  In particular,
\begin{equation}
\la \cU_0 \ra=\la \cV_0 \ra =\la \cW_0 \ra =0,
\end{equation}
and consequently the normalized CPOs are
\begin{equation}
\cU=\cU_0, \quad \cV=\cV_0, \quad \cW=\cW_0.
\end{equation}
There is no further ``gauge ambiguity'' in this case.

Doing simple Wick contractions, we obtain
\begin{gather}
\la \cU\star \cV\ra= {1\over 2\ell^4}, \quad \la \cW\star \cW\ra=-{15\over 8\ell^6}, \quad \la \cU\star\cW\ra =\la \cV\star\cW\ra =0, \nonumber \\
\la \cU\star \cU\star \cU\ra =\la \cV\star \cV\star \cV\ra ={3\over \ell^6}, \quad \la \cU\star \cV\star \cW\ra =-{5\sqrt{6}\over 2\ell^7}, \\
\la \cU^2\star \cV^2 \ra={15\over 2\ell^8}, \quad  \la \cU\cV \star \cU\cV\ra = {12^2\times 4!^2+8!\times 2 \over 12^2 (2\ell)^8} = {21\over 4\ell^8}. \nonumber
\end{gather}
Thus, comparing to \eqref{D4algebraS3}, we have
\begin{equation}
\zeta={4\sqrt{2}\over \ell}, \quad A=-{179\over 32}, \quad \A_1={3\over 16}, \quad \A_2={1\over 2048}, \quad \A_4={5\over 8}.
\end{equation}
In particular, the $S_3$ symmetry is obvious in the TQM.

\subsection{Affine \texorpdfstring{$D_4$}{D4} Quiver} \label{D4details}

Recall that the Higgs branch chiral ring generators are given by
\begin{align}
\cU_0 &= e^{i\pi/6}\tilde{Q}_1 Q_3\tilde{Q}_3 Q_1 + e^{-i\pi/6}\tilde{Q}_2 Q_3\tilde{Q}_3 Q_2, \nonumber \\
\cV_0 &= e^{-i\pi/6}\tilde{Q}_1 Q_3\tilde{Q}_3 Q_1 + e^{i\pi/6}\tilde{Q}_2 Q_3\tilde{Q}_3 Q_2, \\
\cW_0 &= 3^{3/4}i\tilde{Q}_1 Q_2\tilde{Q}_2 Q_3\tilde{Q}_3 Q_1 \nonumber
\end{align}
in terms of gauge-invariant combinations of the hypermultiplets.

We adopt the normalization \eqref{zd4} for the $S^3$ partition function of the affine $D_4$ quiver, which we can write as
\begin{equation}
Z_{D_4} = \int du\prod_A d\sigma_A\sh(2u)^2 Z_{\sigma, u} = \frac{1}{120\pi}
\label{aD4TQM}
\end{equation}
where
\begin{equation}
Z_{\sigma,u}=\int \prod_A DQ^i_A D\tilde Q_{iA}
\exp\left(
4\pi r\int d\varphi
\left[
\sum_A
\tilde Q_{iA}
(\pa_\varphi \D^i_j+\sigma_A \D^i_j+ u t^i{}_j)
Q^j_A
\right]
\right)
\end{equation}
and $t=\sigma_3$. Thus the propagators are
\begin{equation}
\la Q_{iA}(\varphi_1) \tilde Q_B^j (\varphi_2)\ra_{\sigma,u} = -\D_{AB}\D_i^j{\operatorname{sgn}(\varphi_{12})+\Th(\sigma_A \pm u)  \over 8\pi r}e^{- (\sigma_A \pm u)\varphi_{12}},
\end{equation}
where $\varphi_{12}\equiv \varphi_1-\varphi_2$ and the $\pm$ sign is $+$ when $i=j=1$ and $-$ when $i=j=2$. We emphasize that the 1D TQM path integral has an explicit $S_4$ symmetry permuting the $A$ indices (as explained before, only an $S_3$ subgroup acts faithfully on CPOs).  At coincident points, we use the symmetrized expression
\begin{equation}
\la Q_{iA}(\varphi) \tilde Q_B^j (\varphi)\ra_{\sigma,u}=-\D_{AB}\D_i^j { \Th (\sigma_A \pm u)  \over 8\pi r},
\end{equation}
and in computing correlation functions, we always assume the $\varphi_{i}$ are ordered as\footnote{Thus our conventions are that operator insertions in the expression $\langle\cO_1\star \cdots\star \cO_n\rangle$ are understood to be in ascending order; compare to \eqref{CBmatrixmodel} and \eqref{su2gluing}.}
\begin{equation}
\varphi_1 <\varphi_2 <\varphi_3 < \cdots.
\end{equation}
Note that (incomplete) self-contractions of a composite operator can also contribute to connected correlators.  The correlators that we compute below are normalized by \eqref{aD4TQM}.
 
\paragraph{Computation of TQM correlators.}

To compute the (normalized) two-point function $\la \cU\star \cV\ra$, we need to compute $\la \cU_0\star \cV_0\ra$ as well as the one-point functions $\la \cU_0\ra $ and $\la \cV_0 \ra $.

We start by recording the Wick contractions in the 1D TQM on the Higgs branch:
\begin{equation}
\la (\tilde Q_1 Q_3 \tilde Q_3 Q_1) (\varphi_1) (\tilde Q_1 Q_3 \tilde Q_3 Q_1) (\varphi_2) \ra_{\sigma} = \frac{1}{(2\ell)^4}(I_{c}+I_{s}+I_{ss})
\end{equation}
for $\varphi_1<\varphi_2$, where cross-contractions give
\begin{align}
I_{c} &= ((1+\Th(\sigma_1-u))(1-\Th(\sigma_3-u)) + (u\leftrightarrow -u))\times (\sigma_1\leftrightarrow \sigma_3) \nonumber \\
&= 16\left(\frac{1}{\ch(\sigma_1-u)\ch(\sigma_3-u)} + \frac{1}{\ch(\sigma_1+u)\ch(\sigma_3+u)}\right)^2
\end{align}
and self-contractions give
\begin{equation}
I_s=(\Th(\sigma_1-u)^2 (\Th(\sigma_3-u)^2-1) +(u\leftrightarrow -u))+(\sigma_1\leftrightarrow \sigma_3)
\end{equation}
as well as
\begin{equation}
I_{ss}=(\Th(\sigma_1-u)\Th(\sigma_3-u) + (u\leftrightarrow -u))^2.
\end{equation}
Doing the matrix integral, we get
\begin{equation}
\la (\tilde Q_1 Q_3 \tilde Q_3 Q_1) (\varphi_1) (\tilde Q_1 Q_3 \tilde Q_3 Q_1) (\varphi_2) \ra = \frac{\pi ^4-30}{20 \pi ^4 \ell^4}.
\end{equation}
Similarly,
\begin{align}
&\la (\tilde Q_1 Q_3 \tilde Q_3 Q_1)(\varphi_1) (\tilde Q_2 Q_3 \tilde Q_3 Q_2)(\varphi_2) \ra_{\sigma} \nonumber \\
&\phantom{==} \propto (\Th(\sigma_1-u)\Th(\sigma_2-u)(\Th(\sigma_3-u)^2-1) +(u\leftrightarrow -u)) \nonumber \\
&\phantom{==} + (\Th(\sigma_1-u)\Th(\sigma_3-u) + (u\leftrightarrow -u))(\Th(\sigma_2-u)\Th(\sigma_3-u) + (u\leftrightarrow -u))
\end{align}
has only self-contractions.  We get
\begin{equation}
\la (\tilde Q_1 Q_3 \tilde Q_3 Q_1)(\varphi_1) (\tilde Q_2 Q_3 \tilde Q_3 Q_2)(\varphi_2) \ra = \frac{45+\pi ^4}{60 \pi ^4 \ell^4}.
\end{equation}
We also have the one-point functions 
\begin{equation}
\la \tilde Q_1 Q_3 \tilde Q_3 Q_1\ra=  \la \tilde Q_2 Q_3 \tilde Q_3 Q_2\ra = \frac{1}{6 \ell^2}.
\end{equation}
Thus the connected two-point functions ($\langle\mathcal{O}_1\mathcal{O}_2\rangle_c = \langle\mathcal{O}_1\mathcal{O}_2\rangle - \langle\mathcal{O}_1\rangle\langle\mathcal{O}_2\rangle$) are
\begin{gather}
\la (\tilde Q_1 Q_3 \tilde Q_3 Q_1)  (\varphi_1) (\tilde Q_1 Q_3 \tilde Q_3 Q_1) (\varphi_2) \ra_c = \la (\tilde Q_2 Q_3 \tilde Q_3 Q_2)  (\varphi_1) (\tilde Q_2 Q_3 \tilde Q_3 Q_2)  (\varphi_2) \ra_c \nonumber \\
= -2\la (\tilde Q_1 Q_3 \tilde Q_3 Q_1) (\varphi_1) (\tilde Q_2 Q_3 \tilde Q_3 Q_2) (\varphi_2) \ra_c = \frac{2 \pi ^4-135}{90 \pi ^4 \ell^4}.
\end{gather}
Putting everything together, we have
\begin{equation}
\cU=\cU_0-\la \cU_0\ra,\quad \cV=\cV_0-\la \cV_0 \ra
\end{equation}
where
\begin{align}
\la \cU\star \cV \ra &= 2\la \tilde Q_1 Q_3 \tilde Q_3 Q_1\star \tilde Q_1 Q_3 \tilde Q_3 Q_1 \ra_c+\la \tilde Q_1 Q_3 \tilde Q_3 Q_1\star \tilde Q_2 Q_3 \tilde Q_3 Q_2\ra_c =\frac{2 \pi ^4-135}{60 \pi ^4 \ell^4}, \\
\la \cU\star \cU \ra &= \la \tilde Q_1 Q_3 \tilde Q_3 Q_1\star \tilde Q_1 Q_3 \tilde Q_3 Q_1 \ra_c+2\la \tilde Q_1 Q_3 \tilde Q_3 Q_1\star \tilde Q_2 Q_3 \tilde Q_3 Q_2 \ra_c = 0.
\end{align}
Similarly, one can check explicitly that
\begin{equation}
\la \cU\star \cW \ra=\la \cV\star \cW\ra=0.
\end{equation}
At the level of the TQM, this is a simple consequence of the exact $S_3$ symmetry of \eqref{aD4TQM}.

We summarize the results of similar computations for various correlators below:
\begin{gather}
\la \cU\star \cV\ra = \frac{2 \pi ^4-135}{60 \pi ^4 \ell ^4}, \quad \la \cU\star \cU\star \cU\ra = \la \cV\star \cV\star \cV\ra = \frac{32 \pi ^4-2835}{420 \sqrt{3} \pi ^4 \ell ^6}, \nonumber \\[5 pt]
\la \cW\star \cW\ra = \frac{3 \sqrt{3} \left(\pi ^4-105\right)}{140 \pi ^4 \ell ^6}, \quad \la \cU\star \cV\star \cW\ra = \frac{3 \sqrt[4]{3} \left(\pi ^4-105\right)}{140 \pi ^4 \ell ^7}, \label{similarresults} \\[5 pt]
\la \cU^2 \star \cV^2\ra = -\frac{\left(\pi ^4-105\right) \left(32 \pi ^4-2835\right)}{490 \pi ^4 \left(2 \pi ^4-135\right) \ell ^8}, \quad \la \cU\cV \star \cU\cV\ra = -\frac{14175+12180 \pi ^4-128 \pi ^8}{2800 \pi ^8 \ell ^8}. \nonumber
\end{gather}
These correspond to \eqref{specificsu2} with $\zeta = 1$.  Note that we have performed Gram-Schmidt diagonalization to define the composite operators.

\subsection{Affine \texorpdfstring{$D_n$}{Dn} Quiver} \label{Dndetails}

\paragraph{Partition function.}

We start by simplifying the Higgs branch matrix model of the affine $D_n$ quiver, while also reviewing the mirror equivalence to SQCD at the level of $S^3$ partition functions.  The Cauchy determinant formula
\begin{equation}
\frac{\prod_{i<j} \sh(x_i - x_j)\sh(y_i - y_j)}{\prod_{i, j} \ch(x_i - y_j)} = \sum_{\rho\in S_N} \frac{(-1)^\rho}{\prod_{i=1}^N \ch(x_i - y_{\rho(i)})} = \det M,
\label{cauchy}
\end{equation}
where $M_{ij} = 1/\ch(x_i - y_j)$ and $i, j = 1, \ldots, N$, proves useful for removing ``$\sh$'' factors from the integrand.

We first check that accounting for the volume factor of two (for $PSU$ versus $SU$ gauge group) is necessary to match the partition function to that of $SU(2)$ SQCD.\footnote{Alternatively, the affine $D_n$ quiver can be realized by gluing a non-affine $D_3$ quiver to a $D_{n-3}$ quiver by gauging the $SU(2)$ flavor node(s). Each $D_k$ quiver takes the form
\[
\xymatrix{
U(1)\ar@{-}[r] & U(2) \ar@{-}[r] & \cdots \ar@{-}[r] & U(2) \ar@{-}[r] & \boxed{SU(2)} \\
& U(1) \ar@{-}[u]
}
\]
with $k-2$ $U(2)$ gauge nodes.}  Relabeling $\sigma_{1, 3} = u_0^{1, 2}$ and $\sigma_{2, 4} = u_{n-2}^{1, 2}$, we write
\begin{align*}
Z_{D_n} &= \frac{1}{2^{n-4}}\int \left(\prod_{I=0}^{n-2}\prod_{i=1}^2 du_I^i\right)\delta(u_1^1 + u_1^2)\frac{\prod_{I=1}^{n-3} \sh(u_I^1 - u_I^2)^2}{\prod_{I=0}^{n-3}\prod_{i, j = 1}^2 \ch(u_I^i - u_{I+1}^j)} \\
&= \frac{1}{2^{n-4}}\int \left(\prod_{I=0}^{n-2}\prod_{i=1}^2 du_I^i\right)\frac{\delta(u_1^1 + u_1^2)}{\sh(u_0^1 - u_0^2)\sh(u_{n-2}^1 - u_{n-2}^2)}\frac{\prod_{I=0}^{n-3} \sh(u_I^1 - u_I^2)\sh(u_{I+1}^1 - u_{I+1}^2)}{\prod_{I=0}^{n-3}\prod_{i, j = 1}^2 \ch(u_I^i - u_{I+1}^j)}
\end{align*}
and then use \eqref{cauchy} in the form
\begin{equation*}
\frac{\prod_{i<j} \sh(u_I^i - u_I^j)\sh(u_{I+1}^i - u_{I+1}^j)}{\prod_{i, j} \ch(u_I^i - u_{I+1}^j)} = \sum_{\rho\in S_2} \frac{(-1)^\rho}{\prod_{i=1}^2 \ch(u_I^i - u_{I+1}^{\rho(i)})}
\end{equation*}
to get
\begin{equation*}
Z_{D_n} = \frac{1}{2^{n-4}}\int \left(\prod_{I=0}^{n-2}\prod_{i=1}^2 du_I^i\right)\frac{\delta(u_1^1 + u_1^2)}{\sh(u_0^1 - u_0^2)\sh(u_{n-2}^1 - u_{n-2}^2)}\prod_{I=0}^{n-3}\sum_{\rho_I\in S_2} \frac{(-1)^{\rho_I}}{\prod_{i=1}^2 \ch(u_I^i - u_{I+1}^{\rho_I(i)})}.
\end{equation*}
Now note that the integrand is even under swapping the integration variables $u_I^1$ and $u_I^2$ for $I = 0, \ldots, n - 2$; by swapping these variables in turn, this becomes simply
\begin{align}
Z_{D_n} &= \frac{1}{2^{n-4}}\int \left(\prod_{I=0}^{n-2}\prod_{i=1}^2 du_I^i\right)\frac{\delta(u_1^1 + u_1^2)}{\sh(u_0^1 - u_0^2)\sh(u_{n-2}^1 - u_{n-2}^2)}\prod_{I=0}^{n-3}\frac{2}{\prod_{i=1}^2 \ch(u_I^i - u_{I+1}^i)} \nonumber \\
&= 4\int \left(\prod_{I=0}^{n-2}\prod_{i=1}^2 du_I^i\right)\frac{\delta(u_1^1 + u_1^2)}{\sh(u_0^1 - u_0^2)\sh(u_{n-2}^1 - u_{n-2}^2)\prod_{I=0}^{n-3}\prod_{i=1}^2 \ch(u_I^i - u_{I+1}^i)}. \label{afterswapping}
\end{align}
Using \eqref{basicfourier} and simplifying, we get
\begin{align}
Z_{D_n} &= 4\int \left(\prod_{I=0}^{n-2}\prod_{i=1}^2 du_I^i\right)\left(\prod_{I=0}^{n-3}\prod_{i=1}^2 ds_I^i\right)\frac{\delta(u_1^1 + u_1^2)}{\sh(u_0^1 - u_0^2)\sh(u_{n-2}^1 - u_{n-2}^2)}\prod_{I=0}^{n-3}\prod_{i=1}^2 \frac{e^{2\pi is_I^i(u_I^i - u_{I+1}^i)}}{\ch(s_I^i)} \nonumber \\
&= \cdots \nonumber \\
&= \frac{1}{2}\int \frac{ds_0^1\, ds_1^1}{\ch(s_0^1)^2\ch(s_1^1)^{2(n - 3)}}\Th(s_0^1)\Th(s_1^1)\delta(s_0^1 - s_1^1) = \frac{1}{4}\int ds\, \frac{\sh(s)^2}{\ch(s/2)^{2n}}, \label{zdnsimplified}
\end{align}
which coincides with the partition function of $SU(2)$ with $n$ flavors up to our conventional factor of $1/r^2$ (see \eqref{su2partition}).\footnote{Another typo in \cite{Benvenuti:2011ga} can be found in their formula (3.3).  The correct version is
\begin{equation}
\frac{1}{2}\int dx\, \frac{\sh(2x)^2}{\prod_{i=1}^{N_f} \ch(x - m_i)\ch(x + m_i)} = (-1)^{N_f + 1}\sum_{i=1}^{N_f} \frac{m_i\sh(2m_i)}{\prod_{j\neq i} (\sh(m_i)^2 - \sh(m_j)^2)},
\end{equation}
the $(-1)^{N_f + 1}$ on the RHS having been overlooked.  The LHS is the partition function of $SU(2)$ SQCD with $N_f$ flavors and mass parameters, which reduces to our expression \eqref{su2partition} (up to the $1/r^2$) when the $m_i = 0$.  However, the above equality holds only when the mass parameters are distinct.}

\paragraph{Computation of TQM correlators.}

We now compute the one- and two-point functions $\langle \cZ\rangle$ and $\langle \cZ\star \cZ\rangle$.  Recall that
\begin{equation}
\cZ\equiv -\tilde{Q}_1 Q_3\tilde{Q}_3 Q_1 = -(\tilde{Q}_1)^i(Q_3)_i(\tilde{Q}_3)^j(Q_1)_j.
\end{equation}
It is helpful to note that integration by parts can be used to simplify $\langle \cZ\cdots \cZ\rangle_\text{HB}$: in the matrix model \eqref{zdn} for $Z_{D_n}$, an insertion of the form
\begin{equation}
\ch(\sigma - u)\ch(\sigma + u)\partial_\sigma\left[\frac{\Th(\sigma - u)^p\Th(\sigma + u)^q}{\ch(\sigma - u)\ch(\sigma + u)}\right]
\end{equation}
(where $\sigma$ is $\sigma_1$ or $\sigma_3$ and $u\equiv u_1^1 = -u_1^2$) is a total derivative.  In particular, taking $(p, q)$ to be $(0, 0)$ or $(1, 0)$ shows that the expressions
\begin{equation}
\Th(\sigma - u) + \Th(\sigma + u), \quad 1 - 2\Th(\sigma - u)^2 - \Th(\sigma - u)\Th(\sigma + u)
\end{equation}
are total derivatives.  This observation is useful because before simplification, $\langle \cZ\cdots \cZ\rangle_\text{HB}$ is a polynomial in $\Th(\sigma\pm u)$, while dropping total derivatives allows it to be written as a polynomial in $\Th(\sigma - u)$; this facilitates manipulation of the resulting integrals because it allows for shifts of $\sigma$ by $u$, thus decoupling the $u$ integral.

To set the stage, we determine the expression for the partition function after integrating over all Higgs branch variables except for $u_0^i$ (i.e., $\sigma_{1, 3}$) and $u_1^i$.  This yields a simplified Higgs branch matrix model with all scalar VEVs integrated out, apart from those relevant to an insertion of $\langle \cZ\cdots \cZ\rangle_\text{HB}$.  Starting from the first line of \eqref{zdnsimplified}, we derive that:
\begin{align}
Z_{D_n} &= -2i\int \frac{ds\, e^{2\pi is(u_1^1 - u_1^2)}\Th(s)}{\ch(s)^{2n - 6}}\left(\prod_{i=1}^2 du_0^i\, du_1^i\right)\frac{\delta(u_1^1 + u_1^2)}{\sh(u_0^1 - u_0^2)\ch(u_0^1 - u_1^1)\ch(u_0^2 - u_1^2)} \nonumber \\
&= -2i\int \frac{ds\, e^{4\pi isu}\Th(s)}{\ch(s)^{2n - 6}}\left(\prod_{i=1}^2 du_0^i\right)\frac{du}{\sh(u_0^1 - u_0^2)\ch(u_0^1 - u)\ch(u_0^2 + u)}.
\end{align}
Let us write
\begin{equation}
\langle{\cdots}\rangle = \frac{1}{Z_{D_n}}Z_{D_n}[\langle{\cdots}\rangle_\text{HB}].
\end{equation}
We now compute by taking Wick contractions that
\begin{equation}
\langle \cZ\rangle_\text{HB} = -\left(\frac{1}{8\pi r}\right)^2(\Th(\sigma_1 + u_1^1)\Th(\sigma_3 + u_1^1) + \Th(\sigma_1 + u_1^2)\Th(\sigma_3 + u_1^2)).
\end{equation}
Setting $u\equiv u_1^1 = -u_1^2$ and integrating by parts allows us to write
\begin{equation}
\langle \cZ\rangle_\text{HB}\sim -2\left(\frac{1}{8\pi r}\right)^2\Th(\sigma_1 - u)\Th(\sigma_3 - u).
\end{equation}
We also have
\begin{equation}
\langle \cZ\star \cZ\rangle_\text{HB} = \left(\frac{1}{8\pi r}\right)^4(I_c + I_s + I_{ss})
\end{equation}
where, as in the $D_4$ case,
\begin{align}
I_c &= ((1 + \Th(\sigma_1 + u_1^1))(1 - \Th(\sigma_3 + u_1^1)) + (u_1^1\leftrightarrow u_1^2))\times (\sigma_1\leftrightarrow \sigma_3), \\
I_s &= (\Th(\sigma_1 + u_1^1)^2(\Th(\sigma_3 + u_1^1)^2 - 1) + (u_1^1\leftrightarrow u_1^2)) + (\sigma_1\leftrightarrow \sigma_3), \\
I_{ss} &= (\Th(\sigma_1 + u_1^1)\Th(\sigma_3 + u_1^1) + (u_1^1\leftrightarrow u_1^2))^2.
\end{align}
Setting $u\equiv u_1^1 = -u_1^2$ and integrating by parts gives
\begin{equation}
I_c + I_s + I_{ss}\sim 4\left[3 - 2\prod_{A = 1, 3} \Th(\sigma_A - u) - 16\sum_{A = 1, 3}\frac{1}{\ch(\sigma_A - u)^2} + 96\prod_{A = 1, 3}\frac{1}{\ch(\sigma_A - u)^2}\right].
\end{equation}
To evaluate the multidimensional integrals for $\langle \cZ\rangle$ and $\langle \cZ\star \cZ\rangle$, our main tool for simplification is to take a Fourier transform whenever an argument of ``$\sh$,'' ``$\ch$,'' or ``$\Th$'' involves a combination of two variables: this allows us to decouple the single-variable integrals.  For instance, we have
\begin{align*}
\langle \cZ\rangle &= -\frac{2i}{Z_{D_n}}\int \frac{ds\, e^{4\pi isu}\Th(s)}{\ch(s)^{2n - 6}}\left(\prod_{i=1}^2 du_0^i\right)\frac{du}{\sh(u_0^1 - u_0^2)\ch(u_0^1 - u)\ch(u_0^2 + u)}\langle \cZ\rangle_\text{HB} \\
&= \frac{4i}{Z_{D_n}}\left(\frac{1}{8\pi r}\right)^2\int \frac{ds\, e^{4\pi isu}\Th(s)}{\ch(s)^{2n - 6}}\left(\prod_{i=1}^2 du_0^i\right)du\, \frac{\Th(u_0^1 - u)\Th(u_0^2 - u)}{\sh(u_0^1 - u_0^2)\ch(u_0^1 - u)\ch(u_0^2 + u)} \\
&= \frac{2i}{Z_{D_n}}\left(\frac{1}{8\pi r}\right)^2\int ds\, \frac{\Th(s)}{\ch(s)^{2n - 5}}\left(\prod_{i=1}^2 du_0^i\right)\frac{\Th(u_0^1)\Th(u_0^2)e^{-2\pi isu_0^2}}{\sh(u_0^1 - u_0^2)\ch(u_0^1)},
\end{align*}
where we have shifted $u_0^i\to u_0^i + u$ and integrated over $u$ using \eqref{basicfourier}.  Taking the Fourier transform of the $1/\sh(u_0^1 - u_0^2)$, and iterating this process as necessary (possibly with the help of various identities from Appendix \ref{integralidentities}), leaves us with nested single-variable integrals that can be evaluated sequentially to yield a single integral:
\begin{equation}
\langle \cZ\rangle = \frac{1}{4Z_{D_n}}\left(\frac{1}{8\pi r}\right)^2\int ds\, \frac{\sh(s)^2}{\ch(s/2)^{2n}}(s^2 - 1).
\end{equation}
For $\langle \cZ\star \cZ\rangle$, we must evaluate three additional integrals (call them $\mathcal{I}_1, \mathcal{I}_2, \mathcal{I}_{12}$).  First, we have
\begin{align}
\mathcal{I}_1 &\equiv \int \frac{ds\, e^{4\pi isu}\Th(s)}{\ch(s)^{2n - 6}}\left(\prod_{i=1}^2 du_0^i\right)\frac{du}{\sh(u_0^1 - u_0^2)\ch(u_0^1 - u)\ch(u_0^2 + u)}\left[\frac{1}{\ch(u_0^1 - u)^2}\right] \nonumber \\
&= \frac{1}{2}\int ds\, \frac{\Th(s)}{\ch(s)^{2n - 5}}\left(\prod_{i=1}^2 du_0^i\right)\frac{e^{-2\pi isu_0^2}}{\sh(u_0^1 - u_0^2)\ch(u_0^1)^3} \nonumber \\
&= \frac{i}{64}\int ds\, \frac{\sh(s)^2}{\ch(s/2)^{2n}}(s^2 + 1). \label{resultI1}
\end{align}
Second, we have
\begin{align}
\mathcal{I}_2 &\equiv \int \frac{ds\, e^{4\pi isu}\Th(s)}{\ch(s)^{2n - 6}}\left(\prod_{i=1}^2 du_0^i\right)\frac{du}{\sh(u_0^1 - u_0^2)\ch(u_0^1 - u)\ch(u_0^2 + u)}\left[\frac{1}{\ch(u_0^2 - u)^2}\right] \nonumber \\
&= \frac{1}{2}\int ds\, \frac{\Th(s)}{\ch(s)^{2n - 5}}\left(\prod_{i=1}^2 du_0^i\right)\frac{e^{-2\pi isu_0^2}}{\sh(u_0^1 - u_0^2)\ch(u_0^1)\ch(u_0^2)^2} \nonumber \\
&= \frac{i}{64}\int ds\, \frac{\sh(s)^2}{\ch(s/2)^{2n}}(s^2 + 1) - \frac{i}{16\pi}\int ds\, \frac{s\sh(s)}{\ch(s/2)^{2n - 2}}. \label{resultI2}
\end{align}
Third, we have
\begin{align}
\mathcal{I}_{12} &\equiv \int \frac{ds\, e^{4\pi isu}\Th(s)}{\ch(s)^{2n - 6}}\left(\prod_{i=1}^2 du_0^i\right)\frac{du}{\sh(u_0^1 - u_0^2)\ch(u_0^1 - u)\ch(u_0^2 + u)}\left[\frac{1}{\prod_{i=1}^2 \ch(u_0^i - u)^2}\right] \nonumber \\
&= \frac{1}{2}\int ds\, \frac{\Th(s)}{\ch(s)^{2n - 5}}\left(\prod_{i=1}^2 du_0^i\right)\frac{e^{-2\pi isu_0^2}}{\sh(u_0^1 - u_0^2)\ch(u_0^1)^3\ch(u_0^2)^2} \nonumber \\
&= \frac{i}{3072}\int ds\, \frac{\sh(s)^2}{\ch(s/2)^{2n}}(s^4 + 10s^2 + 9) - \frac{i}{96\pi}\int ds\, \frac{s\sh(s)}{\ch(s/2)^{2n - 2}}. \label{resultI12}
\end{align}
Combining the results \eqref{resultI1}, \eqref{resultI2}, \eqref{resultI12}, and
\begin{gather}
\int \frac{ds\, e^{4\pi isu}\Th(s)}{\ch(s)^{2n - 6}}\left(\prod_{i=1}^2 du_0^i\right)\frac{du}{\sh(u_0^1 - u_0^2)\ch(u_0^1 - u)\ch(u_0^2 + u)}\left[\Th(u_0^1 - u)\Th(u_0^2 - u)\right] \nonumber \\
= -\frac{i}{16}\int ds\, \frac{\sh(s)^2}{\ch(s/2)^{2n}}(s^2 - 1)
\end{gather}
(which we deduce from our result for $\langle \cZ\rangle$) gives
\begin{align}
\langle \cZ\star \cZ\rangle &= -\frac{2i}{Z_{D_n}}\int \frac{ds\, e^{4\pi isu}\Th(s)}{\ch(s)^{2n - 6}}\left(\prod_{i=1}^2 du_0^i\right)\frac{du}{\sh(u_0^1 - u_0^2)\ch(u_0^1 - u)\ch(u_0^2 + u)}\langle \cZ\star \cZ\rangle_\text{HB} \nonumber \\
&= \frac{1}{4Z_{D_n}}\left(\frac{1}{8\pi r}\right)^4\int ds\, \frac{\sh(s)^2}{\ch(s/2)^{2n}}(s^2 - 1)^2.
\end{align}
Combining the above gives \eqref{resultsZandZZ}.

\section{Quantized Coulomb Branches for \texorpdfstring{$AD$}{AD}} \label{ADcoulomb}

Here, we consider some realizations of deformation quantizations of the $\mathbb{C}^2/\Gamma_{A_N}$ and $\mathbb{C}^2/\Gamma_{D_N}$ singularities by Lagrangian quantum field theories, namely the Coulomb branches of 3D $\mathcal{N} = 4$ $U(1)$ and $SU(2)$ gauge theories with arbitrary matter representations.  In these cases, the choice of basis is strongly constrained by $U(1)$ and $\mathbb{Z}_2$ flavor symmetries, respectively.

We expect both the Coulomb branch chiral ring (the ``classical'' Coulomb branch) and its Poisson structure to depend only on $N$, because $N$ determines the holomorphic symplectic form.\footnote{The fact that the first subleading term in the star product is determined by the Coulomb branch also follows from a less transparent topological descent argument \cite{Beem:2016cbd}.}  We also expect the number of distinct quantizations realized by these theories to be related to partitions of $N$.  An interesting question that one might ask, which we do not attempt to answer here, is: do there exist examples of different Lagrangian theories with the same ``quantum'' Coulomb branch, to higher orders in $\hbar\sim 1/r$ beyond $O(\hbar^1)$?

Nondegenerate short star products for quotient singularities, including Kleinian singularities, have been classified in \cite{Etingof:2019guc}.  For example, even nondegenerate short star products for $A_n$ singularities depend on $n_e + n_s$ parameters where the first $n_e = \lfloor(n + 1)/2\rfloor$ parameters determine the corresponding quantum algebra $\mathcal{A}$ up to isomorphism (i.e., the period of the quantization) and the remaining $n_s = n_e + ((-1)^n - 1)/2$ parameters determine maps from the associated graded algebra $\operatorname{gr}(\mathcal{A})$ (the ``commutative limit'' of the associative algebra $\mathcal{A}$) into $\mathcal{A}$, corresponding to physical gauge fixings.  This agrees with the counting of free parameters in \cite{Beem:2016cbd} for the examples of $A_{n\leq 4}$ \emph{before} imposing unitarity (i.e., positivity), which is a stronger condition than nondegeneracy.  In the examples below, fixing a Lagrangian SCFT should be understood as fixing a particular value of the period for the quantization.

\subsection{\texorpdfstring{$U(1)$}{U(1)}} \label{U1coulomb}

Consider $U(1)$ for some set of charges $\{q\}$ with multiplicities $\{N_q\}$, where $q\in \mathbb{Z}\setminus\{0\}$ and $N_q\in \mathbb{Z}_{\geq 0}$ (uncharged matter does not contribute, but we may consider the pure case).  The shift operators for the Coulomb branch chiral ring generators are:
\begin{equation}
\mathcal{M}^{\pm 1} = \prod_q \left[\frac{(-1)^{(|q|\pm q)/2}}{r^{|q|/2}}\left(\frac{1 - qB}{2} + iq\sigma\right)_{(|q|\pm q)/2}\right]^{N_q}e^{\mp(\frac{i}{2}\partial_\sigma + \partial_B)}, \quad \Phi = \frac{1}{r}\left(\sigma + \frac{i}{2}B\right).
\end{equation}
We compute that
\begin{equation}
\mathcal{M}^{\mp 1}\star \mathcal{M}^{\pm 1} = \prod_q (-iq\Phi)^{|q|N_q} + O\left(\frac{1}{r}\right).
\end{equation}
Setting $N = \sum_q |q|N_q$ and
\begin{equation}
\mathcal{X} = \frac{1}{(4\pi)^{N/2}C^{1/2}}\mathcal{M}^{-1}, \quad \mathcal{Y} = \frac{1}{(4\pi)^{N/2}C^{1/2}}\mathcal{M}^1, \quad \mathcal{Z} = -\frac{i}{4\pi}\Phi, \quad C\equiv \prod_q q^{|q|N_q}
\end{equation}
(this normalization being natural from the point of view of correlation functions), we find that $\mathcal{X}\mathcal{Y} = \mathcal{Z}^N$ in the chiral ring.  Accounting for sign, we obtain
\begin{equation}
\sum_{P\in \{\text{partitions of $N$}\}} 2^{\#\operatorname{parts}(P)} > p(N)
\end{equation}
distinct quantizations from these theories for fixed $N$.  At finite $r$ and to subleading order in $1/r$, we compute that
\begin{align}
\mathcal{M}^{\mp 1}\star \mathcal{M}^{\pm 1} &= \prod_q \left[\frac{1}{r^{|q|}}\left(-iqr\Phi + \frac{|q|\pm q - 1}{2}\right)_{|q|}\right]^{N_q} \\
&= \left[\prod_q (-iq\Phi)^{|q|N_q}\right]\left[1 + \frac{i}{r\Phi}\sum_q N_q\left(\pm\frac{|q|}{2} + q - \operatorname{sgn}(q)\right)\right] + O\left(\frac{1}{r^2}\right),
\end{align}
so that
\begin{equation}
[\mathcal{M}^{-1}, \mathcal{M}^1]_\star = \frac{i}{r\Phi}\left(\sum_q |q|N_q\right)\left[\prod_q (-iq\Phi)^{|q|N_q}\right] + O\left(\frac{1}{r^2}\right).
\end{equation}
Equivalently,
\begin{equation}
[\mathcal{X}, \mathcal{Y}]_\star = \frac{1}{r}P(\mathcal{Z}) = \frac{N}{4\pi r}\mathcal{Z}^{N-1} + O\left(\frac{1}{r^2}\right).
\end{equation}
Hence the Poisson structure, like the chiral ring, depends only on $N$ (as expected).  These quantizations are distinguished by the coefficients of the subleading terms in the polynomial $P(\mathcal{Z})$ (computing the commutator is simpler than directly computing three-point functions because various gauge-fixing ambiguities cancel in the former).

The structure constants for the deformation quantizations corresponding to the Higgs branch of the affine $A_{2, 3}$ quivers were originally bootstrapped in \cite{Beem:2016cbd} and later derived from localization in \cite{Dedushenko:2016jxl}.  By using the above techniques for the Coulomb branch of the mirror dual, we obtain these results and more with very little effort.

\subsection{\texorpdfstring{$SU(2)$}{SU(2)}}

Consider $SU(2)$ SQCD with matter specified by some set of spins $\{j\}$ with multiplicities $\{N_j\}$, where $j\in \frac{1}{2}\mathbb{Z}_{> 0}$ and $N_j\in \mathbb{Z}_{\geq 0}$ (uncharged matter does not contribute, but we may consider the pure case).  In conventions where the weights of $SU(2)$ are half-integers and monopole charges are even integers ($b\in 2\mathbb{Z}$), we have
\begin{equation}
M^b = \frac{\prod_j \left[\prod_{m_j} \frac{(-1)^{(m_j b)_+}}{r^{|m_j b|/2}}(\frac{1}{2} + irm_j\Phi)_{(m_j b)_+}\right]^{N_j}}{\frac{1}{r^{|b|}}(ir\operatorname{sgn}(b)\Phi)_{|b|}}e^{-b(\frac{i}{2}\partial_\sigma + \partial_B)}, \quad \Phi^2 = \frac{1}{r^2}\left(\sigma + \frac{i}{2}B\right)^2,
\end{equation}
where $m_j\in \{-j, -j + 1, \ldots, j\}$.  Indicating the commutative limit with an $\infty$ subscript (for $r\to\infty$), we have
\begin{align}
M_\infty^b &= (-i\operatorname{sgn}(b)\Phi)^{|b|(\frac{1}{2}\sum_j S_j N_j - 1)}\left(\prod_j\prod_{m_j>0} m_j^{|b|m_j N_j}\right)e[b], \\
S_j &\equiv \sum_{m_j=-j}^j |m_j| = \begin{cases} j(j + 1) & \text{if $j\in \mathbb{Z}$}, \\ (j + 1/2)^2 & \text{if $j\in \mathbb{Z} + \frac{1}{2}$}. \end{cases}
\end{align}
Set $N = \sum_j S_j N_j$.  Then in particular, we see that
\begin{equation}
\Delta(\mathcal{M}^2) = N - 2, \quad \Delta(\Phi\mathcal{M}^2) = N - 1, \quad \Delta(\Phi^2) = 2.
\end{equation}
On dimensional grounds, the bubbling coefficient for $\mathcal{M}_\infty^2$ is a monomial in $\Phi$ for $N\geq 2$, which we can eliminate by a change of basis.  So for $N\geq 2$:
\begin{align}
\mathcal{M}_\infty^2 &= M_\infty^2 + M_\infty^{-2} = \left(\prod_j\prod_{m_j>0} m_j^{2m_j N_j}\right)(i\Phi)^{N - 2}((-1)^{N}e[2] + e[-2]), \\
\Phi\mathcal{M}_\infty^2 &= \Phi(M_\infty^2 - M_\infty^{-2}) = \left(\prod_j\prod_{m_j>0} m_j^{2m_j N_j}\right)\Phi(i\Phi)^{N - 2}((-1)^{N}e[2] - e[-2]).
\end{align}
Using $e[2]e[-2] = 1$ gives
\begin{equation}
\Phi^2(\mathcal{M}_\infty^2)^2 - (\Phi\mathcal{M}_\infty^2)^2 = 4\left(\prod_j\prod_{m_j>0} m_j^{2m_j N_j}\right)^2(\Phi^2)^{N-1}.
\end{equation}
Then setting
\begin{equation}
\mathcal{X} = C^{-1}\Phi\mathcal{M}_\infty^2, \quad \mathcal{Y} = -iC^{-1}\mathcal{M}_\infty^2, \quad \mathcal{Z} = \Phi^2, \quad C\equiv 2\left(\prod_j\prod_{m_j>0} m_j^{2m_j N_j}\right)
\end{equation}
yields the equation of a $D_N$ singularity:
\begin{equation}
\mathcal{X}^2 + \mathcal{Z}\mathcal{Y}^2 + \mathcal{Z}^{N-1} = 0.
\end{equation}
For $N = 1$ (i.e., $N_{1/2} = 1$), one can show that the relevant bubbling coefficient vanishes by a polynomiality computation at finite $r$ \cite{Dedushenko:2018icp}, but let us not assume this.  We have
\begin{align}
\mathcal{M}_\infty^2 &= \left(M_\infty^2 + \frac{c}{\Phi}\right) + \left(M_\infty^{-2} - \frac{c}{\Phi}\right) = -\frac{1}{2i\Phi}(e[2] - e[-2]), \\
\Phi\mathcal{M}_\infty^2 &= \Phi\left(M_\infty^2 + \frac{c}{\Phi}\right) - \Phi\left(M_\infty^{-2} - \frac{c}{\Phi}\right) = -\frac{1}{2i}(e[2] + e[-2]) + 2c,
\end{align}
so that
\begin{equation}
\Phi^2(\mathcal{M}_\infty^2)^2 - (\Phi\mathcal{M}_\infty^2 - 2c)^2 = 1.
\end{equation}
Equivalently,
\begin{equation}
\mathcal{X} = \Phi\mathcal{M}_\infty^2, \quad \mathcal{Y} = -i\mathcal{M}_\infty^2, \quad \mathcal{Z} = \Phi^2, \quad (\mathcal{X} - 2c)^2 + \mathcal{Z}\mathcal{Y}^2 + 1 = 0.
\end{equation}
Unless $c = 0$, this is a nonsingular deformation of a $D_1$ singularity (as can be seen from the nonvanishing of the partial derivatives at $(0, 0, 0)$).  For $N = 0$ (the pure case), we have
\begin{align}
\mathcal{M}_\infty^2 &= \left(M_\infty^2 + \frac{c}{\Phi^2}\right) + \left(M_\infty^{-2} + \frac{c}{\Phi^2}\right) = -\frac{1}{\Phi^2}(e[2] + e[-2]) + \frac{2c}{\Phi^2}, \\
\Phi\mathcal{M}_\infty^2 &= \Phi\left(M_\infty^2 + \frac{c}{\Phi^2}\right) - \Phi\left(M_\infty^{-2} + \frac{c}{\Phi^2}\right) = -\frac{1}{\Phi}(e[2] - e[-2]),
\end{align}
and therefore
\begin{equation}
(\Phi^2\cdot \mathcal{M}_\infty^2 - 2c)^2 - \Phi^2(\Phi\mathcal{M}_\infty^2)^2 = 4.
\end{equation}
Equivalently,
\begin{equation}
\mathcal{X} = \Phi\mathcal{M}_\infty^2, \quad \mathcal{Y} = -i\mathcal{M}_\infty^2, \quad \mathcal{Z} = \Phi^2, \quad \mathcal{Z}(\mathcal{X}^2 + \mathcal{Z}\mathcal{Y}^2 + 4ic\mathcal{Y}) + 4(1 - c^2) = 0.
\end{equation}
The degree of the relation is reduced when $c = \pm 1$ (the sign ambiguity is present even when using polynomiality \cite{Dedushenko:2018icp}):
\begin{equation}
\mathcal{X} = \Phi\mathcal{M}_\infty^2, \quad \mathcal{Y} = \pm 4\mathcal{M}_\infty^2, \quad \mathcal{Z} = \Phi^2, \quad \mathcal{X}^2 + \mathcal{Z}\mathcal{Y}^2 + \mathcal{Y} = 0,
\end{equation}
where we have slightly redefined the variables.  This gives an alternative way to fix $c^2$.  Note that the theory is good for $N\geq 3$, we expect the $D_N$ equation to hold for $N\geq 1$, and we expect it to be modified as above for $N = 0$.  The possibilities for bad theories are simply $\{j\} = \{\}$ ($N = 0$), $\{j\} = \{1/2\}$ ($N = 1$), and $\{j\} = \{1/2, 1/2\}, \{1\}$ ($N = 2$).

\section{Higgs Branch Chiral Rings for \texorpdfstring{$DE$}{DE}} \label{DEhiggs}

In this appendix, we derive the Higgs branch chiral rings of the $D$- and $E$-type quivers considered in the main text.  We discuss the $D_n$ chiral ring in some detail, since a comprehensive derivation seems to be missing from the literature (see \cite{Lindstrom:1999pz}, Section 5 of \cite{Collinucci:2016hpz}, and Appendix A.1 of \cite{Collinucci:2017bwv} for earlier discussions).  In the cases of $E_{6, 7, 8}$, we also fill in some details regarding existing derivations (useful references include \cite{Lindstrom:1999pz} and Appendix A.2 of \cite{Collinucci:2017bwv}).

Note that for 3D $\mathcal{N} = 4$ theories containing only vector multiplets and hypermultiplets, there exists no distinction between the D-term and F-term relations in 3D $\mathcal{N} = 2$ language because the auxiliary fields combine into an $SU(2)_R$ triplet (equivalently, the K\"ahler potential fixes the superpotential).  Hence we may equivalently write the D-term relations in the TQM, which take the form
\begin{equation}
(\tilde{Q}\mathcal{R}(T)Q)(\varphi) = 0
\label{Dterm}
\end{equation}
for all $T\in \mathfrak{g}$ \cite{Dedushenko:2016jxl}, or derive the F-term relations from the superpotential, as we do below.\footnote{While \eqref{Dterm} holds at the level of the chiral ring, it may be modified by contact terms at the level of correlation functions.  Additionally, the RHS of \eqref{Dterm} receives contributions from FI parameters, which we have set to zero.}

\subsection{Affine \texorpdfstring{$D_4$}{D4} Quiver} \label{D4chiralring}

The affine $D_4$ quiver contains hypermultiplets $(Q_A)_i$, $(\tilde{Q}_A)^i$ with $A = 1, \ldots, 4$ and $i = 1, 2$.  The superpotential is 
\begin{equation}
W=\Phi_{ij} \sum_A Q_A^i \tilde Q_A^j+\sum_A \phi_A Q_A^i \tilde Q_A^j \epsilon_{ij}
\end{equation}
where $\Phi$ and $\phi_A$ are adjoint chirals for the $SU(2)$ and $U(1)$ gauge nodes, respectively.  We introduce the notation $\langle AB\rangle\equiv \tilde{Q}_A Q_B$; then the F-term relations give
\begin{equation}
\sum_A A\rangle\langle A = \langle AA\rangle = 0.
\end{equation}
For fixed $A$, we have the four relations
\begin{equation}
\sum_{B\neq A} \langle AB\rangle\langle BA\rangle = 0.
\end{equation}
Hence out of the six candidate chiral ring generators with $\Delta = 2$, namely
\begin{equation}
\langle AB\rangle\langle BA\rangle \text{ with } A < B,
\end{equation}
only two are independent.  We also see that out of the eight candidate chiral ring generators with $\Delta = 3$, namely
\begin{equation}
\langle AB\rangle\langle BC\rangle\langle CA\rangle \text{ with } A < B < C \text{ or } A < C < B,
\end{equation}
only one of them is independent because any two such operators are equal by one of the twelve relations
\begin{equation}
\sum_C \langle AC\rangle\langle CB\rangle = 0
\end{equation}
for fixed $A, B$ with $A\neq B$ (here, the order of $A$ and $B$ matters).  The properly normalized chiral ring generators may be taken to be
\begin{align}
\cZ &= \sqrt{3}(\langle 13\rangle\langle 31\rangle + \langle 23\rangle\langle 32\rangle), \\
\cY &= \sqrt{3}i(\langle 13\rangle\langle 31\rangle - \langle 23\rangle\langle 32\rangle), \\
\cX &= 2\cdot 3^{3/4}i\langle 12\rangle\langle 23\rangle\langle 31\rangle.
\end{align}
They satisfy the chiral ring relation for $D_4$ because
\begin{align}
\cX^2 + \cZ\cY^2 + \cZ^3 &= 12\sqrt{3}\langle 23\rangle\langle 31\rangle(\langle 13\rangle^2\langle 31\rangle\langle 32\rangle + \langle 32\rangle^2\langle 23\rangle\langle 13\rangle - \langle 12\rangle^2\langle 23\rangle\langle 31\rangle) \nonumber \\
&= 12\sqrt{3}\langle 23\rangle\langle 31\rangle(-\langle 13\rangle\langle 31\rangle\langle 14\rangle\langle 42\rangle - \langle 23\rangle\langle 32\rangle\langle 14\rangle\langle 42\rangle - \langle 14\rangle\langle 43\rangle\langle 34\rangle\langle 42\rangle) \nonumber \\
&= 12\sqrt{3}\langle 23\rangle\langle 31\rangle\langle 14\rangle\langle 42\rangle(-\langle 31\rangle\langle 13\rangle - \langle 32\rangle\langle 23\rangle - \langle 34\rangle\langle 43\rangle) = 0.
\end{align}
Moreover, the $S_3$ generators \eqref{s3generators} act as
\begin{equation}
r_{\mZ_2} : \begin{cases} \la 12\ra \la 23 \ra \la 31\ra \mapsto \la 21\ra \la 13 \ra \la 32\ra = -\la 24\ra \la 43 \ra \la 32\ra = \la 24\ra \la 41 \ra \la 12\ra = -\la 23\ra \la 31 \ra \la 12\ra, \\ \la 13\ra \la 31 \ra \mapsto \la 23\ra \la 32 \ra \end{cases}
\end{equation}
and
\begin{equation}
s_{\mZ_3} : \begin{cases} \la 12\ra \la 23 \ra \la 31\ra \mapsto \la 23\ra \la 31 \ra \la 12\ra, \\ \la 13\ra \la 31 \ra \mapsto \la 21\ra \la 12 \ra = \la 34\ra \la 43 \ra = -\la 13\ra \la 31 \ra-\la 23\ra \la 32 \ra, \\ \la 23\ra \la 32 \ra \mapsto \la 31\ra \la 13 \ra, \end{cases}
\end{equation}
giving the expected \eqref{s3action}.

\subsection{Affine \texorpdfstring{$D_{n > 4}$}{D(n > 4)} Quiver} \label{Dnchiralring}

For the affine $D_n$ quiver, we have adjoint chirals $\phi_A$ and $\Phi_I$ ($I = 1, \ldots, n - 3$) for the $U(1)$ and $U(2)$ nodes, respectively.  The superpotential is
\begin{align}
W &= ((\tilde{Q}_1)^i(\Phi_1)_i{}^j(Q_1)_j - \phi_1(\tilde{Q}_1)^i(Q_1)_i) + ((\tilde{Q}_3)^i(\Phi_1)_i{}^j(Q_3)_j - \phi_3(\tilde{Q}_3)^i(Q_3)_i) \nonumber \\
&\phantom{==} + ((\tilde{Q}_2)^i(\Phi_{n-3})_i{}^j(Q_2)_j - \phi_2(\tilde{Q}_2)^i(Q_2)_i) + ((\tilde{Q}_4)^i(\Phi_{n-3})_i{}^j(Q_4)_j - \phi_4(\tilde{Q}_4)^i(Q_4)_i) \nonumber \\
&\phantom{==} + \textstyle \sum_{I=1}^{n-4} ((\tilde{K}_I)_k{}^i(\Phi_I)_i{}^j(K_I)_j{}^k - (K_I)_k{}^i(\Phi_{I+1})_i{}^j(\tilde{K}_I)_j{}^k).
\end{align}
The signs keep track of orientation in the $\mathcal{N} = 2$ sense (the legs are unoriented in the $\mathcal{N} = 4$ sense).  The F-term relations are
\begin{align}
(\tilde{Q}_A)^i(Q_A)_i &= 0 \quad (A = 1, 2, 3, 4), \\
(Q_1)_i(\tilde{Q}_1)^j + (Q_3)_i(\tilde{Q}_3)^j + (K_1)_i{}^k(\tilde{K}_1)_k{}^j &= 0, \\
(Q_2)_i(\tilde{Q}_2)^j + (Q_4)_i(\tilde{Q}_4)^j - (\tilde{K}_{n-4})_i{}^k(K_{n-4})_k{}^j &= 0, \\
(\tilde{K}_I)_i{}^k(K_I)_k{}^j - (K_{I+1})_i{}^k(\tilde{K}_{I+1})_k{}^j &= 0 \quad (I = 1, \ldots, n - 5).
\end{align}
It should be kept in mind that the $U(2)$ indices are associated with different nodes.  Below, gauge indices are appropriately contracted between pairs of hypers when suppressed.

To justify our description of the Higgs branch chiral ring in \eqref{hbdnfirst}--\eqref{hbdnlast}, we first list some useful equivalences between chiral ring elements, which are reflected in correlation functions.\footnote{In the process, we fix several mistakes in (A.3) of \cite{Collinucci:2017bwv}.}  From the F-term relations, we derive
\begin{align}
\tilde{Q}_2(\tilde{K}_{n-4}K_{n-4})^a Q_2 &= \tilde{Q}_4(\tilde{K}_{n-4}K_{n-4})^a Q_4 \nonumber \\
&= \tilde{Q}_2(Q_2\tilde{Q}_2 + Q_4\tilde{Q}_4)^a Q_2 = \tilde{Q}_4(Q_2\tilde{Q}_2 + Q_4\tilde{Q}_4)^a Q_4 \nonumber \\
&= \begin{cases} 0 & a\in 2\mathbb{Z}, \\ (\tilde{Q}_2 Q_4\tilde{Q}_4 Q_2)^{(a + 1)/2} & a\in 2\mathbb{Z} + 1 \end{cases} \label{24powersimp}
\end{align}
and
\begin{align}
\tilde{Q}_1(K_1\tilde{K}_1)^a Q_1 &= \tilde{Q}_3(K_1\tilde{K}_1)^a Q_3 \nonumber \\
&= (-1)^a\tilde{Q}_1(Q_1\tilde{Q}_1 + Q_3\tilde{Q}_3)^a Q_1 = (-1)^a\tilde{Q}_3(Q_1\tilde{Q}_1 + Q_3\tilde{Q}_3)^a Q_3 \nonumber \\
&= \begin{cases} 0 & a\in 2\mathbb{Z}, \\ -(\tilde{Q}_1 Q_3\tilde{Q}_3 Q_1)^{(a + 1)/2} & a\in 2\mathbb{Z} + 1. \end{cases} \label{13powersimp}
\end{align}
Similarly, we derive that
\begin{align}
\tilde{Q}_2(\tilde{K}_{n-4}K_{n-4})^a Q_4\tilde{Q}_4 Q_2 &= \tilde{Q}_4(\tilde{K}_{n-4}K_{n-4})^a Q_2\tilde{Q}_2 Q_4 \nonumber \\
&= \begin{cases} (\tilde{Q}_2 Q_4\tilde{Q}_4 Q_2)^{a/2 + 1} & a\in 2\mathbb{Z}, \\ 0 & a\in 2\mathbb{Z} + 1, \end{cases} \\
\tilde{Q}_1(K_1\tilde{K}_1)^a Q_3\tilde{Q}_3 Q_1 &= \tilde{Q}_3(K_1\tilde{K}_1)^a Q_1\tilde{Q}_1 Q_3 \nonumber \\
&= \begin{cases} (\tilde{Q}_1 Q_3\tilde{Q}_3 Q_1)^{a/2 + 1} & a\in 2\mathbb{Z}, \\ 0 & a\in 2\mathbb{Z} + 1. \end{cases} \label{1331powersimp}
\end{align}
Moreover, we see that
\begin{equation}
\tilde{Q}_A\tilde{K}_{n-4}\cdots \tilde{K}_1(K_1\tilde{K}_1)^a K_1\cdots K_{n-4}Q_{A'} = \tilde{Q}_A(\tilde{K}_{n-4}K_{n-4})^{n + a - 4}Q_{A'}
\label{24sandwich}
\end{equation}
for $A, A'\in \{2, 4\}$ and
\begin{equation}
\tilde{Q}_A K_1\cdots K_{n-4}(\tilde{K}_{n-4}K_{n-4})^a\tilde{K}_{n-4}\cdots \tilde{K}_1 Q_{A'} = \tilde{Q}_A(K_1\tilde{K}_1)^{n + a - 4}Q_{A'}
\label{13sandwich}
\end{equation}
for $A, A'\in \{1, 3\}$ (these operators by themselves are not gauge-invariant unless $A = A'$).  Finally, rearranging and squaring both sides of the F-term equations for the trivalent $U(2)$ nodes gives
\begin{align*}
(Q_1)_i(\tilde{Q}_1)^j + (Q_3)_i(\tilde{Q}_3)^j = -(K_1)_i{}^k(\tilde{K}_1)_k{}^j &\implies 2\tilde{Q}_1 Q_3\tilde{Q}_3 Q_1 = \Tr((K_1\tilde{K}_1)^2), \\
(Q_2)_i(\tilde{Q}_2)^j + (Q_4)_i(\tilde{Q}_4)^j = (\tilde{K}_{n-4})_i{}^k(K_{n-4})_k{}^j &\implies 2\tilde{Q}_2 Q_4\tilde{Q}_4 Q_2 = \Tr((\tilde{K}_{n-4}K_{n-4})^2).
\end{align*}
But $\Tr((K_I\tilde{K}_I)^2) = \Tr((\tilde{K}_I K_I)^2)$ and $\Tr((\tilde{K}_I K_I)^2) = \Tr((K_{I+1}\tilde{K}_{I+1})^2)$ (the latter for $I = 1, \ldots, n - 5$), implying that
\begin{equation}
\tilde{Q}_1 Q_3\tilde{Q}_3 Q_1 = \tilde{Q}_2 Q_4\tilde{Q}_4 Q_2
\label{1331is2442}
\end{equation}
in the chiral ring.\footnote{This conclusion also holds for $n = 4$, from squaring both sides of
\[
(Q_1)_i(\tilde{Q}_1)^j + (Q_3)_i(\tilde{Q}_3)^j = -(Q_2)_i(\tilde{Q}_2)^j - (Q_4)_i(\tilde{Q}_4)^j.
\]}

Now consider the $\mathbb{Z}_2$ action on \eqref{hbdnfirst}--\eqref{hbdnlast}.  The ``$U(1)$ Schouten identity'' implies that
\begin{equation}
\tilde{Q}_A K_1\cdots K_{n-4}Q_{A'}\tilde{Q}_{A'}\tilde{K}_{n-4}\cdots \tilde{K}_1 Q_A = \tilde{Q}_{A'}\tilde{K}_{n-4}\cdots \tilde{K}_1 Q_A\tilde{Q}_A K_1\cdots K_{n-4}Q_{A'}
\end{equation}
where $A\in \{1, 3\}$ and $A'\in \{2, 4\}$.  From \eqref{13sandwich} and \eqref{13powersimp}, we have
\begin{equation}
\tilde{Q}_3 K_1\cdots K_{n-4}(Q_2\tilde{Q}_2 + Q_4\tilde{Q}_4)\tilde{K}_{n-4}\cdots \tilde{K}_1 Q_3 = \begin{cases} 0 & n\in 2\mathbb{Z} + 1, \\ -(\tilde{Q}_1 Q_3\tilde{Q}_3 Q_1)^{n/2 - 1} & n\in 2\mathbb{Z}. \end{cases}
\end{equation}
From \eqref{13sandwich} and \eqref{1331powersimp}, we also have
\begin{equation}
\tilde{Q}_1 K_1\cdots K_{n-4}(Q_2\tilde{Q}_2 + Q_4\tilde{Q}_4)\tilde{K}_{n-4}\cdots \tilde{K}_1 Q_3\tilde{Q}_3 Q_1 = \begin{cases} (\tilde{Q}_1 Q_3\tilde{Q}_3 Q_1)^{(n - 1)/2} & n\in 2\mathbb{Z} + 1, \\ 0 & n\in 2\mathbb{Z}. \end{cases}
\end{equation}
So we see that the $\mathbb{Z}_2$ symmetry that takes $2\leftrightarrow 4$ (i.e., $(Q_2, \tilde{Q}_2)\leftrightarrow (Q_4, \tilde{Q}_4)$) acts as
\begin{equation}
\mathbb{Z}_2 : (\cX, \cY, \cZ)\mapsto (-\cX, -\cY, \cZ)
\end{equation}
regardless of whether $n\in 2\mathbb{Z}$ or $n\in 2\mathbb{Z} + 1$.  Equivalently, the $\mathbb{Z}_2$ symmetry can be implemented by swapping $1\leftrightarrow 3$ (i.e., $(Q_1, \tilde{Q}_1)\leftrightarrow (Q_3, \tilde{Q}_3)$).  To see this, note that \eqref{24powersimp}, \eqref{24sandwich}, and \eqref{1331is2442} imply that
\begin{equation}
\tilde{Q}_2\tilde{K}_{n-4}\cdots \tilde{K}_1(Q_1\tilde{Q}_1 + Q_3\tilde{Q}_3)K_1\cdots K_{n-4}Q_2 = \begin{cases} 0 & n\in 2\mathbb{Z} + 1, \\ -(\tilde{Q}_1 Q_3\tilde{Q}_3 Q_1)^{n/2 - 1} & n\in 2\mathbb{Z}. \end{cases}
\end{equation}
Moreover, combining
\begin{equation}
(Q_1)_i(\tilde{Q}_1)^k(Q_3)_k(\tilde{Q}_3)^j + (Q_3)_i(\tilde{Q}_3)^k(Q_1)_k(\tilde{Q}_1)^j = (K_1)_i{}^k(\tilde{K}_1)_k{}^\ell(K_1)_\ell{}^m(\tilde{K}_1)_m{}^j
\end{equation}
with \eqref{24powersimp}, \eqref{24sandwich}, and \eqref{1331is2442} gives
\begin{equation}
\tilde{Q}_2\tilde{K}_{n-4}\cdots \tilde{K}_1 Q_1\tilde{Q}_3 K_1\cdots K_{n-4}Q_2\tilde{Q}_1 Q_3 + (1\leftrightarrow 3) = \begin{cases} (\tilde{Q}_1 Q_3\tilde{Q}_3 Q_1)^{(n - 1)/2} & n\in 2\mathbb{Z} + 1, \\ 0 & n\in 2\mathbb{Z}. \end{cases}
\end{equation}
Hence the $\mathbb{Z}_2$ symmetry that takes $1\leftrightarrow 3$ acts in exactly the same way as that which takes $2\leftrightarrow 4$.  We use $1\leftrightarrow 3$ by convention.

Next, consider the chiral ring relation.  First let $n\in 2\mathbb{Z}$ and set
\begin{equation}
\cY\equiv \cY' + (-\cZ)^{n/2 - 1}.
\end{equation}
Defining the orientation-reversed operator
\begin{equation}
\bar{\cY}'\equiv \cY'|_{1\leftrightarrow 3} = 2\tilde{Q}_1 K_1\cdots K_{n-4}Q_2\tilde{Q}_2\tilde{K}_{n-4}\cdots \tilde{K}_1 Q_1,
\end{equation}
we see that
\begin{equation}
\cY' + \bar{\cY}' = -2(-\cZ)^{n/2 - 1}, \quad \cY'\bar{\cY}' \cZ = \cX^2.
\end{equation}
Thus we get
\begin{align}
\cZ\cY'^2 = \cZ\cY'(-2(-\cZ)^{n/2 - 1} - \bar{\cY}') &\Longleftrightarrow \cX^2 + \cZ\cY'^2 - 2\cY'(-\cZ)^{n/2} = 0 \nonumber \\
&\Longleftrightarrow \cX^2 + \cZ\cY^2 = \cZ^{n-1},
\end{align}
as desired.  Now let $n\in 2\mathbb{Z} + 1$ and set
\begin{equation}
\cX\equiv \cX' - (-\cZ)^{(n - 1)/2}.
\end{equation}
Defining the orientation-reversed operator
\begin{equation}
\bar{\cX}'\equiv \cX'|_{1\leftrightarrow 3} = 2\tilde{Q}_3 K_1\cdots K_{n-4}Q_2\tilde{Q}_2\tilde{K}_{n-4}\cdots \tilde{K}_1 Q_1\tilde{Q}_1 Q_3,
\end{equation}
we see that
\begin{equation}
\cX' + \bar{\cX}' = 2(-\cZ)^{(n - 1)/2}, \quad \cX'\bar{\cX}' = \cZ\cY^2.
\end{equation}
Thus we get
\begin{align}
\cX'^2 = \cX'(2(-\cZ)^{(n - 1)/2} - \bar{\cX}') &\Longleftrightarrow \cX'^2 - 2\cX'(-\cZ)^{(n - 1)/2} + \cZ\cY^2 = 0 \nonumber \\
&\Longleftrightarrow \cX^2 + \cZ\cY^2 = \cZ^{n-1},
\end{align}
as desired.

To conclude, we remark that the basis \eqref{hbdnfirst}--\eqref{hbdnlast} (which we refer to as the ``alternate basis'') differs from the earlier one that we used when $n = 4$, namely \eqref{ourbasis}, and that the $\mathbb{Z}_2$ acts differently in the two cases.  When $n = 4$, we have in the alternate basis that
\begin{align}
\cZ &= -\sqrt{3}\tilde{Q}_1 Q_3\tilde{Q}_3 Q_1, \nonumber \\
\cY &= \sqrt{3}i(2\tilde{Q}_2 Q_3\tilde{Q}_3 Q_2 + \tilde{Q}_1 Q_3\tilde{Q}_3 Q_1), \label{alternatebasis} \\
\cX &= 2\cdot 3^{3/4}i\tilde{Q}_1 Q_2\tilde{Q}_2 Q_3\tilde{Q}_3 Q_1, \nonumber
\end{align}
where we have rescaled the generators so that they satisfy $\cX^2 + \cZ\cY^2 + \cZ^3 = 0$ and so that $\cX$ is the same as in \eqref{ourbasis}.  At the level of the chiral ring, this alternate basis maps to
\begin{equation}
(\mathcal{Z}_C, \mathcal{Y}_C, \mathcal{X}_C) = \left(-C^2\left(\frac{1}{8}\Phi^2 + \mathcal{M}_\infty^2\right), iC^2\left(\frac{3}{8}\Phi^2 - \mathcal{M}_\infty^2\right), C^3\Phi\mathcal{M}_\infty^2\right)
\end{equation}
on the Coulomb branch of $SU(2)$ SQCD with $N_f = 4$, where we have set $C\equiv 3^{1/4}(4\pi)^{-1}$.  A short calculation with the corresponding commutative shift operators shows that these operators likewise satisfy $\mathcal{X}_C^2 + \mathcal{Z}_C\mathcal{Y}_C^2 + \mathcal{Z}_C^3 = 0$.  By the same reasoning as in Section \ref{Nf4}, the enhanced $S_3$ symmetry requires that\footnote{At the level of the quantized chiral ring, we know that
\begin{equation}
\tilde{Q}_1 Q_2\tilde{Q}_2 Q_3\tilde{Q}_3 Q_1\leftrightarrow -\frac{1}{128\pi^3}\left(i\Phi\mathcal{M}^2 + \frac{1}{r}\mathcal{M}^2\right),
\end{equation}
as well as \eqref{1331} and \eqref{2332}.  These correspondences are consistent with \eqref{alternatebasis} if we define the Higgs branch variables $\cX, \cY, \cZ$ at the quantum level simply by subtracting their one-point functions.  These $1/r$ corrections ensure that the one-point functions of $\mathbb{Z}_2$-odd operators are zero, in the alternate basis.}
\begin{equation}
(\mathcal{Z}_C, \mathcal{Y}_C, \mathcal{X}_C) = \left(-C^2\left(\frac{1}{8}\widehat{\Phi^2} + \mathcal{M}^2\right), iC^2\left(\frac{3}{8}\widehat{\Phi^2} - \mathcal{M}^2\right), C^3\left(\Phi\mathcal{M}^2 - \frac{i}{r}\mathcal{M}^2\right)\right)
\end{equation}
at the quantum level, where we have defined $\widehat{\Phi^2}\equiv \Phi^2 - 1/3r^2$ (which satisfies $\langle\widehat{\Phi^2}\rangle = 0$).  In the alternate basis, the $\mathbb{Z}_2$ symmetry therefore acts as:
\begin{equation}
\left(\widehat{\Phi^2}, \mathcal{M}^2, \Phi\mathcal{M}^2\right)\mapsto \left(-\frac{1}{2}\widehat{\Phi^2} + 4\mathcal{M}^2, \frac{1}{2}\mathcal{M}^2 + \frac{3}{16}\widehat{\Phi^2}, -\Phi\mathcal{M}^2 + \frac{3i}{2r}\left(\mathcal{M}^2 + \frac{1}{8}\widehat{\Phi^2}\right)\right).
\end{equation}
This should be contrasted with the $\mathbb{Z}_2$ symmetry acting on \eqref{ourbasis}, which is more natural from the Coulomb branch point of view in that it simply flips the signs of monopoles.  The $\mathbb{Z}_2$ is only ambiguous when $n = 4$ because it can be conjugated by elements of $S_3$: otherwise, it is unique.

\subsection{Affine \texorpdfstring{$E_n$}{En} Quivers}

We now turn to the $E$-type quiver theories.  In all cases, the fundamental ``meson'' operators satisfy $\smash{M_{(I)}^{\ell_I}} = 0$ where $\ell_I$ is the length of leg $I$.  To derive the chiral ring relation for $E_6$, we need only the $U(1)$ Schouten identity: following \cite{Collinucci:2017bwv}, the trick is to write the generators containing squares of mesons as $U(1)\times U(1)$ bifundamentals.  For $E_{7, 8}$, we instead employ the $U(2)$ Schouten identity: following \cite{Lindstrom:1999pz}, we define auxiliary operators with only $U(2)$ indices uncontracted.  We present the derivations for $E_{6, 7, 8}$ in decreasing amounts of detail.

Deriving a Schouten identity for tensors of given rank involves antisymmetrizing over an appropriate number of indices and then contracting a subset of these indices.  For instance, the Schouten identity for two-component vectors follows from contracting any two indices in $x^{[i}y^j z^{k]} = 0$.  A Schouten identity for matrices \cite{Lindstrom:1999pz} following from $M_{k_1}{}^{[i_1}N_{k_2}{}^{i_2}K_{k_3}{}^{i_3]} = 0$ is
\begin{equation}
\Tr(\{M, N\}K) = \sum_\text{cyc} \Tr(MN)\Tr(K) - \Tr(M)\Tr(N)\Tr(K),
\label{22schouten}
\end{equation}
where the indices range over $\{1, 2\}$.

\subsubsection{\texorpdfstring{$E_6$}{E6}} \label{e6chiralring}

Our conventions are as in Section \ref{starshaped}.  In this case, the symmetry acts as $\mathbb{Z}_2 : (\cX, \cY, \cZ)\mapsto (-\cX, \cY, -\cZ)$.  The $U(2)$ and $U(1)$ D-term relations imply that, for each $T[SU(3)]$ leg,
\begin{equation}
Q_j^C\tilde Q^i_C + q_j \tilde q^i=0, \quad q_i \tilde q^i=0.
\end{equation}
For a given leg, one can verify using these relations that, for example, the $\Delta = 2$ CPOs
\begin{equation}
q_i\tilde q^j Q_j^A\tilde Q^i_B -{1\over 3}q_i\tilde q^j Q_j^C\tilde Q^i_C \D^A_B
\end{equation}
are equivalent to $-M^A{}_C M^C{}_B$ in the chiral ring.  Since the trace part vanishes in the chiral ring, we may simply write $M^A{}_B = Q_i^A\tilde{Q}_B^i$.

We first summarize some useful relations.  Writing the $M_{(I)}{}^A{}_B$ as matrices, we have
\begin{equation}
M_{(1)} + M_{(2)} + M_{(3)} = 0
\label{sumzero}
\end{equation}
from the $SU(3)$ D-term relation.  We see from the D-term relations for each leg that
\begin{equation}
\tr(M_{(I)}^p) = 0
\label{tracepower}
\end{equation}
for integers $p\geq 1$ and $I = 1, 2, 3$.  Moreover, we have
\begin{equation}
M_{(I)}^3 = 0.
\label{moretracepower}
\end{equation}
Indeed,
\begin{align}
\cdots M_{(I)}{}^A{}_B M_{(I)}{}^B{}_C M_{(I)}{}^C{}_D\cdots &= \cdots (Q_{(I)})_i^A(\tilde{Q}_{(I)})_B^i(Q_{(I)})_j^B(\tilde{Q}_{(I)})_C^j(Q_{(I)})_k^C(\tilde{Q}_{(I)})_D^k\cdots \nonumber \\
&= \cdots (Q_{(I)})_i^A(\tilde{q}_{(I)})^i(q_{(I)})_j(\tilde{q}_{(I)})^j(q_{(I)})_k(\tilde{Q}_{(I)})_D^k\cdots \nonumber \\
&= 0,
\end{align}
since $(q_{(I)})_j(\tilde{q}_{(I)})^j = 0$.

Let us now enumerate the nontrivial chiral ring elements of small dimension (compare to \cite{Collinucci:2017bwv}).  The $p = 1$ case of \eqref{tracepower} rules out chiral ring elements at $\Delta = 1$.  From \eqref{sumzero} and \eqref{tracepower}, we also have
\begin{equation}
\tr(M_{(I)}M_{(J)}) = -\tr(M_{(I)}M_{(K)}) = \tr(M_{(J)}M_{(K)}) = -\tr(M_{(J)}M_{(I)}) \implies \tr(M_{(I)}M_{(J)}) = 0,
\end{equation}
ruling out chiral ring elements at $\Delta = 2$.  At $\Delta = 3$, $\tr(M_{(I)}^2 M_{(J)})$ is nontrivial while
\begin{equation}
\tr(M_{(I)}M_{(J)}M_{(K)}) = -\tr(M_{(J)}^2 M_{(K)} + M_{(J)}M_{(K)}^2) = 0,
\end{equation}
giving a single candidate for the chiral ring generator $\cZ$ (up to normalization):
\begin{align}
\tr(M_{(1)}^2 M_{(2)}) &= \tr(M_{(2)}^2 M_{(3)}) = \tr(M_{(3)}^2 M_{(1)}) \nonumber \\
&= -\tr(M_{(1)}^2 M_{(3)}) = -\tr(M_{(2)}^2 M_{(1)}) = -\tr(M_{(3)}^2 M_{(2)}).
\end{align}
At $\Delta = 4$, \eqref{moretracepower} implies that
\begin{equation}
\tr(M_{(1)}^2 M_{(2)}^2) = \tr(M_{(1)}^2 M_{(3)}^2) = \tr(M_{(2)}^2 M_{(3)}^2),
\label{Vequivalence}
\end{equation}
giving a single candidate for the chiral ring generator $\cY$.  This is the only candidate because
\begin{gather}
\tr((M_{(I)}M_{(J/K)})^2) = -\tr(M_{(I)}M_{(J)}M_{(I)}M_{(K)}) = -\tr(M_{(I)}^2 M_{(J)}^2 + M_{(I)}^2 M_{(K)}^2), \\
\tr(M_{(I)}^2 M_{(J)}M_{(K)}) = \tr((M_{(J)}M_{(K)})^2 + M_{(J)}^2 M_{(K)}^2).
\end{gather}
At $\Delta = 5$, there are no nontrivial chiral ring elements.  Indeed, with two types of $M_{(I)}$, there is only one pattern of contraction:
\begin{equation}
\tr(M_{(I)}^2 M_{(J)}M_{(I)}M_{(J)}).
\end{equation}
With three types, we have the possibilities
\begin{gather}
\tr(M_{(I)}^2 M_{(J)}M_{(I)}M_{(K)}), \quad \tr(M_{(I)}^2 M_{(J)}^2 M_{(K)}), \nonumber \\
\tr(M_{(I)}^2 M_{(J)}M_{(K)}M_{(J)}), \quad \tr(M_{(I)}M_{(J)}M_{(I)}M_{(J)}M_{(K)}).
\end{gather}
But we have
\begin{gather}
\tr(M_{(I)}^2 M_{(J)}M_{(I)}M_{(K)}) = -\tr(M_{(I)}^2 M_{(J)}M_{(I)}M_{(J)}) = \tr(M_{(I)}^2 M_{(J)}M_{(K)}M_{(J)}) \nonumber \\
= -\tr(M_{(I)}^2 M_{(K)}^2 M_{(J)}) = 0, \\
\tr(M_{(I)}M_{(J)}M_{(I)}M_{(J)}M_{(K)}) = -\tr(M_{(I)}^2 M_{(J)}M_{(I)}M_{(J)}) - \tr(M_{(J)}^2 M_{(I)}M_{(J)}M_{(I)}) \nonumber \\
= -0 - 0 = 0.
\end{gather}
At $\Delta = 6$, the possible contractions involving two types of $M_{(I)}$ are
\begin{equation}
\tr(M_{(I)}^2 M_{(J)}^2 M_{(I)}M_{(J)}), \quad \tr((M_{(I)}^2 M_{(J)})^2), \quad \tr((M_{(I)}M_{(J)})^3),
\end{equation}
and those involving three types can all be written as linear combinations of those involving two types using \eqref{sumzero}.  Restricting our attention to two types, we derive that
\begin{equation}
\tr(M_{(I)}^2 M_{(J)}^2 M_{(I)}M_{(J)}) + \tr(M_{(I)}^2 M_{(K)}^2 M_{(I)}M_{(K)}) = -\tr((M_{(I)}^2 M_{(J/K)})^2)
\label{recallsum}
\end{equation}
and
\begin{equation}
\tr((M_{(I)}M_{(J)})^3) = -\tr((M_{(I)}M_{(K)})^3) = \tr((M_{(J)}M_{(K)})^3) = -\tr((M_{(I)}M_{(J)})^3) = 0,
\end{equation}
so it suffices to consider contractions of the form $\tr(M_{(I)}^2 M_{(J)}^2 M_{(I)}M_{(J)})$.  But
\begin{equation}
\tr(M_{(I)}^2 M_{(J)}^2 M_{(I)}M_{(J)}) = \tr(M_{(I)}^2 M_{(J)}^2 M_{(K)}^2),
\end{equation}
so we are left with only two independent chiral ring elements at $\Delta = 6$:
\begin{equation}
\tr(M_{(1)}^2 M_{(2)}^2 M_{(3)}^2), \quad \tr(M_{(1)}^2 M_{(3)}^2 M_{(2)}^2).
\end{equation}
One linear combination of them should give the square of the generator at $\Delta = 3$, and the other should give the new generator $\cX$ at $\Delta = 6$.

To derive the chiral ring relation (compare to \cite{Lindstrom:1999pz}, but without FI parameters), set
\begin{align}
\cW &\equiv \tr(M_{(1)}^2 M_{(2)}), \\
\cV &\equiv \tr(M_{(1)}^2 M_{(2)}^2), \\
\cU &\equiv \tr(M_{(1)}^2 M_{(2)}^2 M_{(3)}^2), \\
\bar{\cU} &\equiv \tr(M_{(1)}^2 M_{(3)}^2 M_{(2)}^2).
\end{align}
Let
\begin{equation}
(IJ)\equiv (\tilde{q}_{(I)})^i(Q_{(I)})_i^A(\tilde{Q}_{(J)})_A^j(q_{(J)})_j.
\end{equation}
Using $(M_{(I)}^2)^A{}_B = (q_{(I)})_j(\tilde{q}_{(I)})^i(Q_{(I)})_i^A(\tilde{Q}_{(I)})_B^j$ and rearranging, we have
\begin{align}
\tr(M_{(I)}^2 M_{(J)}^2) &= (IJ)(JI), \\
\tr(M_{(I)}^2 M_{(J)}^2 M_{(K)}^2) &= -(IK)(KJ)(JI).
\end{align}
Hence we derive that
\begin{align}
\tr(M_{(1)}^2 M_{(2)}^2 M_{(3)}^2)\tr(M_{(1)}^2 M_{(3)}^2 M_{(2)}^2) &= (12)(21)(13)(31)(23)(32) \nonumber \\
&= \tr(M_{(1)}^2 M_{(2)}^2)\tr(M_{(1)}^2 M_{(3)}^2)\tr(M_{(2)}^2 M_{(3)}^2),
\end{align}
meaning (by virtue of \eqref{Vequivalence})
\begin{equation}
\cU\bar{\cU} = \cV^3.
\label{productformula}
\end{equation}
Now let
\begin{equation}
(IJK)\equiv (\tilde{q}_{(I)})^i(Q_{(I)})_i^A(\tilde{Q}_{(J)})_A^j(Q_{(J)})_j^B(\tilde{Q}_{(K)})_B^k(q_{(K)})_k.
\end{equation}
Recall \eqref{recallsum}, which is equivalent to
\begin{align}
\tr(M_{(I)}^2 M_{(J)}^2 M_{(K)}^2) + \tr(M_{(I)}^2 M_{(K)}^2 M_{(J)}^2) &= -\tr((M_{(I)}^2 M_{(J)})^2) = -\tr((M_{(I)}^2 M_{(K)})^2) \nonumber \\
&= -\tr((M_{(J)}^2 M_{(I)})^2) = -\tr((M_{(J)}^2 M_{(K)})^2) \nonumber \\
&= -\tr((M_{(K)}^2 M_{(I)})^2) = -\tr((M_{(K)}^2 M_{(J)})^2),
\end{align}
and note that
\begin{align}
\tr(M_{(1)}^2 M_{(2)}) = -(121), \quad \tr((M_{(1)}^2 M_{(2)})^2) = (121)^2.
\end{align}
So we get
\begin{equation}
\tr((M_{(1)}^2 M_{(2)})^2) = \tr(M_{(1)}^2 M_{(2)})^2,
\end{equation}
which implies that
\begin{equation}
\cU + \bar{\cU} = -\cW^2.
\label{sumformula}
\end{equation}
Combining \eqref{productformula} and \eqref{sumformula} gives
\begin{equation}
\cU^2 + \cU\cW^2 + \cV^3 = 0,
\end{equation}
and making the change of variables
\begin{equation}
\cU = i\cX - \cZ^2, \quad \cV = -\cY, \quad \cW = \sqrt{2}\cZ
\end{equation}
gives
\begin{equation}
\cX^2 + \cY^3 + \cZ^4 = 0
\end{equation}
where
\begin{equation}
\cZ = \frac{1}{\sqrt{2}}\tr(M_{(1)}^2 M_{(2)}), \quad \cY = -\tr(M_{(1)}^2 M_{(2)}^2), \quad \cX = -i[\tr(M_{(1)}^2 M_{(2)}^2 M_{(3)}^2) + \cZ^2].
\end{equation}
In this presentation, the $\mathbb{Z}_2$ symmetry acts as $(1)\leftrightarrow (2)$.

\subsubsection{\texorpdfstring{$E_7$}{E7}} \label{e7chiralring}

For each $T[SU(4)]$ leg, we denote the bifundamental hypers by
\begin{equation}
((q_{12})_i, (\tilde{q}_{12})^i), \quad ((q_{23})_i^A, (\tilde{q}_{23})_A^i), \quad ((q_{34})_A^N, (\tilde{q}_{34})_N^A),
\end{equation}
where $i = 1, 2$; $A = 1, 2, 3$; and $N = 1, 2, 3, 4$.  The subscripts indicate the ranks of the gauge nodes.  Accounting for orientation, the D/F-term equations are (in our conventions)
\begin{equation}
(q_{12})_i(\tilde{q}_{12})^i = 0, \quad (q_{12})_i(\tilde{q}_{12})^j + (q_{23})_i^A(\tilde{q}_{23})_A^j = 0, \quad (q_{23})_i^A(\tilde{q}_{23})_B^i - (\tilde{q}_{34})_N^A(q_{34})_B^N = 0,
\label{e7conventions}
\end{equation}
and we have the mesons
\begin{equation}
M^M{}_N\equiv (q_{34})_A^M(\tilde{q}_{34})_N^A,
\end{equation}
which are traceless in the chiral ring by the D/F-term relations.  We can also write
\begin{align}
(M^2)^M{}_N &= (q_{23})_i^A(\tilde{q}_{23})_B^i(q_{34})_A^M(\tilde{q}_{34})_N^B, \\
(M^3)^M{}_N &= -(q_{12})_i(\tilde{q}_{12})^j(q_{23})_j^A(\tilde{q}_{23})_B^i(q_{34})_A^M(\tilde{q}_{34})_N^B.
\end{align}
Higher powers vanish, as do all traces of powers: $\tr(M^p) = 0$ for $p\geq 1$.

For the leg of length two, we have the D/F-term relation
\begin{equation}
q_i^M\tilde{q}_M^j = 0
\end{equation}
and the meson
\begin{equation}
M^M{}_N\equiv q_i^M\tilde{q}_N^i,
\end{equation}
whose trace and higher powers vanish.

For the quiver as a whole, we have
\begin{equation}
M_{(1)} + M_{(2)} + M_{(3)} = 0
\end{equation}
by the $SU(4)$ D-term relation.

To proceed, define (as in \cite{Lindstrom:1999pz}) the traceless $U(2)$ matrices\footnote{The minus sign in $K$ is a consequence of our conventions \eqref{e7conventions}.}
\begin{align}
\cM^i{}_j &\equiv (\tilde{q}_{12(1)})^i(q_{12(1)})_j, \\
\cN^i{}_j &\equiv (\tilde{q}_{23(1)})_A^i(q_{23(1)})_j^B(\tilde{q}_{34(1)})_M^A(q_{34(1)})_B^N(M_{(3)})^M{}_N, \\
\cK^i{}_j &\equiv -(\tilde{q}_{23(1)})_A^i(q_{23(1)})_j^B(\tilde{q}_{34(1)})_M^A(q_{34(1)})_B^N(M_{(2)}^3)^M{}_N.
\end{align}
Then we have by construction that
\begin{equation}
\tr(\cM\cN) = \cZ, \quad \tr(\cM\cK) = -\cY,
\end{equation}
and we compute using the D/F-term relations that
\begin{equation}
\tr(\cN^2) = -2\cY, \quad \tr(\cN\cK) = -\cZ^2, \quad \cY\tr(\cM\cN\cK\cN) = \cX^2.
\end{equation}
We now write
\begin{equation}
\tr(\cM\cN\cK\cN) = \tr(\{\cM, \cN\}\cK\cN) - \frac{1}{2}\tr(\cM\cK\{\cN, \cN\})
\end{equation}
and use the $2\times 2$ Schouten identity \eqref{22schouten} as well as $\tr(\cM) = \tr(\cN) = \tr(\cK) = 0$ to get
\begin{equation}
\tr(\cM\cN\cK\cN) = \tr(\cM\cN)\tr(\cK\cN) - \frac{1}{2}\tr(\cM\cK)\tr(\cN^2),
\end{equation}
which implies that
\begin{equation}
\cX^2 + \cY^3 + \cY\cZ^3 = 0,
\end{equation}
as desired.

\subsubsection{\texorpdfstring{$E_8$}{E8}} \label{e8chiralring}

In this case, we define the traceless $U(2)$ matrices
\begin{align}
\cA^i{}_j &\equiv (M_{(1)})^M{}_N(\tilde{q}_{46(2)})_M^A(q_{46(2)})_B^N(\tilde{q}_{24(2)})_A^i(q_{24(2)})_j^B, \\
\cB^i{}_j &\equiv (M_{(1)}^3)^M{}_N(\tilde{q}_{46(2)})_M^A(q_{46(2)})_B^N(\tilde{q}_{24(2)})_A^i(q_{24(2)})_j^B, \\
\cC^i{}_j &\equiv -(M_{(1)}^5)^M{}_N(\tilde{q}_{46(2)})_M^A(q_{46(2)})_B^N(\tilde{q}_{24(2)})_A^i(q_{24(2)})_j^B,
\end{align}
with the minus sign due to our conventions for $U(1)$ nodes; then by construction, we have $\cY = \tr(\cA\cC)$ and $\cX = \tr(\cA\cB\cC)$.  We also compute that
\begin{equation}
\tr(\cA\cB) = 0, \quad \tr(\cB\cC) = \cZ^2, \quad \tr(\cA^2) = -2\cZ, \quad \tr(\cB^2) = -2\cY.
\end{equation}
Now consider the following expression, which we simplify by writing in terms of anticommutators, using the $2\times 2$ Schouten identity \eqref{22schouten}, and using that individual traces of $\cA, \cB, \cC$ vanish:
\begin{align}
\tr(\cA\cB\cC\cA\cB\cC) &= \tr(\{\cA, \cB\}\cC\cA\cB\cC) - \tr(\cB\{\cA, \cC\}\cA\cB\cC) + \frac{1}{2}\tr(\cB\cC\{\cA, \cA\}\cB\cC) \\
&= \tr(\cA\cB)\tr(\cA\cB\cC^2) - \tr(\cA\cC)\tr(\cA\cB\cC\cB) + \frac{1}{2}\tr(\cA^2)\tr((\cB\cC)^2).
\end{align}
We also have that
\begin{equation}
\tr(\cA\cB\cC\cB) = \tr(\cA\cB\{\cB, \cC\}) - \frac{1}{2}\tr(\cA\{\cB, \cB\}\cC) = \tr(\cA\cB)\tr(\cB\cC) - \frac{1}{2}\tr(\cA\cC)\tr(\cB^2),
\end{equation}
which, in combination with $\tr(\cA\cB) = 0$, implies that
\begin{equation}
\tr(\cA\cB\cC\cA\cB\cC) = \frac{1}{2}\tr(\cA\cC)^2\tr(\cB^2) + \frac{1}{2}\tr(\cA^2)\tr((\cB\cC)^2).
\end{equation}
By the 1D Schouten identity, we have $\tr((\cA\cB\cC)^2) = \tr(\cA\cB\cC)^2 = \cX^2$ as well as $\tr((\cB\cC)^2) = \tr(\cB\cC)^2 = \cZ^4$, so we arrive at
\begin{equation}
\cX^2 + \cY^3 + \cZ^5 = 0,
\end{equation}
as desired.

\bibliographystyle{JHEP}
\bibliography{ademirror}

\providecommand{\href}[2]{#2}\begingroup\raggedright\begin{thebibliography}{10}

\bibitem{Giombi:2011kc}
S.~Giombi, S.~Minwalla, S.~Prakash, S.~P. Trivedi, S.~R. Wadia, and X.~Yin,
  {\it {Chern-Simons Theory with Vector Fermion Matter}},  {\em Eur. Phys. J.}
  {\bf C72} (2012) 2112, [\href{http://arxiv.org/abs/1110.4386}{{\tt
  arXiv:1110.4386}}].

\bibitem{Aharony:2012nh}
O.~Aharony, G.~Gur-Ari, and R.~Yacoby, {\it {Correlation Functions of Large N
  Chern-Simons-Matter Theories and Bosonization in Three Dimensions}},  {\em
  JHEP} {\bf 12} (2012) 028, [\href{http://arxiv.org/abs/1207.4593}{{\tt
  arXiv:1207.4593}}].

\bibitem{Aharony:2012ns}
O.~Aharony, S.~Giombi, G.~Gur-Ari, J.~Maldacena, and R.~Yacoby, {\it {The
  Thermal Free Energy in Large N Chern-Simons-Matter Theories}},  {\em JHEP}
  {\bf 03} (2013) 121, [\href{http://arxiv.org/abs/1211.4843}{{\tt
  arXiv:1211.4843}}].

\bibitem{Aharony:2015mjs}
O.~Aharony, {\it {Baryons, monopoles and dualities in Chern-Simons-matter
  theories}},  {\em JHEP} {\bf 02} (2016) 093,
  [\href{http://arxiv.org/abs/1512.00161}{{\tt arXiv:1512.00161}}].

\bibitem{Karch:2016sxi}
A.~Karch and D.~Tong, {\it {Particle-Vortex Duality from 3d Bosonization}},
  {\em Phys. Rev.} {\bf X6} (2016), no.~3 031043,
  [\href{http://arxiv.org/abs/1606.01893}{{\tt arXiv:1606.01893}}].

\bibitem{Murugan:2016zal}
J.~Murugan and H.~Nastase, {\it {Particle-vortex duality in topological
  insulators and superconductors}},  {\em JHEP} {\bf 05} (2017) 159,
  [\href{http://arxiv.org/abs/1606.01912}{{\tt arXiv:1606.01912}}].

\bibitem{Seiberg:2016gmd}
N.~Seiberg, T.~Senthil, C.~Wang, and E.~Witten, {\it {A Duality Web in 2+1
  Dimensions and Condensed Matter Physics}},  {\em Annals Phys.} {\bf 374}
  (2016) 395--433, [\href{http://arxiv.org/abs/1606.01989}{{\tt
  arXiv:1606.01989}}].

\bibitem{Intriligator:1994rx}
K.~A. Intriligator, N.~Seiberg, and S.~H. Shenker, {\it {Proposal for a simple
  model of dynamical SUSY breaking}},  {\em Phys. Lett.} {\bf B342} (1995)
  152--154, [\href{http://arxiv.org/abs/hep-ph/9410203}{{\tt hep-ph/9410203}}].

\bibitem{Brodie:1998vv}
J.~H. Brodie, P.~L. Cho, and K.~A. Intriligator, {\it {Misleading anomaly
  matchings?}},  {\em Phys. Lett.} {\bf B429} (1998) 319--326,
  [\href{http://arxiv.org/abs/hep-th/9802092}{{\tt hep-th/9802092}}].

\bibitem{Intriligator:2005if}
K.~A. Intriligator, {\it {IR free or interacting? A Proposed diagnostic}},
  {\em Nucl. Phys.} {\bf B730} (2005) 239--251,
  [\href{http://arxiv.org/abs/hep-th/0509085}{{\tt hep-th/0509085}}].

\bibitem{Intriligator:1996ex}
K.~A. Intriligator and N.~Seiberg, {\it {Mirror symmetry in three-dimensional
  gauge theories}},  {\em Phys. Lett.} {\bf B387} (1996) 513--519,
  [\href{http://arxiv.org/abs/hep-th/9607207}{{\tt hep-th/9607207}}].

\bibitem{deBoer:1996mp}
J.~de~Boer, K.~Hori, H.~Ooguri, and Y.~Oz, {\it {Mirror symmetry in
  three-dimensional gauge theories, quivers and D-branes}},  {\em Nucl. Phys.}
  {\bf B493} (1997) 101--147, [\href{http://arxiv.org/abs/hep-th/9611063}{{\tt
  hep-th/9611063}}].

\bibitem{deBoer:1996ck}
J.~de~Boer, K.~Hori, H.~Ooguri, Y.~Oz, and Z.~Yin, {\it {Mirror symmetry in
  three-dimensional theories, SL(2,Z) and D-brane moduli spaces}},  {\em Nucl.
  Phys.} {\bf B493} (1997) 148--176,
  [\href{http://arxiv.org/abs/hep-th/9612131}{{\tt hep-th/9612131}}].

\bibitem{Hanany:1996ie}
A.~Hanany and E.~Witten, {\it {Type IIB superstrings, BPS monopoles, and
  three-dimensional gauge dynamics}},  {\em Nucl. Phys.} {\bf B492} (1997)
  152--190, [\href{http://arxiv.org/abs/hep-th/9611230}{{\tt hep-th/9611230}}].

\bibitem{Kachru:2016rui}
S.~Kachru, M.~Mulligan, G.~Torroba, and H.~Wang, {\it {Bosonization and Mirror
  Symmetry}},  {\em Phys. Rev.} {\bf D94} (2016), no.~8 085009,
  [\href{http://arxiv.org/abs/1608.05077}{{\tt arXiv:1608.05077}}].

\bibitem{Kachru:2016aon}
S.~Kachru, M.~Mulligan, G.~Torroba, and H.~Wang, {\it {Nonsupersymmetric
  dualities from mirror symmetry}},  {\em Phys. Rev. Lett.} {\bf 118} (2017),
  no.~1 011602, [\href{http://arxiv.org/abs/1609.02149}{{\tt
  arXiv:1609.02149}}].

\bibitem{Cremonesi:2013lqa}
S.~Cremonesi, A.~Hanany, and A.~Zaffaroni, {\it {Monopole operators and Hilbert
  series of Coulomb branches of $3d$ $\mathcal{N} = 4$ gauge theories}},  {\em
  JHEP} {\bf 01} (2014) 005, [\href{http://arxiv.org/abs/1309.2657}{{\tt
  arXiv:1309.2657}}].

\bibitem{Cremonesi:2014kwa}
S.~Cremonesi, A.~Hanany, N.~Mekareeya, and A.~Zaffaroni, {\it {Coulomb branch
  Hilbert series and Hall-Littlewood polynomials}},  {\em JHEP} {\bf 09} (2014)
  178, [\href{http://arxiv.org/abs/1403.0585}{{\tt arXiv:1403.0585}}].

\bibitem{Hanany:2016pfm}
A.~Hanany and M.~Sperling, {\it {Algebraic properties of the monopole
  formula}},  {\em JHEP} {\bf 02} (2017) 023,
  [\href{http://arxiv.org/abs/1611.07030}{{\tt arXiv:1611.07030}}].

\bibitem{Hanany:2016gbz}
A.~Hanany and R.~Kalveks, {\it {Quiver Theories for Moduli Spaces of Classical
  Group Nilpotent Orbits}},  {\em JHEP} {\bf 06} (2016) 130,
  [\href{http://arxiv.org/abs/1601.04020}{{\tt arXiv:1601.04020}}].

\bibitem{Hanany:2017ooe}
A.~Hanany and R.~Kalveks, {\it {Quiver Theories and Formulae for Nilpotent
  Orbits of Exceptional Algebras}},  {\em JHEP} {\bf 11} (2017) 126,
  [\href{http://arxiv.org/abs/1709.05818}{{\tt arXiv:1709.05818}}].

\bibitem{Hanany:2018xth}
A.~Hanany and D.~Miketa, {\it {Nilpotent orbit Coulomb branches of types AD}},
  {\em JHEP} {\bf 02} (2019) 113, [\href{http://arxiv.org/abs/1807.11491}{{\tt
  arXiv:1807.11491}}].

\bibitem{Chester:2014mea}
S.~M. Chester, J.~Lee, S.~S. Pufu, and R.~Yacoby, {\it {Exact Correlators of
  BPS Operators from the 3d Superconformal Bootstrap}},  {\em JHEP} {\bf 03}
  (2015) 130, [\href{http://arxiv.org/abs/1412.0334}{{\tt arXiv:1412.0334}}].

\bibitem{Beem:2016cbd}
C.~Beem, W.~Peelaers, and L.~Rastelli, {\it {Deformation quantization and
  superconformal symmetry in three dimensions}},  {\em Commun. Math. Phys.}
  {\bf 354} (2017), no.~1 345--392,
  [\href{http://arxiv.org/abs/1601.05378}{{\tt arXiv:1601.05378}}].

\bibitem{Nekrasov:2002qd}
N.~A. Nekrasov, {\it {Seiberg-Witten prepotential from instanton counting}},
  {\em Adv. Theor. Math. Phys.} {\bf 7} (2003), no.~5 831--864,
  [\href{http://arxiv.org/abs/hep-th/0206161}{{\tt hep-th/0206161}}].

\bibitem{Nekrasov:2003rj}
N.~Nekrasov and A.~Okounkov, {\it {Seiberg-Witten theory and random
  partitions}},  {\em Prog. Math.} {\bf 244} (2006) 525--596,
  [\href{http://arxiv.org/abs/hep-th/0306238}{{\tt hep-th/0306238}}].

\bibitem{Nekrasov:2010ka}
N.~Nekrasov and E.~Witten, {\it {The Omega Deformation, Branes, Integrability,
  and Liouville Theory}},  {\em JHEP} {\bf 09} (2010) 092,
  [\href{http://arxiv.org/abs/1002.0888}{{\tt arXiv:1002.0888}}].

\bibitem{Yagi:2014toa}
J.~Yagi, {\it {$\Omega$-deformation and quantization}},  {\em JHEP} {\bf 08}
  (2014) 112, [\href{http://arxiv.org/abs/1405.6714}{{\tt arXiv:1405.6714}}].

\bibitem{Bullimore:2015lsa}
M.~Bullimore, T.~Dimofte, and D.~Gaiotto, {\it {The Coulomb Branch of 3d
  ${\mathcal{N}= 4}$ Theories}},  {\em Commun. Math. Phys.} {\bf 354} (2017),
  no.~2 671--751, [\href{http://arxiv.org/abs/1503.04817}{{\tt
  arXiv:1503.04817}}].

\bibitem{Bullimore:2016hdc}
M.~Bullimore, T.~Dimofte, D.~Gaiotto, J.~Hilburn, and H.-C. Kim, {\it {Vortices
  and Vermas}},  {\em Adv. Theor. Math. Phys.} {\bf 22} (2018) 803--917,
  [\href{http://arxiv.org/abs/1609.04406}{{\tt arXiv:1609.04406}}].

\bibitem{Dedushenko:2016jxl}
M.~Dedushenko, S.~S. Pufu, and R.~Yacoby, {\it {A one-dimensional theory for
  Higgs branch operators}},  {\em JHEP} {\bf 03} (2018) 138,
  [\href{http://arxiv.org/abs/1610.00740}{{\tt arXiv:1610.00740}}].

\bibitem{Dedushenko:2017avn}
M.~Dedushenko, Y.~Fan, S.~S. Pufu, and R.~Yacoby, {\it {Coulomb Branch
  Operators and Mirror Symmetry in Three Dimensions}},  {\em JHEP} {\bf 04}
  (2018) 037, [\href{http://arxiv.org/abs/1712.09384}{{\tt arXiv:1712.09384}}].

\bibitem{Dedushenko:2018icp}
M.~Dedushenko, Y.~Fan, S.~S. Pufu, and R.~Yacoby, {\it {Coulomb Branch
  Quantization and Abelianized Monopole Bubbling}},  {\em JHEP} {\bf 10} (2019)
  179, [\href{http://arxiv.org/abs/1812.08788}{{\tt arXiv:1812.08788}}].

\bibitem{Chester:2014fya}
S.~M. Chester, J.~Lee, S.~S. Pufu, and R.~Yacoby, {\it {The $ \mathcal{N}=8 $
  superconformal bootstrap in three dimensions}},  {\em JHEP} {\bf 09} (2014)
  143, [\href{http://arxiv.org/abs/1406.4814}{{\tt arXiv:1406.4814}}].

\bibitem{Agmon:2017xes}
N.~B. Agmon, S.~M. Chester, and S.~S. Pufu, {\it {Solving M-theory with the
  Conformal Bootstrap}},  {\em JHEP} {\bf 06} (2018) 159,
  [\href{http://arxiv.org/abs/1711.07343}{{\tt arXiv:1711.07343}}].

\bibitem{Agmon:2019imm}
N.~B. Agmon, S.~M. Chester, and S.~S. Pufu, {\it {The M-theory Archipelago}},
  \href{http://arxiv.org/abs/1907.13222}{{\tt arXiv:1907.13222}}.

\bibitem{Chang:2019dzt}
C.-M. Chang, M.~Fluder, Y.-H. Lin, S.-H. Shao, and Y.~Wang, {\it {3d N=4
  Bootstrap and Mirror Symmetry}},  \href{http://arxiv.org/abs/1910.03600}{{\tt
  arXiv:1910.03600}}.

\bibitem{Etingof:2019guc}
P.~Etingof and D.~Stryker, {\it {Short star-products for filtered
  quantizations, I}},  \href{http://arxiv.org/abs/1909.13588}{{\tt
  arXiv:1909.13588}}.

\bibitem{Gaiotto:2019mmf}
D.~Gaiotto and T.~Okazaki, {\it {Sphere correlation functions and Verma
  modules}},  \href{http://arxiv.org/abs/1911.11126}{{\tt arXiv:1911.11126}}.

\bibitem{Kapustin:2006pk}
A.~Kapustin and E.~Witten, {\it {Electric-Magnetic Duality And The Geometric
  Langlands Program}},  {\em Commun. Num. Theor. Phys.} {\bf 1} (2007) 1--236,
  [\href{http://arxiv.org/abs/hep-th/0604151}{{\tt hep-th/0604151}}].

\bibitem{Gaiotto:2008ak}
D.~Gaiotto and E.~Witten, {\it {S-Duality of Boundary Conditions In N=4 Super
  Yang-Mills Theory}},  {\em Adv. Theor. Math. Phys.} {\bf 13} (2009), no.~3
  721--896, [\href{http://arxiv.org/abs/0807.3720}{{\tt arXiv:0807.3720}}].

\bibitem{Smith80aclass}
S.~P. Smith, {\it A class of algebras similar to the enveloping algebra of
  sl2},  {\em Transactions of the American Mathematical Society} (1980)
  335--366.

\bibitem{Levy}
P.~Levy, {\it {Isomorphism Problems of Noncommutative Deformations of Type $D$
  Kleinian Singularities}},  {\em Transactions of the American Mathematical
  Society} {\bf 361} (2009), no.~5 2351--2375.

\bibitem{namikawa2010}
Y.~Namikawa, {\it Poisson deformations of affine symplectic varieties, ii},
  {\em Kyoto J. Math.} {\bf 50} (2010), no.~4 727--752.

\bibitem{Collinucci:2016hpz}
A.~Collinucci, S.~Giacomelli, R.~Savelli, and R.~Valandro, {\it {T-branes
  through 3d mirror symmetry}},  {\em JHEP} {\bf 07} (2016) 093,
  [\href{http://arxiv.org/abs/1603.00062}{{\tt arXiv:1603.00062}}].

\bibitem{Collinucci:2017bwv}
A.~Collinucci, S.~Giacomelli, and R.~Valandro, {\it {T-branes, monopoles and
  S-duality}},  {\em JHEP} {\bf 10} (2017) 113,
  [\href{http://arxiv.org/abs/1703.09238}{{\tt arXiv:1703.09238}}].

\bibitem{Benvenuti:2011ga}
S.~Benvenuti and S.~Pasquetti, {\it {3D-partition functions on the sphere:
  exact evaluation and mirror symmetry}},  {\em JHEP} {\bf 05} (2012) 099,
  [\href{http://arxiv.org/abs/1105.2551}{{\tt arXiv:1105.2551}}].

\bibitem{Borokhov:2002cg}
V.~Borokhov, A.~Kapustin, and X.-k. Wu, {\it {Monopole operators and mirror
  symmetry in three-dimensions}},  {\em JHEP} {\bf 12} (2002) 044,
  [\href{http://arxiv.org/abs/hep-th/0207074}{{\tt hep-th/0207074}}].

\bibitem{Dumas}
F.~Dumas, {\it {An introduction to noncommutative polynomial invariants}},
  {\em CIMPA: Homological methods and representations of non-commutative
  algebras} (2006).

\bibitem{Lindstrom:1999pz}
U.~Lindstrom, M.~Rocek, and R.~von Unge, {\it {HyperKahler quotients and
  algebraic curves}},  {\em JHEP} {\bf 01} (2000) 022,
  [\href{http://arxiv.org/abs/hep-th/9908082}{{\tt hep-th/9908082}}].

\bibitem{Gulotta:2011si}
D.~R. Gulotta, C.~P. Herzog, and S.~S. Pufu, {\it {From Necklace Quivers to the
  F-theorem, Operator Counting, and T(U(N))}},  {\em JHEP} {\bf 12} (2011) 077,
  [\href{http://arxiv.org/abs/1105.2817}{{\tt arXiv:1105.2817}}].

\bibitem{Nishioka:2011dq}
T.~Nishioka, Y.~Tachikawa, and M.~Yamazaki, {\it {3d Partition Function as
  Overlap of Wavefunctions}},  {\em JHEP} {\bf 08} (2011) 003,
  [\href{http://arxiv.org/abs/1105.4390}{{\tt arXiv:1105.4390}}].

\bibitem{Tachikawa:2015bga}
Y.~Tachikawa, {\it {A review of the $T_N$ theory and its cousins}},  {\em PTEP}
  {\bf 2015} (2015), no.~11 11B102,
  [\href{http://arxiv.org/abs/1504.01481}{{\tt arXiv:1504.01481}}].

\bibitem{Dey:2014tka}
A.~Dey, A.~Hanany, P.~Koroteev, and N.~Mekareeya, {\it {Mirror Symmetry in
  Three Dimensions via Gauged Linear Quivers}},  {\em JHEP} {\bf 06} (2014)
  059, [\href{http://arxiv.org/abs/1402.0016}{{\tt arXiv:1402.0016}}].

\bibitem{Kapustin:2010xq}
A.~Kapustin, B.~Willett, and I.~Yaakov, {\it {Nonperturbative Tests of
  Three-Dimensional Dualities}},  {\em JHEP} {\bf 10} (2010) 013,
  [\href{http://arxiv.org/abs/1003.5694}{{\tt arXiv:1003.5694}}].

\bibitem{Dey:2011pt}
A.~Dey, {\it {On Three-Dimensional Mirror Symmetry}},  {\em JHEP} {\bf 04}
  (2012) 051, [\href{http://arxiv.org/abs/1109.0407}{{\tt arXiv:1109.0407}}].

\bibitem{Dey:2013nf}
A.~Dey and J.~Distler, {\it {Three Dimensional Mirror Symmetry and Partition
  Function on $S^3$}},  {\em JHEP} {\bf 10} (2013) 086,
  [\href{http://arxiv.org/abs/1301.1731}{{\tt arXiv:1301.1731}}].

\bibitem{Assel:2015hsa}
B.~Assel, N.~Drukker, and J.~Felix, {\it {Partition functions of 3d $\hat
  D$-quivers and their mirror duals from 1d free fermions}},  {\em JHEP} {\bf
  08} (2015) 071, [\href{http://arxiv.org/abs/1504.07636}{{\tt
  arXiv:1504.07636}}].

\bibitem{Tachikawa:2019dvq}
Y.~Tachikawa and G.~Zafrir, {\it {Reflection groups and 3d $\mathcal{N}\ge $ 6
  SCFTs}},  \href{http://arxiv.org/abs/1908.03346}{{\tt arXiv:1908.03346}}.

\bibitem{Beem:2013sza}
C.~Beem, M.~Lemos, P.~Liendo, W.~Peelaers, L.~Rastelli, and B.~C. van Rees,
  {\it {Infinite Chiral Symmetry in Four Dimensions}},  {\em Commun. Math.
  Phys.} {\bf 336} (2015), no.~3 1359--1433,
  [\href{http://arxiv.org/abs/1312.5344}{{\tt arXiv:1312.5344}}].

\bibitem{Pan:2019bor}
Y.~Pan and W.~Peelaers, {\it {Schur correlation functions on $S^3\times S^1$}},
   {\em JHEP} {\bf 07} (2019) 013, [\href{http://arxiv.org/abs/1903.03623}{{\tt
  arXiv:1903.03623}}].

\bibitem{Dedushenko:2019yiw}
M.~Dedushenko and M.~Fluder, {\it {Chiral Algebra, Localization, Modularity,
  Surface defects, And All That}},  \href{http://arxiv.org/abs/1904.02704}{{\tt
  arXiv:1904.02704}}.

\bibitem{Dedushenko:2019mzv}
M.~Dedushenko, {\it {From VOAs to short star products in SCFT}},
  \href{http://arxiv.org/abs/1911.05741}{{\tt arXiv:1911.05741}}.

\bibitem{Pan:2019shz}
Y.~Pan and W.~Peelaers, {\it {Deformation quantizations from vertex operator
  algebras}},  \href{http://arxiv.org/abs/1911.09631}{{\tt arXiv:1911.09631}}.

\bibitem{Dedushenko:2019mnd}
M.~Dedushenko and Y.~Wang, {\it {4d/2d $\rightarrow $ 3d/1d: A song of
  protected operator algebras}},  \href{http://arxiv.org/abs/1912.01006}{{\tt
  arXiv:1912.01006}}.

\bibitem{Dimofte:2018abu}
T.~Dimofte and N.~Garner, {\it {Coulomb Branches of Star-Shaped Quivers}},
  {\em JHEP} {\bf 02} (2019) 004, [\href{http://arxiv.org/abs/1808.05226}{{\tt
  arXiv:1808.05226}}].

\bibitem{Minahan:1996fg}
J.~A. Minahan and D.~Nemeschansky, {\it {An N=2 superconformal fixed point with
  E(6) global symmetry}},  {\em Nucl. Phys.} {\bf B482} (1996) 142--152,
  [\href{http://arxiv.org/abs/hep-th/9608047}{{\tt hep-th/9608047}}].

\bibitem{Minahan:1996cj}
J.~A. Minahan and D.~Nemeschansky, {\it {Superconformal fixed points with E(n)
  global symmetry}},  {\em Nucl. Phys.} {\bf B489} (1997) 24--46,
  [\href{http://arxiv.org/abs/hep-th/9610076}{{\tt hep-th/9610076}}].

\end{thebibliography}\endgroup

\end{document}